\documentclass[a4paper,11pt]{article}
\usepackage{jheppub}
\usepackage{lineno}

\usepackage{braket}
\usepackage{mathtools}
\usepackage[svgnames]{xcolor}
\usepackage{amssymb}
\usepackage{amsmath}
\usepackage{subcaption}
\usepackage{bbm}
\usepackage{hyperref}
\usepackage{graphicx}
\usepackage{float}
\usepackage{tikz}
\usepackage{comment}
\usepackage{enumitem}
\usepackage{xcolor}
\usepackage{xspace}
\usepackage{booktabs}

\usepackage{etoolbox}

\usetikzlibrary{calc,arrows.meta,decorations.markings}

\def\quiverL{3.6}     % distance between node centres (cm)
\def\quiverR{0.66}    % node radius (cm)
\def\quiversep{0.1}   % spacing between parallel arrows (cm)

\newcommand{\bundle}[3]{%
  \foreach \k in {1,...,#3}{%
    \pgfmathsetmacro{\d}{(\k-(#3+1)/2)*\quiversep}%
    \pgfmathsetmacro{\h}{sqrt(\quiverR*\quiverR-\d*\d)}%
    \draw[quiverarrow]
      ($ ($(#1)!\h cm!(#2)$) !\d cm!90:(#2) $) --
      ($ ($(#2)!\h cm!(#1)$) !\d cm!-90:(#1) $);
  }%
}

\newcommand{\Kahler}{K\"ahler\xspace}

\newlength{\hodgecellwd}
\newcommand{\hodgecell}[1]{\makebox[\hodgecellwd][c]{\ensuremath{#1}}}

\title{
BPS Chaos and Fortuity in Higgs Branches I: \\ The Gepner Point
}

\author[a]{Micha Berkooz,}
\author[a]{Trivko Kukolj,}
\author[b,c]{Ohad Mamroud}
\author[a]{and Josef Seitz}

\affiliation[a]{Department of Particle Physics and Astrophysics, Weizmann Institute of Science, Rehovot
7610001, Israel}
\affiliation[b]{SISSA, via Bonomea 265, 34136 Trieste, Italy}
\affiliation[c]{Laboratory for Theoretical Fundamental Physics, Institute of Physics,
\'Ecole Polytechnique F\'ed\'erale de Lausanne (EPFL), CH-1015 Lausanne, Switzerland}

\emailAdd{micha.berkooz@weizmann.ac.il}
\emailAdd{trivko.kukolj@weizmann.ac.il}
\emailAdd{ohad.mamroud@epfl.ch}
\emailAdd{josef-emanuel.seitz@weizmann.ac.il}

\abstract{The BPS states in Calabi--Yau compactifications of type II string theory can sometimes be described using quiver quantum mechanics whose Higgs branch is a complete intersection manifold. Their degeneracy resembles that of BPS black holes. A small fraction of these states corresponds to multi-center supergravity brane configurations and can be continued to the (near-Higgs) Coulomb branch. We identify them as monotone states. The overwhelming majority of states, known as pure-Higgs states, are fortuitous states, which concentrate at a particular $R$-charge. This division has a precise mathematical meaning as two different types of cohomologies of the Higgs branch target space. We argue that a natural probe of BPS chaos in these models is the Berry/Chern curvature. In the case of the Gepner model, we use localization techniques in an auxiliary two-dimensional CFT to compute the curvature, its RMT properties, and deviations from it.}

\begin{document}
\maketitle
\flushbottom

\section{Introduction}\label{intro}

In a theory of quantum gravity, a black hole is a thermodynamic object with a large number of microstates which share similar quantum numbers. In some cases, string theory can give a concrete realization of the microscopic theory in some limits, which provides physical intuition for the microstates. When the black holes are supersymmetric, this can be made more quantitative: certain asymptotically flat supersymmetric black holes can be understood as bound states of D-branes (\cite{Strominger:1996sh,Maldacena:1997de} and considerable subsequent work, see e.g. the review \cite{Peet:2000hn}). At strong coupling, the IR-description is that of a black hole with macroscopic entropy. At weak coupling, an object with the same charges can be constructed using branes wrapped around internal directions. The low-energy description is given by the effective field theory for fields living on the branes, and the (log of the) number of ground states of that theory matches the black hole entropy. 

Sometimes, the low-energy effective description has many parameters that are hard to, or cannot, be determined explicitly and which fluctuate considerably over the moduli space of the string background. In this case, one can hope to deal with the system using some statistical methods. In particular we would like to gain insight into the following two issues:
\begin{itemize}
    \item The effective low-energy models are often many-body interacting frustrated systems. The statistics of close-by energy levels of observables, in cases where they are understood, satisfy Wigner--Dyson $\beta$-ensemble RMT universality \cite{Cotler:2016fpe}, as is expected from any non-integrable system \cite{Bohigas1984CharacterizationOC}. However, this cannot be the full story. Such matrix models do not have, or are essentially fine-tuned to have \cite{Jafferis:2022uhu,Jafferis:2022wez}, an intrinsic notion of ``simple operators'' which have a maximal chaos exponent \cite{Maldacena:2015waa}, which appears naturally in the gravitational picture.
    For concrete string-theoretic black holes, we would like to understand what makes the microscopic structure of both the Hamiltonian and the observables special (for example, structures similar to SYK, but concretely embedded in string theory). 
    
    \item  In addition to states corresponding to black holes, there are other states with the same quantum numbers, corresponding to multi-graviton states at weak coupling. The expectation in the literature is that these two classes of states will differ in how the expectation values of simple operators in the theory behave when projected onto either class. For supersymmetric states, this distinctive feature goes by the name of BPS chaos \cite{Lin:2022rzw,Lin:2022zxd,Chen:2024oqv,Chen:2026vml}. A related issue is that of the classification of monotone and fortuitous states \cite{Chang:2024zqi,Chang:2022mjp}.
\end{itemize}
We would like to see how these phenomena appear in some popular stringy models of black holes. We concretely address these issues in the D-brane models, and we provide some hints as to deviations from RMT. 

A construction that exhibits both black hole microstates and other states is provided by multi-center brane configurations \cite{Denef:2002ru,Bena:2006kb,Denef:2007vg,Bena:2012hf,Manschot:2012rx,Lee:2012sc,deBoer:2008zn,deBoer:2009un,Martinec:2015pfa,Balasubramanian:2006gi,Gimon:2009gk,deBoer:2008ss,Bena:2011zw}. Consider type IIB string theory on $\mathbb R^{1,3} \times CY_3$, with D-branes wrapping different three-cycles in the internal space. In one description, the branes are viewed as four-dimensional particles, and they form bound states due to electromagnetic and gravitational interactions. In another description, the field theory of open strings on the branes reduces in the IR to a quiver quantum mechanics (see also \cite{Douglas:1996sw}). For example, consider the case where there are three branes intersecting each other a large number of times.\footnote{We note that there might be other brane configurations with the same total charge, and so the quiver QM does not exhaust all possible states of such charge in the theory, and represents only a specific sector. That sector could, in principle, mix with others. In fact, such other sectors are expected to exist, as there is no known choice of internal manifold $CY_3$ where the entropy of a quiver QM reproduces the full black hole entropy of such charges. Nonetheless, we concentrate on it as it already exhibits macroscopic entropy and a notion of fortuity. See \cite{Shi:2026rbr} for a related discussion. \label{foo:caution}} The quiver quantum mechanics therefore has a large number of bifundamental fields $x$, $y$, $z$, and a superpotential 
\begin{equation}
    W(x,y,z) \equiv \sum_{i = 1}^{a}\sum_{j = 1}^{b}\sum_{k = 1}^{c}f_{ijk}x_i y_j z_k\,,
    \label{eq:quiver-chord-superpotential-intro}
\end{equation}
where $a,b,c$ denote the number of intersections between pairs of branes, which are proportional to their charges.\footnote{In \cite{Anninos:2016szt}, a disordered version of the model without the gauge redundancy was considered.} Its $R$-symmetry is $U(1)_R \times SU(2)_J$, where the Abelian part is emergent, and the non-Abelian part is identified with the spatial isometries of the non-compact $\mathbb R^{1,3}$ in which the branes live. In a further IR limit, $g_{YM}^2 \rightarrow \infty$ and the low-energy description is a nonlinear sigma model whose target space is the classical space of Higgs branch supersymmetric vacua $X$, which is a complete intersection space. The parameters $f_{ijk}$ describe the (complex structure) moduli of $X$. 

\medskip
\noindent The points that we make in this paper are the following:

\paragraph{Fortuity in quiver QM.} The supersymmetric states are cohomologies of $X$ (the Higgs branch moduli space), and they come in two kinds.  The first kind descend directly from the cohomologies of the space in which $X$ is embedded, and are called \emph{ambient} cohomologies. The second kind depend on the moduli in a more sensitive way, and are called \emph{primitive} (or \emph{variant}) cohomologies \cite{Bena:2012hf}. 

In the context of the quiver quantum mechanics, these correspond to two different types of states. The ambient cohomologies correspond to states that can be described on the Higgs branch using a set of collective coordinates that capture the locations of the branes in spacetime. They map to gravitational bound states of relatively large separations that solve Denef's equations \cite{Denef:2002ru}, and are (imprecisely\footnote{See Section~\ref{subsec:higgs branch fortuity}.}) referred to as \emph{Coulomb} states (and should be more accurately called near-Higgs Coulomb states). The primitive cohomologies correspond to states which do not admit an interpretation as well-separated bound state configurations. One uses them to model single-center black hole microstates, when there is a scaling regime in Denef's equations. They are usually referred to as \emph{pure-Higgs} states.

The number of Coulomb states grows polynomially in the total brane charges, and they come in nontrivial representations of the $SU(2)_J$ $R$-charge symmetry of the quiver quantum mechanics, which is identified with the angular momenta in the four non-compact dimensions in which the branes live.\footnote{The configurations are static but carry nonzero angular momentum because of dyonic interactions.} On the other hand, the number of pure-Higgs states grows exponentially, and they exhibit $R$-charge concentrations, as they are all in the singlet of $SU(2)_J$.

Moreover, the ambient/Coulomb and primitive/pure-Higgs states can also be regarded as \emph{monotone} and \emph{fortuitous} states, respectively, as also noted by \cite{Chryssanthacopoulos:2026mni}. In holographic conformal field theories, the holographic limit involves taking the number of degrees of freedom, $N$, to be large, and therefore concerns a sequence of theories with increasing $N$. It was then noted that there are often two types of supersymmetric states.\footnote{More precisely, the mapping applies to the $Q$-cohomologies, where $Q$ is the supercharge. These cohomologies are in one-to-one correspondence with the supersymmetric states, but the representatives may change in a complicated way as we change $N$.} The first kind are the states that remain supersymmetric for any value of $N$, called monotone states. The space of these states in the $N$-th theory in the sequence can be naturally embedded into those with higher $N$s. The second kind are fortuitous states \cite{Chang:2024zqi}, whose supersymmetry depends on a relation between the degrees of freedom we use to define the theory, such as trace relations in gauge theories \cite{Chang:2022mjp,Choi:2022caq,Kim:2026mch,Kibe:2026qid,Benjamin:2026fortuity} (see also e.g. \cite{Belin:2025hsg,Behan:2025hbx,Choi:2025pqr,Choi:2026faq,deMelloKoch:2025cec,Fliss:2026rad,Kim:2025vup,Chen:2025sum,Chae:2026vgn,Chryssanthacopoulos:2026mni}). These relations often hold only below some critical $N$, and so these fortuitous states cannot be naturally mapped into theories with larger $N$. In the case of the primitive/pure-Higgs states, the relevant relations are not trace relations, but rather polynomial relations in the Jacobian ring associated with the cohomology, as we describe below. In various examples, the monotone states were interpreted to describe horizonless geometries \cite{Chang:2024zqi,Chang:2025rqy,Hughes:2025tdy,Choi:2025pqr,Djukic:2026ejk,Djukic:2026phb}. Black hole microstates, on the other hand, are believed to be fortuitous states.\footnote{In $\mathcal N = 4$ super Yang--Mills, the entropy of $1/16$-BPS black holes is much larger than that of the monotone states in the theory, so it is assumed that fortuitous states are needed to account for it \cite{Chang:2024zqi}.}

\paragraph{BPS chaos and the Berry curvature.} We can further ask about BPS chaos, and whether the chaotic properties of the ambient/Coulomb/monotone states differ from those of the primitive/pure-Higgs/fortuitous ones. It was recently suggested that a natural observable that could probe BPS chaos is the mixing of the BPS states under adiabatic changes of the couplings of the theory \cite{Chen:2026vml}, i.e. the non-Abelian Berry curvature of the BPS subspace.

In our quiver quantum mechanics, the superpotential couplings $f_{ijk}$ provide a natural set of parameters to vary. In the Higgs description, they induce deformations of the complex structure of $X$. 
The BPS spaces correspond to the cohomology $H^\bullet(X)$. The Chern connection tracks these spaces as the complex structure is changed, and is the mathematical realization of the Berry connection. 
Our physical question then becomes a mathematical one: how does the Chern curvature of the cohomology bundle over the moduli space of complete intersections differ between the ambient and the primitive cohomologies?

As we review below, the Berry/Chern curvature $F$ has a simple form, resembling the tt$^*$ structure of \cite{Cecotti:1991me,Papadodimas:2009eu,Baggio:2017aww},
\begin{equation}
    F_{I\bar J} = [\theta_I , \theta_{\bar J}^\dagger] \,.
    \label{eq:F_tt*}
\end{equation}
Here $I, \bar J$ denote the directions by which we deform the model and along which we compute the curvature. As we explain in Section~\ref{subsec: 2D CFT calculations}, the $\theta$'s are identified with the chiral ring structure constants of an auxiliary 2d $\mathcal N = (2,2)$ CFT with the same superpotential as our QM. The $f_{ijk}$ are then exactly marginal couplings, and the moduli space is a 2d conformal manifold. The Berry curvature criterion we study then agrees with \cite{Chen:2026vml}.

The curvature \eqref{eq:F_tt*} turns out to be a commutator of two sparse rectangular matrices. If there is BPS chaos, we expect some features in the spectral statistics that are reminiscent of other structured chaotic systems like SYK. 
In the triangular quiver, the computation of $F$ is rather involved, and we leave it for future work \cite{future-black-hole-paper}. However, the notion of Coulomb and pure-Higgs states extends to many other quiver QM theories, and to other supersymmetric quantum mechanical nonlinear sigma models. In Section~\ref{sec: the (hyper)quintic at the Gepner point} we study a simpler example as a case study, where the target space is a Calabi--Yau hypersurface $X$ at a point in moduli space with large discrete symmetry, specified by the superpotential
\begin{equation}
    \mathcal{W}_0(x)= \frac{1}{n+2} \sum_{i=1}^{n+2} x_i^{n+2} \,.
\end{equation}
In this case, we can find an analytic expression for the Berry curvature. Via exact diagonalization, we find level repulsion and a ramp in the spectral form factor whose length increases with the system size. We also find features of the spectral statistics beyond RMT that can be explained by the sparsity, and model them using what we call deformed Wishart ensembles \cite{Alt_2017,hachem2005empiricaldistributioneigenvaluesgram,Hachem_2007}.

\paragraph{Deviations from RMT.} One of our motivations for studying BPS chaos in these models is that physical systems are often many-body systems, like SYK or our quiver QM, where there are a number of fundamental constituents out of which the Hilbert space is constructed. ``Simple'' operators, used to construct the Hamiltonian or observables, are defined with respect to these basic constituents, and constitute a small fraction of the allowed Hermitian operators on the Hilbert space. Often the Hamiltonian is a sum of simple operators with different couplings. A large class of interesting many-body models is that of ``embedded ensembles'' \cite{kota2014embedded,Benet:2002br} where the couplings are random independent Gaussian variables.\footnote{These models often have a double scaling limit, where a chord structure appears \cite{erdHos2014phase,Berkooz:2018qkz,Berkooz:2018jqr,Berkooz:2024lgq}. 
}

The resemblance of simple operators, or their projections\footnote{In some cases, the projection itself is the origin of the randomness \cite{Belin:2026ssc}.} onto a subspace, to a random matrix is usually diagnosed via their level statistics by checking (usually numerically) whether their eigenvalues exhibit level repulsion. For nearby eigenvalues this is almost a universal feature for any non-integrable system \cite{Bohigas1984CharacterizationOC}. One can then ask how far the spectral rigidity extends throughout the spectrum. The system is considered to be strongly chaotic if the rigidity extends to an $O(1)$ fraction of the spectrum, and weakly chaotic otherwise \cite{Gharibyan:2018jrp}. Because of the small number of parameters in their Hamiltonian, many-body systems often exhibit additional features (corresponding to global fluctuations of all eigenvalues) that mask the rigidity at large energy separations and short time scales \cite{Berkooz:2020fvm,Jia:2019orl,Caceres:2022kyr,Altland:2017eao,Altland:2024ubs}.

While these features can be suppressed in certain observables, as when adding a Gaussian filter to compute the spectral form factor \cite{Gharibyan:2018jrp}, they are caused by underlying physics and affect other observables. For example, systems with gravitational duals exhibit a maximal quantum Lyapunov exponent for simple operators \cite{Maldacena:2015waa}. In many-body systems, this is related to the fact that the degrees of freedom that constitute the simple operators are the same ones that constitute the Hamiltonian, and so the two operators are naturally correlated in a complicated way. Such correlations are very difficult to achieve in the framework of Wigner--Dyson random matrix theory (even for multi-matrix ones \cite{Jafferis:2022uhu,Jafferis:2022wez}). So for the purpose of black hole physics, going beyond RMT is essential.

In this work we study examples inspired by actual string-theoretic black holes, and we identify what is special about them in comparison to, say, the Wigner--Dyson $\beta$-ensembles. We will also see below that for the quintic-like models of Section~\ref{sec: the (hyper)quintic at the Gepner point} we naturally run into a Wishart embedded ensemble.

\subsection{Outline of the paper}

In Section~\ref{sec:motivation} we review the brane construction motivating our work, the quiver quantum mechanics arising from it, and the two types of supersymmetric states it has: the near-Higgs Coulomb states and the pure-Higgs states. We then point out the analogy with monotone and fortuitous states, respectively, for such quantum-mechanical models. We also briefly introduce the quintic-like models, which are based on a Calabi--Yau hypersurface inside $\mathbb{CP}^{n+1}$.

We begin Section~\ref{sec:complete intersection spaces and their cohomologies} with a summary of formulas required for the computations later in the paper. We then provide a more extensive and pedagogical review of the BPS states of a nonlinear sigma model whose target space is a complete intersection manifold, and of the Berry/Chern curvature on the moduli space of complex structures. In the case where the complete intersection manifold $X$ is Calabi--Yau, motion on the moduli space of complex structures is related to multiplication in the ring of cohomologies, and one can think of the involved data as related to the chiral ring of a 2d $\mathcal{N}=(2,2)$ SCFT. From the physics point of view, localization arguments together with the Calabi--Yau/Landau--Ginzburg correspondence then allow us to derive formulas for the Chern curvature.

In Section~\ref{sec: the (hyper)quintic at the Gepner point}, we discuss the calculation of the Berry curvature and its structure in the simplest case, which is the quintic-like model at a point in moduli space with large discrete symmetry. The auxiliary 2d CFT is a product of minimal models, which is a generalization of the Gepner point for the quintic threefold. We are able to find an analytic expression for the Chern curvature, and numerically compute its spectral properties for deformations along random directions. 

In the spectral form factor and number variance, we find BPS chaos as well as signatures beyond RMT, typical of SYK-like systems. The spectral density exhibits several features which we are able to explain using a semi-analytic approach based on deformed Wishart ensembles.

\section{Fortuity in Quiver Black Holes}\label{sec:motivation}

Our eventual goal is to study the statistics of black hole observables which are concretely built in string theory. Since we will be revisiting models that were constructed some time ago, we will present in this section an extended introduction to these models. We can then rephrase what is known about them in a more current language, and be more precise about our goals in this paper.

\subsection{String theory constructions and quiver quantum mechanics}\label{subsec: quiver QM}

To study a system containing both black hole microstates and perhaps more general bound states, we consider type IIB string theory on $\mathbb R^{1,3} \times CY_3$, where $CY_3$ is a Calabi--Yau threefold with volume $V_6 \equiv L_X^6$ and (normalized) holomorphic three-form $\widehat \Omega$. We denote the string length by $\ell_s$ and the asymptotic string coupling by $g_s$. We will work at weak string coupling, $g_s \ll 1$, and take $\ell_s \ll L_X$. We also add D3-branes wrapped around compact special Lagrangian three-cycles $\Gamma_p \subset CY_3$ at positions $\mathbf{x}_p$ in the non-compact space, so they are separated by $x_{pq} = |\mathbf{x}_p - \mathbf{x}_q|$. We would like to find the supersymmetric ground states of the system, and the appropriate effective description depends on how well separated the branes are.

\paragraph{Gravitating bound states.} At large separation $x_{pq} \gg L_X$, the appropriate low-energy bulk theory is four-dimensional $\mathcal N = 2$ supergravity with $n_v = h^{2,1}(CY_3)$ vector multiplets, and the branes appear there as point particles. These particles are generically dyonic, with electric and magnetic charges determined by the homology classes of the wrapped cycles. The corresponding central charges in the supersymmetry algebra are $Z_p=\int_{\Gamma_p}\widehat{\Omega}$, and they determine the BPS masses through $M_p=|Z_p|/\ell_P$, with $\ell_P \propto g_s \ell_s^4 / L_X^{3}$ the four-dimensional Planck length \cite{Becker:1995kb,Denef:2000nb}. States in the theory can also carry angular momentum, coming from the $SU(2)_J$ isometries of the four-dimensional non-compact space.
These multi-centered configurations can form supersymmetric bound states, where the brane particles are in specific wave functions. The spectrum of bound states depends on the charges and couplings, and each has a typical set of positions and distances $x_{pq}$. When a set of triangle inequalities is fulfilled, the moduli space of multi-centered configurations and their bound states includes continuous families of ``scaling solutions'', where the separations between centers (and the configuration's angular momentum) approach zero with finite limiting ratios. These solutions develop an increasingly deep throat \cite{Denef:2002ru,Denef:2007vg,Bena:2006kb,Bena:2007qc}, which motivated studying their relation to single-center black holes \cite{Bena:2012hf}.

\paragraph{The open string picture.} To better understand the microscopics of the model we bring the branes close together in the non-compact directions, $x_{pq} \ll \ell_s$, where the same system can be described using the light open strings ending on the wrapped branes \cite{Denef:2002ru}. The Calabi--Yau compactification preserves eight real supercharges, of which each isolated BPS brane preserves four. The branes are particles in 4d and hence the low-energy theory is a 0+1d $\mathcal N = 4$ supersymmetric quantum mechanics. The light bosonic worldvolume modes include the three scalars describing the transverse displacements of each brane in the non-compact $\mathbb R^3$, which become three real scalars $\mathbf{x}_p(t)$ describing the position of the wrapped brane, and the worldline gauge field $A_p(t)$ which enforces Gauss's law. Together with a two-component complex fermion they form a one-dimensional $\mathcal N = 4$ $U(1)$ vector multiplet. We take the cycles $\Gamma_p$ to be special Lagrangian three-spheres, so that the branes have no continuous internal deformation or Wilson line moduli and so no additional light internal degrees of freedom.\footnote{If there are such light internal degrees of freedom, they also form 1d $\mathcal N = 4$ multiplets and should be added to the theory. This is the case when $CY_3$ is the six-torus $T^6$, for example (see \cite{Shi:2026rbr}). In the quiver they correspond to adding arrows between the node and itself.}

Strings connecting different branes can supply additional light modes localized at the intersections of their wrapped cycles. Generically, the   modes form bifundamental one-dimensional chiral multiplets, each containing a complex scalar. Their masses are also affected by the central charge $Z_p$ of the branes, and the open string picture is controlled \cite{Denef:2002ru} when the central charges are approximately aligned, $|\delta_{pq}| \equiv | \arg(Z_p/Z_q) | \ll 1$.  
The chiral multiplets, together with the vector multiplets associated with the individual branes, define a quiver quantum mechanics: each brane corresponds to a node, and each intersection mode corresponds to an arrow between the appropriate nodes. The signed intersection number of the wrapped three-cycles, or equivalently the pairing of their electromagnetic charges, denoted by $\lvert\langle\Gamma_p,\Gamma_q\rangle\rvert$, counts these arrows. 

\begin{figure}[t]
\centering
\begin{tikzpicture}[
  quivernode/.style={circle, draw, fill=black!6, minimum size=2*\quiverR cm,
                     inner sep=0pt, font=\large},
  quiverarrow/.style={line width=0.4pt, postaction={decorate},
    decoration={markings, mark=at position 0.5 with
                {\arrow{Latex[length=1.1mm,width=0.75mm]}}}},
]

\node[quivernode] (G1) at (0, {\quiverL*sqrt(3)/2}) {$\Gamma_1$};
\node[quivernode] (G2) at ({-\quiverL/2}, 0)       {$\Gamma_2$};
\node[quivernode] (G3) at ({\quiverL/2}, 0)        {$\Gamma_3$};

\bundle{G1}{G2}{4}
\bundle{G2}{G3}{5}
\bundle{G3}{G1}{6}

\path ($ ($(G1)!0.5!(G2)$) !0.35cm!-90:(G2) $)
  node[anchor=east] {$a = |\langle \Gamma_1, \Gamma_2 \rangle|$};
\path ($ ($(G2)!0.5!(G3)$) !0.5cm!-90:(G3) $)
  node[anchor=north] {$b = |\langle \Gamma_2, \Gamma_3 \rangle|$};
\path ($ ($(G3)!0.5!(G1)$) !0.35cm!-90:(G1) $)
  node[anchor=west] {$c = |\langle \Gamma_3, \Gamma_1 \rangle|$};

\end{tikzpicture}
\caption{The simplest quiver diagram with a closed loop. Each node corresponds to a $U(1)$ gauge group and is specified by a special Lagrangian 3-cycle $\Gamma_i$. As the arrows form a closed loop, the three inner products $\langle \Gamma_1, \Gamma_2 \rangle$, $\langle \Gamma_2, \Gamma_3 \rangle$, $\langle \Gamma_3, \Gamma_1 \rangle$ all have the same sign.}
\label{fig:quiver}
\end{figure}
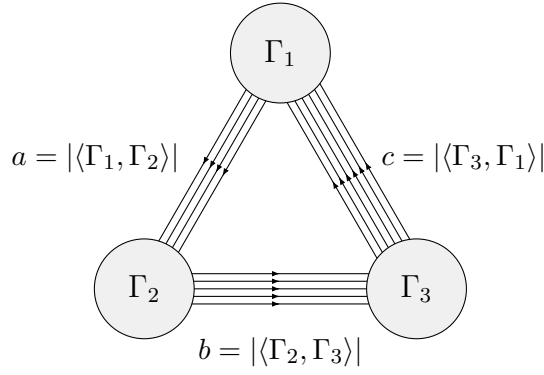

The resulting quantum mechanics has four real supercharges, and its action takes the form obtained by dimensionally reducing a four-dimensional $\mathcal N = 1$ quiver gauge theory \cite{Denef:2002ru}, whose $R$-symmetry includes the flat-space angular momentum $SU(2)_J$. The theory also has Fayet--Iliopoulos (FI) parameters $\xi_p$, determined by the central charges $Z_p$ evaluated at the asymptotic Calabi--Yau moduli. In the quiver regime of nearly aligned central charges, we use the dimensionless normalization \cite{Bena:2012hf}
\begin{equation}
    \xi_p \equiv \operatorname{Im}\bigl(e^{-i\alpha}Z_p\bigr) = |Z_p|\sin\bigl(\arg (Z_p) - \alpha\bigr) \,, \qquad \alpha \equiv \arg\bigl(\sum_q Z_q\bigr) \,.
    \label{eq:quiver-FI-central-charges}
\end{equation}
Thus $\sum_p\xi_p = 0$, and all FI parameters vanish when the central charges have the same phase, so that the constituent branes preserve the same supersymmetry \cite{Denef:2002ru,Bena:2012hf}.

\paragraph{Closed quiver and macroscopic entropy.} Let us now specialize to a particular case where our construction involves just three branes with homology classes $\Gamma_{1,2,3}$ as in Figure~\ref{fig:quiver}.
The relevant effective theory for the open strings on the branes is a quiver quantum mechanics with three nodes, gauge group $U(1)^3/U(1)_{\mathrm{diag}}$, and an oriented triangular loop. It has $U(1)_R \times SU(2)_J$ $R$-symmetry, where the $SU(2)_J$ is identified with the four-dimensional angular momentum in the non-compact space, and the $U(1)_R$ is emergent for the light open-string modes.\footnote{Examples where it is broken at higher orders can be found in \cite{Douglas:2002fr}.}  Following the conventions of \cite{Bena:2012hf}, we denote the $a = |\langle \Gamma_1, \Gamma_2 \rangle |$ fields from node $1$ to node $2$ by $x_i$, the $b = |\langle \Gamma_2, \Gamma_3 \rangle |$ fields from node $2$ to node $3$ by $y_j$, and the $c = |\langle \Gamma_3, \Gamma_1 \rangle |$ fields from node $3$ to node $1$ by $z_k$. The most general cubic superpotential is
\begin{equation}
    W(x,y,z) \equiv \sum_{i = 1}^{a}\sum_{j = 1}^{b}\sum_{k = 1}^{c}f_{ijk}x_i y_j z_k\,,
    \label{eq:quiver-chord-superpotential}
\end{equation}
where, given a brane configuration and CY geometry, the couplings $f_{ijk}$ can be obtained in principle through explicit computation of disk three-point amplitudes \cite{Douglas:2002fr}. The three FI parameters $\xi_{1,2,3}$ introduced in \eqref{eq:quiver-FI-central-charges} obey $\xi_1 + \xi_2 + \xi_3 = 0$, leaving only two independent parameters. Writing $\lvert x\rvert^2 \equiv \sum_i\lvert x_i\rvert^2$, and similarly for $y$ and $z$, the D-term equations are
\begin{equation}
    \lvert x\rvert^2 - \lvert z\rvert^2 = -\xi_1 \,, 
    \qquad 
    \lvert y\rvert^2 - \lvert x\rvert^2 = -\xi_2 \,, 
    \qquad
    \lvert z\rvert^2 - \lvert y\rvert^2 = -\xi_3 \,,
    \label{eq:quiver-chord-D-terms}
\end{equation}
while the F-term equations are
\begin{equation}
    \frac{\partial W}{\partial x_i} = 0 \,, 
    \quad
    \frac{\partial W}{\partial y_j} = 0 \,, 
    \quad
    \frac{\partial W}{\partial z_k} = 0 \,.
    \label{eq:quiver-chord-F-terms}
\end{equation}

\subsection{Fortuity lives on the Higgs branch}\label{subsec:higgs branch fortuity}

Supersymmetric states of the theory can be described either in terms of vector-multiplet (Coulomb) or chiral-multiplet (Higgs) variables, each valid in a different region of field space \cite{Denef:2002ru,Bena:2012hf}. Where the masses of the chiral fields $\sim x_{pq}$ are large compared to the energies of interest, they can be integrated out, giving a Coulomb-branch description with a one-loop corrected \Kahler metric of the form $\frac{1}{g_{YM}^2}+\frac{\kappa}{x_{pq}^3}$, as can be made precise in the D0--D4 system \cite{Douglas:1996yp,Diaconescu:1997ut}. Its supersymmetric solutions are the solutions of Denef's equations which also control the multi-center BPS configurations discussed above \cite{Denef:2002ru}. The first-order BPS conditions do not involve the metric and therefore exist everywhere on the Coulomb branch. In particular, the scaling solutions exist whenever the triangle inequalities $a+b \geq c$, $b+c \geq a$, $c+a \geq b$ are fulfilled.

In the IR, $g_{YM}^2\rightarrow \infty$, and different scalings of the fields and FI parameters with $g_{YM}$ select different low-energy limits \cite{Bena:2012hf}. Holding $x_{pq}/g_{YM}$ fixed isolates the far Coulomb branch: the metric becomes flat and the vector multiplets are free except for a singularity at the origin. This sector decouples from the Higgs branch. Holding instead $x_{pq}$ fixed removes the tree-level term and only leaves the 1-loop $1/x_{pq}^3$ correction. This is the near-Higgs Coulomb branch of \cite{Bena:2012hf}. Its dynamics is scale-invariant and the small separation region develops an infinitely long throat. Quantizing the BPS solution, one can argue that the throat is being capped \cite{deBoer:2008zn}. However, using the Coulomb degrees of freedom is not justified in this scaling, as the chiral fields become light there. One could at best hope that the description of the supersymmetric ground states survives.

Instead, this limit is more accurately described in Higgs branch variables: the kinetic term of the vector multiplet $1/g_{YM}^2$ vanishes, making these fields auxiliary while the chiral fields remain dynamical. At low energies the theory reduces to a supersymmetric sigma model on the Higgs branch. With an appropriate scaling of the FI parameters, this limit retains multi-center bound states, whose relative positions $\mathbf{x}_{pq}$ are represented by composite operators, and these states survive within the Higgs branch. The Higgs theory also contains additional ``pure-Higgs'' states that are not accounted for by the multi-center Coulomb quantization \cite{Bena:2012hf}.
The naive ``near-Higgs Coulomb'' branch is thus to be seen as part of the Higgs branch. 

Classically, the Higgs branch consists of coincident branes whose bifundamental scalars obey \eqref{eq:quiver-chord-D-terms} and \eqref{eq:quiver-chord-F-terms}, modulo the two effective $U(1)$ gauge transformations. For generic $f_{ijk}$, the simultaneous solutions form three branches, with one set of bifundamental fields vanishing on each of them \cite{Denef:2007vg,Bena:2012hf}. The FI chamber selects which scalar vanishes. For example, when $\xi_1 < 0$ and $\xi_2 < 0$ we have $z = 0$. The other choices are obtained by cyclic relabeling. On this branch the D-term quotient gives $\mathbb{CP}^{a-1}_x \times \mathbb{CP}^{b-1}_y$, and the only remaining F-terms are the $c$ equations $\partial_{z_k}W = 0$. The Higgs branch is consequently the complete intersection
\begin{equation}
    X^c_{ab} \equiv \left\{\partial_{z_k}W = 0\,,\ k = 1,\ldots,c\right\} \subset \mathbb{CP}^{a-1}_x \times \mathbb{CP}^{b-1}_y \,,
    \label{eq:quiver-chord-general-Higgs-branch}
\end{equation}
whose dimension is $n \equiv \dim_{\mathbb{C}}X^c_{ab} = a + b - c - 2$.

Quantization identifies the Higgs branch BPS Hilbert space with the cohomology $H^\bullet(X^c_{ab})$ \cite{Bena:2012hf}. To be more precise, this cohomology is identified with the $Q$-cohomology in the supersymmetric quantum mechanics, where $Q$ is a nilpotent supercharge. Each such $Q$-cohomology has a single BPS state annihilated by both $Q$ and $Q^\dagger$, which is the harmonic representative of a cohomology class in $H^\bullet(X^c_{ab})$. The Higgs branch BPS states decompose into the states that have a Coulomb branch counterpart, i.e., the multi-center solutions, and the pure-Higgs states,
\begin{equation}
    \mathcal{H}_{\mathrm{BPS}} = \mathcal{H}_{\mathrm{C}} \oplus \mathcal{H}_{\mathrm{pH}} \,.
    \label{eq:quiver-chord-BPS-decomposition}
\end{equation}
In the math literature, the $Q$-cohomology classes corresponding to Coulomb states are called the ambient cohomologies, as they are generated by restricting products of the \Kahler classes of $\mathbb{CP}^{a-1}_x \times \mathbb{CP}^{b-1}_y$, modulo the cohomological relations on $X^c_{ab}$. They are hence directly determined by the ambient space $\mathbb{CP}^{a-1}_x \times \mathbb{CP}^{b-1}_y$ in which the complete intersection $X^c_{ab}$ is embedded, and belong to $H^{p,p}(X)$, $p=0,\dots,n$. These classes are largely independent of the precise couplings $f_{ijk}$. The number of Coulomb states grows polynomially when $a,b,c$ are scaled together as 
\begin{equation}
    \log \dim(\mathcal H_C) = \frac23 \log(abc) + O(1) \,.
    \label{eq:quiver-coulomb-entropy}
\end{equation}

The pure-Higgs states form additional cohomologies in the middle degree (because of the Lefschetz hyperplane theorem), $H^{n-p,p}(X)$, and they appear for the same values of $a,b,c$ as those where the scaling solutions appear. These cohomologies are often called (middle) primitive, or variant, cohomologies. As we review below, the pure-Higgs states become cohomologies only due to the relations imposed by the F-term equations \eqref{eq:quiver-chord-F-terms}, and so they sensitively depend on the superpotential couplings $f_{ijk}$. These states are singlets under the $R$-charge $SU(2)_J$. The number of pure-Higgs states grows exponentially in $a,b,c$ \cite{Bena:2012hf}. Defining $A \equiv b + c - a$, $B \equiv c + a - b$, and $C \equiv a + b - c$, when $a,b,c$ are scaled together, the entropies then obey \cite{Bena:2012hf}
\begin{equation}
    \log \dim (\mathcal H_{\mathrm{pH}} ) = \log \Bigl[\frac{(2a)^a (2b)^b (2c)^c}{A^A B^B C^C}\Bigr] - \frac13 \log (abc) + O(1) \,.
    \label{eq:quiver-chord-entropies}
\end{equation}

This distinction has a natural analogue in the classification of BPS states as monotone or fortuitous \cite{Chang:2024zqi}. In a family of theories labeled by $N$, monotone states represent $Q$-cohomology classes inherited from a common space of $Q$-closed expressions and admit continuations to arbitrarily large $N$. Fortuitous states represent additional classes whose $Q$-closure relies on relations specific to finite $N$. In our quiver, the ambient cohomology plays the role of the common cohomology, while restriction to the F-term solutions supplies the additional relations.\footnote{In our case, the relations are those of a polynomial ring, unlike other fortuitous cases where they originate from trace relations of matrices.} Thus, we identify the Coulomb states as monotones and the pure-Higgs states as fortuitous. Besides the fact that the fortuitous states depend on constraints while the monotones do not, they have the following distinguishing features:
\begin{itemize}
    \item As the $SU(2)\sim SO(3)$ $R$-charge corresponds to spatial rotations, the pure-Higgs states are rotationally invariant, while the Coulomb states come in generically nonzero spin representations. The latter are thus easier to probe using gravitational observables.
    \item The pure-Higgs states all share the same value of $SU(2)$ $R$-charge, so they exhibit $R$-charge concentration \cite{Chang:2024lxt}, while the Coulomb (or technically near-Higgs Coulomb) states do not. $R$-charge concentration has also been observed in super JT gravity \cite{Stanford:2017thb,Heydeman:2020hhw}.
    \item The fortuitous states depend on the value of $a,b,c$; there exists no natural sequence of fortuitous cohomologies with $a,b,c$, while for the monotone states, we can formulate such a sequence by restricting the ambient cohomologies written as products of \Kahler classes.
    \item The number of pure-Higgs states grows exponentially in the dimension $n$, which is proportional to the square of the charge in the full string theory. This is the same behavior as for black holes.
\end{itemize}
Furthermore, the monotone states are sensitive to external moduli (here, if the FI parameters flip sign, their number changes), while the fortuitous states are wall-crossing invariants (see also \cite{Manschot:2012rx}, where this phenomenon is discussed more generally).
This resonates with the gravity interpretation: in four dimensions, due to the attractor mechanism, the supergravity solution close to the horizon is independent of asymptotic moduli \cite{Ferrara:1995ih,Strominger:1996kf,Ferrara:1996dd}, while multi-centered bound state solutions can decay across walls of marginal stability \cite{Denef:2000nb}.

\subsection{Fortuity beyond black holes}

This mathematical notion of fortuity, as it is introduced here, extends to many other cases where a moduli space arises as a restriction (i.e. F-term condition) of an ambient projective space (which arises due to D-term constraints). For example, one can consider a simpler one-dimensional $\mathcal N = 4$ $U(1)$ gauged linear sigma model (GLSM), with $n+2$ chiral multiplets $x_i$ of charge $+1$, one chiral multiplet $P$ of charge $-(n+2)$, and superpotential
\begin{equation}
    W_{\mathrm{GLSM}}(x,P;\zeta) \equiv P f_{\zeta}(x) \,, \qquad f_{\zeta}(x) \equiv \sum_I \zeta^I\phi_I(x) \,.
    \label{eq:hypersurface-W}
\end{equation}
Here the $\phi_I(x)$ run over all monomials of total degree $n+2$ in the $n+2$ fields $x_i$, and the $\zeta^I$ are their coefficients, so $f_{\zeta}$ is a general homogeneous polynomial of degree $n+2$. For a positive FI parameter the D- and F-term equations set $P=0$ and identify the Higgs branch with the Calabi--Yau hypersurface $X_{\zeta} = \{f_{\zeta}=0\} \subset \mathbb{CP}^{n+1}$. The low-energy theory on this branch is a supersymmetric nonlinear sigma model with target $X_{\zeta}$. The case $n=3$ gives the quintic threefold.

The BPS states in this simpler model also divide into ambient/monotone and primitive/fortuitous BPS states. The entropies\footnote{These entropies follow from \eqref{eq:hypersurface-Jacobian-Hilbert-series} and \eqref{eq:hypersurface-primitive-Hodge-degeneracy}.} of the two classes grow like $\log(n)$ and $n\log(n)$, respectively. The superpotential couplings still control the complex structure deformations in this model. Hence it mimics some of the mathematical structure of the more realistic quiver QM. 
While it is unrelated to any concrete brane construction of black holes, we will get interesting restrictions on the statistical classes of Berry curvature observables on this model. These hint towards a potentially generic k-local structure in stringy observables. One benefit of using this model is that we can explicitly compute the Berry curvature around a special point, the Gepner point where $f_\zeta = \sum_{i=1}^{n+2} x_i^{n+2}$, via localization. This will be extensively discussed in Section \ref{sec: the (hyper)quintic at the Gepner point}.

\section{BPS chaos in complete intersection manifolds}
\label{sec:complete intersection spaces and their cohomologies}

The system of interacting BPS particles described in the previous section admits a low-energy description in terms of quiver quantum mechanics. On the Higgs branch, the effective theory of the three-node quiver can be restated as a sigma model on a complete intersection manifold $X^c_{ab} \subset \mathbb{CP}^{a-1} \times \mathbb{CP}^{b-1}$, where the parameters $a, b, c$ are determined by the underlying string-theoretic realization. 

Beyond counting arguments, it is nontrivial to study the statistical properties of observables on the Higgs branch vacua in this specific model. In this section, we will first discuss the mathematical structure of the Higgs branch vacua and the Berry curvature in the parameter space for one-dimensional $\mathcal{N}=4$ sigma models whose target space is an arbitrary complete intersection manifold $X_{\zeta}$. These appear fairly generally as moduli spaces of (BPS particles of) supersymmetric gauge theories. The subscript $\zeta$ collectively denotes the coordinates on the corresponding complex structure moduli space of $X$, which we denote by $\mathcal{M}(X)\equiv\mathcal{M}$. When we specialize to the case where the complete intersection is a Calabi--Yau $n$-fold, the problem can be equivalently formulated in the language of 2d $\mathcal{N}=(2,2)$ SCFTs, for which localization formulas are available. We state such formulas for the case of a Calabi--Yau hypersurface.

\vspace{5pt}
\noindent Let us now outline the main points of this section:

\begin{itemize}
    
    \item BPS states on the Higgs branch are in one-to-one correspondence with Dolbeault cohomologies of $X_\zeta$:
    \begin{equation}
        \mathcal{H}_{BPS} \cong \bigoplus_{p,q=0}^n H^{q,p}(X_\zeta).
        \label{eq:H_BPS-cohomology}
    \end{equation}
    As mentioned earlier, the cohomologies of $X_\zeta$ can be divided into two types. The first are ambient cohomologies $H^{p,p}(X_\zeta)$, obtained by restricting the \Kahler forms of some larger ambient space\footnote{For the kinds of manifolds $X$ that we consider, the ambient space $V$ is most generally a product of projective spaces.} $V \supset X_\zeta$ to $X_\zeta$. The second are primitive cohomologies $H^{n-p,p}(X_\zeta)$, which are closed when restricted to $X_\zeta$, but are themselves not cohomologies of $V$. The former are analogous to monotone cohomologies, and the latter analogous to fortuitous cohomologies. 
    \item As the complex structure moduli $\zeta$ are varied along $\mathcal M$, the cohomology representatives change. Equivalently, as the couplings are adiabatically varied, the BPS states mix according to the quantum-mechanical Berry curvature.
    We note that this Berry connection is given by the Chern connection on $\mathcal{M}$,
    which can be defined implicitly in terms of the flat Gauss–Manin connection as follows:
    \begin{equation}
        \nabla^{\mathrm{Ch}} = \nabla^{\mathrm{GM}} - \theta - \theta^\dagger \,.
        \label{eq:GM-Chern-decomposition}
    \end{equation}
    Here, $\theta$ and $\theta^\dagger$ are the off-diagonal elements of the connection, which are called Kodaira--Spencer maps. They change the (anti)holomorphic degree by one:
    \begin{equation}
        \theta_I:H^{q,p}(X_{\zeta}) \longrightarrow H^{q-1,p+1}(X_{\zeta})\,,\qquad \theta^\dagger_{\bar{J}}:H^{q,p}(X_{\zeta}) \longrightarrow H^{q+1,p-1}(X_{\zeta})\,.
        \label{eq:Higgs-field-bidegrees}
    \end{equation}    
    We will use $I$ and $\bar{J}$ to label holomorphic and antiholomorphic directions on $\mathcal{M}$. Choosing local bases $\{\psi_\alpha^{(q,p)}({\zeta})\}_{\alpha \in \{1,...,h^{q,p}\}}$ for the cohomology groups $H^{q,p}(X_{\zeta})$ as they vary over $\mathcal M$, we denote the components of $\theta_I$ and $\theta^\dagger_{\bar J}$ by $C_{I\alpha}{}^\beta$ using:
    \begin{equation}
        \theta_I\psi_\alpha^{(q,p)} \equiv C_{I\alpha}{}^\gamma\psi_\gamma^{(q-1,p+1)} \,, 
        \quad 
        \theta^\dagger_{\bar J}\psi_\alpha^{(q,p)} \equiv G^{\gamma\bar\rho}(C_{J\rho}{}^\sigma)^*G_{\alpha\bar\sigma} \psi_\gamma^{(q+1,p-1)} \,,
        \label{eq:Higgs-field-components}
    \end{equation}
    where $G_{\alpha\bar\beta}$ denotes the canonical Hodge metric of $X_\zeta$, defined below in \eqref{eq:Hodge-metric}. The curvature of the Chern connection is then simply
    \begin{equation}
        F_{I\bar{J}} = \bigl[\theta_I,\theta^\dagger_{\bar{J}}\bigr] = \theta_I \theta^\dagger_{\bar J} - \theta^\dagger_{\bar J} \theta_I \,.
        \label{eq:Chern-curvature-commutator}
    \end{equation}
    In components, this becomes:
    \begin{equation}
        (F_{I\bar J})_\alpha{}^\beta = G^{\rho\bar\mu} G_{\alpha\bar\nu} \, (C_{J\mu}{}^\nu)^* \, C_{I\rho}{}^\beta - C_{I\alpha}{}^\lambda \, G^{\beta\bar\gamma} G_{\lambda\bar\delta} \, (C_{J\gamma}{}^\delta)^* \,.
        \label{eq:Chern-curvature-components}
    \end{equation}
    \item For the case where $X$ is Calabi--Yau, there are some simplifications. First, the complex structure deformations are themselves middle cohomologies $\in H^{n-1,1}$, and the middle cohomologies form a ring. The $C_{I \alpha}{}^{\lambda}$ are then structure constants of the ring multiplication, which are more tractable to compute. For example, for the Calabi--Yau hypersurface 
    \begin{equation}
        {X}_{\zeta}= \mathbb{CP}^{n+1}_x|_{\{\mathcal{W}_0(x;\zeta)=0\}} \,,
    \end{equation}
    (with $\mathcal{W}_0$ being a degree-$(n+2)$ polynomial in the homogeneous coordinates), one can use the Jacobian ring representation, or equivalently an auxiliary 2d $\mathcal{N}=(2,2)$ SCFT and the Calabi--Yau/Landau--Ginzburg correspondence \cite{Witten:1993yc}. For a generic superpotential, we can represent the middle cohomologies as monomials in the homogeneous coordinates $x_i$ of the ambient space $\mathbb{CP}^{n+1}$:
    \begin{equation}
        \psi_{\alpha}^{n-p,p} \leftrightarrow \prod_{i=1}^{n+2} x_i^{\alpha^i_{(p)}} \,, \qquad \sum_{i=1}^{n+2} \alpha^i_{(p)} = p(n+2) \,, \qquad 0 \leq \alpha^i_{(p)} \leq n \, .
    \end{equation}
    Then, the 2d SCFT language can be used to derive a localization formula for the (unnormalized) metric in $\mathrm{tt}^*$ fashion \cite{Cecotti:1991me,Gerchkovitz:2014gta,Ishtiaque:2017trm,Chen:2017fvl}, which reads schematically:
    \begin{equation}\label{schematically 2d localization formula}
        \begin{aligned}
            G_{\alpha_{(p)}\bar\beta_{(p)}} & = \int_{\mathbb{C}^{n + 2}} \prod_{i = 1}^{n + 2}d x_i \, d \bar x_i \, x_i^{\alpha^i_{(p)}}\bar x_i^{\beta^i_{(p)}} e^{-4\pi i \mathcal{W}_0(x) - 4\pi i \overline{\mathcal{W}}_0(\bar{x})}  \,.
        \end{aligned}
    \end{equation}    
    In general, the integral is not convergent and needs to be evaluated using thimbles.
    In the math literature, the $\mathrm{tt}^*$-structure has been explored more recently; see e.g. \cite{hertling2002ttgeometryfrobeniusmanifolds,hertling2008nilpotentorbitsgeneralizationhodge,Tang_2025}.
    There is even further simplification when focusing on the Gepner point, where $\mathcal W_0(x) = \frac{1}{n+2}\sum_{i=1}^{n+2} x_i^{n+2}$. This is the case that we will focus on in Section~\ref{sec: the (hyper)quintic at the Gepner point}. 
    There, the ring multiplication is trivial
    \begin{equation}
        C_{I\alpha}{}^{\lambda} = \delta_{I+\alpha}^{\lambda} \,,
    \end{equation}
    and the integral \eqref{schematically 2d localization formula} can be computed explicitly.

    \item After obtaining expressions for $C$ and $G$, we can then compute the Berry curvature explicitly along random directions in the moduli space, and study its spectral statistics. 
\end{itemize}
For the primitive cohomologies, $\dim \bigl( H^{n-p,p}(X_\zeta)\bigr)$ often grows exponentially as $p$ grows (for $p < n/2$), and it is much larger than the number of complex structure deformations, $\dim \mathcal M$, which essentially determines the number of ``independent parameters''. The result is that the matrices $\theta_I$, $\theta_J^\dagger$ are rectangular, and very sparse. Moreover, when one of the terms in \eqref{eq:Chern-curvature-components} has much larger rank than the other, one of the terms is a small-rank deformation of the other, and in a deformation along a random direction, $F_{I\bar J}$ looks like a sparse Wishart matrix. This will be made more concrete in Section~\ref{sec: the (hyper)quintic at the Gepner point}.

\vspace{5pt}
\noindent The remainder of this section discusses the mathematical background in more detail. Readers familiar with this outline may skip directly to Section~\ref{sec: the (hyper)quintic at the Gepner point}.

\subsection{The space of states}\label{subsec:space of states}

Consider a $(0+1)$-dimensional nonlinear sigma model with target space $X_{\zeta}$. We assume that the model has $\mathcal{N} = 4$ supersymmetry, and that $X_{\zeta}$ is a compact complete intersection \Kahler manifold. The subscript $\zeta$ collectively denotes the complex structure moduli of $X$. We will suppress this notation when it is not necessary.

\paragraph{Complete intersection spaces.} To define $X_{\zeta}$, we consider the following general case: let $f_a$ be holomorphic sections of line bundles over an ambient compact \Kahler manifold $V$. Their common zero locus is: 
\begin{equation}
    X_{\zeta} \equiv \left\{f_1(x;{\zeta}) = \cdots = f_r(x;{\zeta}) = 0\right\} \subset V \,, \qquad n \equiv \dim_{\mathbb{C}}X_{\zeta} = \dim_{\mathbb{C}}V - r \,,
    \label{eq:complete-intersection-definition}
\end{equation}
where we assume $V$ to be a product of projective spaces
\begin{equation}
    V\cong \mathbb{CP}^{n_1} \times \mathbb{CP}^{n_2}\times ... \times \mathbb{CP}^{n_k}.
\end{equation}
The $f_a$ are polynomials in the homogeneous coordinates of $V$, whose coefficients encode the complex structure $\zeta$, as we will explain in more detail in Section~\ref{subsec:Chern connection on the moduli space}. The differentials $d f_a$ are assumed to be independent along $X_{\zeta}$, so that the equations carve out a smooth submanifold of codimension $r$, and it is assumed that each $f_a$ has nonzero degree $d_a^{(i)}$ in each $\mathbb{CP}^{n_i}$ (such that the intersection is ample). The complete intersection is Calabi--Yau when $\sum_a d_a^{(i)} = n_i+1$, as its first Chern class vanishes in that case \cite{Green:1987mn}. 

The Hilbert space of supersymmetric ground states of an $\mathcal{N}=4$ sigma model on such $X$ (denoted by $\mathcal{H}_{BPS}$) is naturally identified with the Dolbeault cohomology\footnote{Quantizing the fermions identifies the Hilbert space with the space of differential forms on $X$ \cite{Witten:1982im} (see e.g. \cite{Hori:2003ic}, Chapter 10.4.3 for a review).} of $X$:
\begin{equation}
    \mathcal{H}_{BPS} \cong \bigoplus_{q,p=0}^n H^{q,p}(X) \,.
\end{equation}
We define $H^{q,p}(X)$ as the cohomology with respect to the antiholomorphic derivative $\bar{\partial}$. For complete intersections of the type \eqref{eq:complete-intersection-definition}, the cohomologies are either obtained by restriction from the ambient space cohomologies, or not. By the Lefschetz theorem, the cohomologies not obtained by restriction all sit in the middle cohomology, $H^{q,p}(X)$ with $p+q=n$. More precisely, we have that
\begin{equation}
    h^{q,p}(X) = 
    \begin{cases}
         h^{q,p}(V)\,, & p+q<n \,, \\
         h^{q,p}(V)+c^{q,p}(X)\,, & p+q=n \,, \\
         h^{n-p,n-q}(V)\,, & p+q > n \,.
    \end{cases}
\end{equation}
The Hodge numbers below the middle are identical to the ambient space Hodge numbers and the cohomology representatives can simply be obtained by restriction. Above the middle, the number can be inferred from Poincaré (or more precisely Serre) duality, and the representatives can be obtained also from the restrictions of ambient space cohomologies with some relations. As the cohomologies of the ambient space are elementary to determine, it is straightforward to count them. The number of additional middle cohomologies $c^{q,p}(X)$ depends on the dimensions of the ambient space as well as the degrees that define the polynomial constraints. As they all sit in the middle degree, their number can be inferred from computing a (twisted) Euler characteristic and subtracting the contributions of the ambient cohomologies, as discussed in the examples below. 

The additional middle cohomologies are known in the literature as primitive/variant, while the cohomologies obtained as restrictions from the ambient spaces are called ambient. As discussed in Section~\ref{sec:motivation}, these two classes can naturally be thought of as fortuitous and monotone.

In particular, there exists a natural $SU(2)$-action on the cohomologies, called the Lefschetz $SU(2)$ (see Appendix \ref{app:mathematical minimum}). From the physics point of view, this is the action of the $SU(2)_R$ symmetry of the quantum-mechanical theory. In the physical theory, the total degree $p+q$ is then the fermion number.\footnote{It is also, after a constant shift, the Cartan charge of the Lefschetz $SU(2)$. The difference $p-q$ is the Hodge grading, and depends on the choice of complex structure. In the quiver realization inherited from four-dimensional $\mathcal{N} = 2$ theories, the Lefschetz $SU(2)$ is identified with spatial angular momentum, while the grading by $p-q$ contains the remaining $R$-symmetry information.} The primitive middle cohomologies are singlets of this $SU(2)$, while the ambient ones come in nonzero spin representations.

Another motivation for the monotone--fortuitous taxonomy is that the general form of the ambient cohomologies does not sensitively depend on the dimensions of the ambient projective spaces nor on the constraints. This is best illustrated in the examples below.

\paragraph{Example 1: A hypersurface in $\mathbb{CP}^{n + 1}$.} Consider the hypersurface $X=\{f(x)=0\}\subset\mathbb{CP}^{n + 1}_x$, where $f(x)$ is a degree-$k$ polynomial. The ambient space $\mathbb{CP}^{n + 1}_x$ only contains cohomologies built from the \Kahler form $\omega\in H^{1,1}(\mathbb{CP}^{n + 1}_x)$ and its powers:
\begin{equation}
    1 \,, \ \omega \,, \ \omega^2 \,, \ \ldots \,, \ \omega^m \,, \qquad H^{p,p}(\mathbb{CP}_x^m) = \mathbb{C}\omega^p \,, \qquad H^{q,p}(\mathbb{CP}_x^m) = 0 \quad \text{for} \quad p \neq q\,.
    \label{eq:projective-space-cohomology}
\end{equation}
For the hypersurface $X \subset \mathbb{CP}_x^{n + 1}$, the ambient cohomologies are simply obtained by restricting to $X$:
\begin{equation}
    H_{\mathrm{amb}}^{p,p}(X) = \mathbb{C}(\omega|_X)^p \,, \qquad p = 0,\ldots,n\,.
    \label{eq:hypersurface-ambient-cohomology}
\end{equation}
Thus there is exactly one ambient cohomology in every diagonal Hodge group $H^{p,p}(X)$. They form a spin-$\frac{n}{2}$ representation of $SU(2)_R$. The number of primitive cohomologies can be obtained from the refined Euler characteristic after subtracting the number of ambient cohomologies. The refined Euler characteristic (also known as the Hirzebruch genus\footnote{From a physics point of view, it can be understood as a refined Witten index of the sigma model.}) is defined by
\begin{equation}
    \chi_q(X) = \sum_{p,r} q^r (-1)^{p+r} h^{r,p}(X) \,,
\end{equation}
and can be computed using the Hirzebruch--Riemann--Roch theorem. In this example, the result can be restated as the following residue integral (see Appendix One of \cite{Hirzebruch1966}):
\begin{equation}\label{hrr hypersurface}
    \chi_q(X)=\oint \frac{dt}{2\pi i} \frac{1}{t^{n+2}} \frac{(1-tq)^k-(1-t)^k}{(1-t)(1-tq)[(1-tq)^k-q(1-t)^k]} \,.
\end{equation}
Since the ambient cohomologies are known explicitly, we can subtract them to recover the primitive contribution only:
\begin{equation}\label{HRR hypersurface primitive}
    \chi_{q}^{\text{prim}}(X)=\oint \frac{dt}{2\pi i} \frac{1}{t^{n+2}} \frac{(1-tq)^{k-1}-(1-t)^{k-1}}{[(1-tq)^k-q(1-t)^k]} \,.
\end{equation}
For $k=n+2$, we can recover this result from a simple counting argument of monomials, as will be discussed in Section~\ref{subsec: 2D CFT calculations}. The primitive contribution is only from middle cohomologies in $H^{n-p,p}(X)$, and so one can read off the Hodge numbers from the coefficient of $(-1)^nq^{n-p}$. A pyramid of the resulting primitive Hodge numbers in the case $k=n+2$ is given in Table~\ref{tab:hodge-numbers-quintic}.

\paragraph{ Example 2: A complete intersection in $\mathbb{CP}^m \times \mathbb{CP}^m$.} 
Consider the case where
\begin{equation}
    X = \{f_{ijk} x^i y^j =0\} \subset \mathbb{CP}^m_{x} \times \mathbb{CP}^m_{y} \,, \quad i,j,k \in \{1,...,m+1\} \,,
\end{equation}
with $x^i$ and $y^j$ being homogeneous coordinates on the two projective spaces. This space precisely corresponds to the Higgs branch of vacua of the quiver QM black hole model described in Section~\ref{sec:motivation}, for the case of $a=b=c=m+1$, and its dimension is $n = m-1$. 

In this case, the ambient space has two independent \Kahler forms, $\omega_1$ and $\omega_2$, inherited from the two factors. Its cohomology is represented by their mixed powers,
\begin{equation}
    \omega_1^r \wedge\omega_2^s \in H^{r + s,r + s}(\mathbb{CP}^m \times \mathbb{CP}^m) \,, \qquad 0 \leq r,s \leq m \,.
    \label{eq:product-projective-space-cohomology}
\end{equation} 
The ambient classes on $X$ are the restrictions of $\omega_1^r \wedge \omega_2^s$ to the complete intersection $X$. Below the middle degree these restricted classes are independent, while at higher degree some linear combinations can become equivalent on $X$. Counting the mixed powers below the middle degree and using Poincar\'{e} duality above it gives:
\begin{equation}
    h_{\mathrm{amb}}^{p,p}(X) = \min\{p + 1,a - 1 - p\} \,, \qquad p = 0,\ldots,a - 2 \,, \qquad \sum_{p = 0}^{a - 2}h_{\mathrm{amb}}^{p,p}(X) = \left\lfloor\frac{a^2}{4}\right\rfloor \,.
    \label{eq:abc-ambient-cohomology-count}
\end{equation}
As in the previous example, one computes the Euler characteristic and gets after a straightforward calculation (see also \cite{Lee:2012naa}, with their $\xi$ equal to $-q$):
{\small
\begin{equation}
    \begin{split}
        \chi_q(X) =&  \oint \frac{dt_1}{2\pi i } \frac{1}{t_1^{m+1}} \oint \frac{dt_2}{2\pi i } \frac{1}{t_2^{m+1}} \frac{1}{(1-q t_1)(1-t_1)(1-q t_2)(1-t_2)}\left(\frac{t_1 + t_2-(1+q)t_1 t_2}{1-q t_1 t_2}\right)^{m+1}.
    \end{split}
\end{equation}
}
From this we may read off the Hodge numbers $h^{n-p,p}(X)$. By subtracting the monotone cohomology counts \eqref{eq:abc-ambient-cohomology-count}, we find the number of primitive cohomologies in each class $H^{n-p,p}(X)$. Using this formula, one can show that the total number of states grows as $\sim 8^n/n$, almost all of which come from a Gaussian centered around $h^{\frac{n}{2},\frac{n}{2}}$ with height $\sim 8^n/n^{\frac{3}{2}}$ and width $\sim \sqrt{n}$.

A similar counting can also be done for the general case where the two $\mathbb{CP}$s are not of the same dimension and the number of constraints is also different \cite{Bena:2012hf,Lee:2012naa}. 

\paragraph{The inner product.}
We can now pick a basis of (harmonic) representatives of the cohomologies $\psi_{\alpha}^{(q,p)}\in H^{q,p}(X)$. The Hilbert space is equipped with a natural inner product given by the Hermitian Hodge metric on each slice, whose components are\footnote{Here, we adopt the convention for the inner product that the first entry of the inner product is linear and the second is antilinear.}:
\begin{equation}
    G_{\alpha\bar{\beta}}\equiv G(\psi_{\alpha}^{(q,p)},\psi_{\beta}^{(q,p)}) = \int_X \psi_{\alpha}^{(q,p)} \wedge *\, \overline{\psi_{\beta}^{(q,p)}}  \,.
    \label{eq:Hodge-metric}
\end{equation}
We denote the components of the inverse by $G^{\alpha\bar\beta}$. This inner product depends on the complex structure and, in general, on the \Kahler moduli. 

Note that we distinguish the unnormalized geometric Hodge metric $G$ used above from the normalized metric $g$. When $X$ is Calabi--Yau, choose a local nowhere-vanishing holomorphic top form $\Omega_{\zeta}$ and define its Hodge norm by\footnote{Despite the notation, $\mathcal{V}_{\Omega}$ is not the \Kahler volume $\operatorname{Vol}_J(X) \equiv \frac{1}{n!}\int_X J^n$: it depends on the complex structure and on the holomorphic normalization of $\Omega_{\zeta}$, but not on the \Kahler moduli.}
\begin{equation}
    \mathcal{V}_{\Omega}(\zeta,\bar \zeta) \equiv i^{n^2}\int_{X_{\zeta}}\Omega_{\zeta}\wedge\overline{\Omega_{\zeta}} \,,
    \label{eq:holomorphic-volume-form-Hodge-norm}
\end{equation}
and the normalized metric as
\begin{equation}
    g_{\alpha\bar\beta} \equiv \frac{G_{\alpha\bar\beta}}{\mathcal{V}_{\Omega}} \,.
    \label{eq:normalized-versus-geometric-Hodge-metric}
\end{equation}
The unnormalized convention is natural for the variation of Hodge structure and keeps the relation to the Gauss--Manin connection (introduced in the next subsection) manifest. The normalized metric is invariant under a holomorphic rescaling $\Omega_{\zeta} \mapsto e^{f(\zeta)}\Omega_{\zeta}$, $ \psi_{\alpha}^{(q,p)}\rightarrow e^{f(\zeta)}\psi_{\alpha}^{(q,p)}$  and is the one naturally produced by the normalized CFT correlation functions introduced in Section~\ref{subsec: 2D CFT calculations}.

\subsection{The Chern connection on the moduli space of complex structures}\label{subsec:Chern connection on the moduli space}

Our goal is to compute the Berry curvature in the parameter space of sigma models with complete intersection target spaces. A natural set of parameters is given by the coefficients of the polynomial equations \eqref{eq:complete-intersection-definition} which determine the embedding of the complete intersection in the ambient space.
In e.g. the quiver quantum mechanics, these correspond to deformations by simple operators in the quantum mechanics: they are deformations of the superpotential. It turns out that such deformations have a natural geometric interpretation: they are complex structure deformations of the manifold $X_{\zeta}$.\footnote{For a complete intersection Calabi--Yau, this statement also holds the other way, up to unitary redefinitions of the ambient coordinates \cite{Green:1987rw}. It relies on the short exact sequence of bundles
\begin{equation}
    0 \rightarrow TX \rightarrow T V_X\rightarrow N_{X}\rightarrow 0
\end{equation}
and the corresponding sequence of cohomologies. There are also Calabi--Yau manifolds that are not, strictly speaking, complete intersections, for which the equality also does not hold \cite{Anderson:2026gvz}. The only other parameters of the sigma model are the volumes of the ambient $\mathbb{CP}$s. Variations of those are \Kahler deformations.} We are thus motivated to study motion on the moduli space of complex structures, which we call $\mathcal{M}(X_\zeta)\equiv \mathcal{M}$.

Complex structure deformations can be understood as deformations of the splitting into holomorphic and antiholomorphic directions. Away from singular loci, these deformations do not change the dimension of any cohomology $H^{q,p}(X)$, but they generically may change the $\partial$-harmonic representatives. Physically, this means that the number of BPS states does not change, but the states themselves could mix within the same total degree\footnote{They cannot mix between different total degrees because the de Rham differential $d$ is invariant under complex structure deformations.} $p+q$. This mixing is described by connections on the bundles of cohomologies over $\mathcal{M}$. 

The Berry connection corresponds to a metric-preserving connection. The metric is compatible with a holomorphic structure because of the one-dimensional $\mathcal{N}=4$ supersymmetry. There is a unique connection satisfying these properties, the Chern connection. In a local holomorphic frame, for a complex structure variation $\partial_{\zeta}$, we have 
\begin{equation}
    (\nabla^{\mathrm{Ch}})^{1,0} = \partial_\zeta + G^{-1}\partial_\zeta G , \qquad (\nabla^{\mathrm{Ch}})^{0,1} = \overline{\partial_\zeta} \,.
    \label{eq:local coordinates Chern}
\end{equation}
To compute the Berry curvature, we therefore need to compute the curvature of the Chern connection. To understand properties of the Chern connection, it is convenient to relate it to a different connection, the Gauss--Manin connection $\nabla^{\mathrm{GM}}$.

We can define it as follows: in a local basis $\{\gamma_a\}$ of fixed de Rham cohomology classes, $\nabla^{\mathrm{GM}}\gamma_a = 0$, and therefore $\nabla^{\mathrm{GM}}(v^a\gamma_a) = d v^a\gamma_a$ and the connection is flat. Thus it is the ordinary derivative in such a basis, but not in a basis adapted to the varying holomorphic decomposition, where it has some nonzero Christoffel symbols. The Gauss--Manin connection preserves the bilinear topological pairing $\tilde{\eta}$ (which implements Poincaré duality),
\begin{equation}
    \tilde{\eta}_{\beta\gamma}\equiv \tilde{\eta}(\psi_{\beta}^{(q,p)},\psi_{\gamma}^{(n-q,n-p)}) = \int_{X_{\zeta}} \psi_\beta^{(q,p)}\wedge\psi^{(n-q,n-p)}_\gamma \,,
    \label{eq:Higgs-field-components-holomorphic}
\end{equation}
but it does not preserve the Hodge structure: the decomposition into bidegrees of $\gamma_a$ changes as the splitting between holomorphic and antiholomorphic coordinates varies. More concretely, a general complex structure deformation can be denoted by a form $\mu^i{}_{\bar{j}}$ which encodes the mixing of holomorphic and antiholomorphic directions. An integrability condition 
enforces that $\mu$ is closed, meaning that, modulo trivial complex structure deformations, $\mu \in H^{0,1}(X,T^{1,0} X)$ \cite{Green:1987mn}. Such a form is known as a Beltrami differential. Consider now a basis of such elements
\begin{equation}
    \mu_I \in H^{0,1}(X,T^{1,0} X)\,, \quad I\in \{1,...,\dim \mathcal{M}\} \,. 
\end{equation}
For an infinitesimal complex structure deformation
$\zeta_0 \rightarrow \zeta_0 + \delta \zeta$, a Gauss--Manin-flat $\gamma_a$ which is $(q,p)$ at the ``old'' complex structure $\zeta_0$ picks up a $(q-1,p+1)$-component 
\begin{equation}
    \theta_I \gamma_a \equiv \delta \zeta^I \mu_I \lrcorner \gamma_a  \,,  
\end{equation}
with the symbol $\lrcorner$ denoting contraction of the upper holomorphic index of  $\mu$ with a holomorphic index of $\gamma_a$. Here we have defined the Kodaira--Spencer map $\theta_I$. One can then prove (via uniqueness, see Appendix \ref{app:mathematical minimum}) that the Chern connection is precisely the bidegree-preserving part of the Gauss--Manin connection, i.e. 
\begin{equation}
    \nabla^{\mathrm{Ch}} = \nabla^{\mathrm{GM}} - \theta - \theta^\dagger \,,
    \label{eq:GM-Chern-decomposition}
\end{equation}
where the components of $\theta$ and $\theta^\dagger$ act as\footnote{Here $\dagger$ denotes the Hermitian conjugate, defined with respect to the inner product \eqref{eq:Hodge-metric}.}
\begin{equation}
    \theta_I:H^{q,p}(X_{\zeta}) \longrightarrow H^{q-1,p+1}(X_{\zeta})\,,\qquad \theta^\dagger_{\bar{J}}:H^{q,p}(X_{\zeta}) \longrightarrow H^{q+1,p-1}(X_{\zeta})\,.
    \label{eq:Higgs-field-bidegrees}
\end{equation}
In the local basis $\{\psi_\alpha^{(q,p)}({\zeta})\}$ for the cohomology groups $H^{q,p}(X_{\zeta})$, we denote the components of $\theta_I$ and $\theta^\dagger_{\bar J}$ by 
\begin{equation}
    \theta_I\psi_\alpha^{(q,p)} \equiv C_{I\alpha}{}^\gamma\psi_\gamma^{(q-1,p+1)} \,, 
    \quad 
    \theta^\dagger_{\bar J}\psi_\alpha^{(q,p)} \equiv G^{\gamma\bar\rho}(C_{J\rho}{}^\sigma)^*G_{\alpha\bar\sigma} \psi_\gamma^{(q+1,p-1)} \,.
    \label{eq:Higgs-field-components}
\end{equation}
It is also useful to introduce the holomorphic Yukawa coupling $C_{I\alpha \gamma}$, which  is defined by the relation 
\begin{equation}
    C_{I\alpha\beta} = \tilde{\eta}_{\beta\gamma}C_{I\alpha}{}^\gamma.
\end{equation}
It is now straightforward to derive an expression for the Chern curvature. With the component convention $F_{I\bar J} \equiv [\nabla^{\mathrm{Ch}}_{\bar J},\nabla^{\mathrm{Ch}}_I]$, flatness of the GM connection leads to:
\begin{equation}
    F_{I\bar{J}} = \bigl[\theta_I,\theta^\dagger_{\bar{J}}\bigr] \,.
    \label{eq:Chern-curvature-commutator}
\end{equation}
In components, this becomes
\begin{equation}
    (F_{I\bar J})_\alpha{}^\beta = G^{\rho\bar\mu} G_{\alpha\bar\nu} \, (C_{J\mu}{}^\nu)^* \, C_{I\rho}{}^\beta - C_{I\alpha}{}^\lambda \, G^{\beta\bar\gamma} G_{\lambda\bar\delta} \, (C_{J\gamma}{}^\delta)^* \,,
    \label{eq:Chern-curvature-components}
\end{equation}
where $F_{I\bar J}\psi_\alpha^{(q,p)} \equiv (F_{I\bar J})_\alpha{}^\beta\psi_\beta^{(q,p)}$. In the first term on the right-hand side of \eqref{eq:Chern-curvature-components}, $\rho,\mu$ label $H^{q+1,p-1}(X_{\zeta})$ and $\nu$ labels $H^{q,p}(X_{\zeta})$. In the second term, $\lambda,\delta$ label $H^{q-1,p+1}(X_{\zeta})$ and $\gamma$ labels $H^{q,p}(X_{\zeta})$. 

The Chern connection on the ambient cohomology is flat along the complex structure moduli at fixed \Kahler parameters. Indeed, an ambient class is the restriction of a fixed closed $(p,p)$-form on $V$, so it stays a $(p,p)$-form for every complex structure deformation as the embedding is holomorphic. The degree-changing pieces $\theta,\theta^{\dagger}$ then have to vanish and the Gauss--Manin and Chern connections agree.\footnote{Additionally, the restrictions are Poincaré-dual to specific integral cycles, which just move along the deformation. This also implies that the restrictions are Gauss--Manin- and therefore Chern-parallel, and even when there are ambient and primitive cohomologies with the same bidegree, they do not mix.} Since the former is flat, this is enough to conclude
\begin{equation}
    \left.F_{I\bar J}\right|_{H_{\mathrm{amb}}(X)} = 0 \,.
    \label{eq:ambient-Chern-curvature-vanishes}
\end{equation}
Since the ambient curvature vanishes, it is justified to focus only on the middle cohomology, which we will mostly do from now on.\footnote{One could also consider \Kahler deformations, for which the opposite is true: the curvature on the primitive cohomology vanishes.}

\subsection{The Calabi--Yau case, the Jacobian ring and the 2d CFT perspective}\label{subsec: 2D CFT calculations}

In Section~\ref{subsec:Chern connection on the moduli space}, we explained the general structure of the Berry curvature for sigma models on complete intersection manifolds. Doing concrete computations is still quite challenging: we need to find explicit cohomology representatives and explicit representatives of the complex structure deformation, and then compute the integral. 

For the case where the complete intersection is a Calabi--Yau $n$-fold, the computation becomes tractable. We will discuss the involved techniques anchored in the example of the hyperquintic
\begin{equation}
    X_{\zeta} = \mathbb{CP}^{n+1}|_{\{\mathcal{W}_0(x)=0\}} \,,
\end{equation}
but most of the statements generalize to other complete intersection Calabi--Yau manifolds with some qualifications and subtleties, some of which will be discussed in \cite{future-black-hole-paper}. We will suppress the $\zeta$-dependence of $\mathcal{W}_0$.

From a physics point of view, the reason why the computation becomes tractable is that we can think of a two-dimensional $\mathcal{N}=(2,2)$ sigma model on $X_{\zeta}$. Because $X_{\zeta}$ is Calabi--Yau, this sigma model flows to an auxiliary IR SCFT in which the data necessary to compute the Berry curvature is encoded. The data of this 2d SCFT can then be extracted using the Calabi--Yau/Landau--Ginzburg correspondence \cite{Witten:1993yc} and various localization formulas \cite{Gerchkovitz:2014gta,Ishtiaque:2017trm,Chen:2017fvl} (see also \cite{Chang:2023gow}).

From a mathematics point of view, the cohomologies are equivalently described by a Jacobian ring. The mathematically precise formulation of the localization formula has been discussed to some extent; see e.g. \cite{hertling2002ttgeometryfrobeniusmanifolds,hertling2008nilpotentorbitsgeneralizationhodge,Tang_2025}.

For a Calabi--Yau $n$-fold, there is a nowhere-vanishing holomorphic top form $\Omega$, unique up to an overall complex rescaling. Contracting the vector index of $\mu$ with $\Omega$ gives an $(n - 1,1)$-form and defines an isomorphism
\begin{equation}
    H^1(X,T^{1,0}X) \xrightarrow{\ \lrcorner\Omega\ } H^{n - 1,1}(X) \,, \qquad [\mu] \longmapsto [\mu\lrcorner\Omega] \,.
    \label{eq:complex-structure-contraction}
\end{equation}
The same contraction applies more generally to Dolbeault cohomology with values in holomorphic polyvector fields,
\begin{equation}
    H^p\!\left(X,\wedge^q T^{1,0}X\right) \xrightarrow{\ \lrcorner\Omega\ } H^{n - q,p}(X) \,,
    \label{eq:polyvector-Dolbeault-identification}
\end{equation}
and vice versa. The cohomologies\footnote{Here and in the following, when we write $H^p(...)$, we mean $H^{0,p}(...)$.} $H^p\!\left(X,\wedge^q T^{1,0}X\right)$ form a ring under wedge multiplication, which we refer to as the cohomology ring. This representation of the cohomologies is especially convenient as the complex structure deformations act via ring multiplication: introduce a basis $x_{\alpha}$ for $H^{p}(X,\wedge^q T^{1,0} X)$ such that
\begin{equation}
    \psi_{\alpha} = x_{\alpha} \lrcorner \, \Omega \,.
\end{equation}
Then we have
\begin{equation}
    \mu_I \lrcorner (x_\alpha \lrcorner \, \Omega) = (\mu_I \wedge x_\alpha)\lrcorner \, \Omega = C_{I\alpha}{}^{\lambda} \, x_{\lambda} \lrcorner \, \Omega \,.
\end{equation}
The $C_{I \alpha}{}^{\lambda}$ are thus the structure constants of the cohomology ring.

We can then also consider the Chern curvature of the cohomology ring bundle.
Let $\mathcal{H}^{r,s}$ denote the Hodge bundle with fiber $H^{r,s}(X_{\zeta})$, let $\mathcal{P}^{q,p}$ denote the bundle with fiber $H^p(X_{\zeta},\wedge^qT^{1,0}X_{\zeta})$, and let $\mathcal{L} \equiv \mathcal{H}^{n,0}$ be the Hodge line bundle. Since $\Omega_{\zeta}$ is only a local frame of $\mathcal{L}$, the bundle-level form of \eqref{eq:polyvector-Dolbeault-identification} is
\begin{equation}
    \mathcal{P}^{q,p} \simeq \mathcal{H}^{n-q,p}\otimes\mathcal{L}^{-1}\,.
    \label{eq:polyvector-Hodge-bundle-identification}
\end{equation}
The metric on $\mathcal{L}$ is $\mathcal{V}_{\Omega}$, so the metric induced on $\mathcal{P}^{q,p}$ and its Chern curvature are
\begin{equation}
    g^{\mathcal{P}} = \frac{G^{\mathcal{H}}}{\mathcal{V}_{\Omega}} \,, 
    \qquad
    F^{\mathcal{P}}_{I\bar J} = F^{\mathcal{H}}_{I\bar J} - \partial_I\partial_{\bar J}\log\mathcal{V}_{\Omega}\,\mathbf{1} 
    = 
    F^{\mathcal{H}}_{I\bar J} + g^{\mathcal{P}}_{I\bar J}\,\mathbf{1} = \mathcal{F}_{I\bar J} \,.
    \label{eq:polyvector-Chern-curvature}
\end{equation}
Thus the curvature naturally carried by the polyvector bundle is $\mathcal{F}$, which can be shown to indeed be the curvature of the connection compatible with the normalized metric $g^{\mathcal{P}}$. We now explain how the data necessary to compute $F$ (or $\mathcal{F}$) is encoded in the auxiliary 2d CFT. 

\paragraph{Geometric and CFT data.}

Consider the two-dimensional $\mathcal{N} = (2,2)$ superconformal field theory obtained at the infrared fixed point of the nonlinear sigma model on $X$. The theory has vector and axial $R$-symmetries $U(1)_V\times U(1)_A$. If an operator has charges $(q_L,q_R)$ under the left- and right-moving $U(1)$ $R$-currents, we define $q_V\equiv q_L+q_R$ and $q_A\equiv q_L-q_R$. Each complex target-space coordinate contributes $c = 3$ to each chiral algebra, and hence $c_L = c_R = 3n$ \cite{Cecotti:1991me}.

For the nonlinear sigma model on $X$, each chiral multiplet contains a complex scalar $\phi^i$, two Weyl fermions $\psi_+^i$ and $\psi_-^i$, and an auxiliary complex scalar $F^i$, together with their complex conjugates. In the Ramond sector the fermions are periodic and have zero modes. Quantizing their zero-mode algebra makes the wavefunctions differential forms on $X$: the supercharges act as $\partial$ and $\bar\partial$ and their adjoints, so that the Ramond ground states are the harmonic forms representing $H^{q,p}(X)$. All R-R ground states have conformal dimensions $(h^R_L,h^R_R) = (\frac{c_L}{24},\frac{c_R}{24})$. The fermionic vacuum is represented by $\Omega\in H^{n,0}(X)$ and has charges $(-\frac{n}{2},-\frac{n}{2})$, since each of the $n$ complex fermion pairs contributes $-1/2$ to the vacuum charge in the corresponding chiral sector. A state in $H^{q,p}(X)$ is obtained by acting with $n-q$ left-moving contraction modes and $p$ right-moving wedge modes, each of which raises the corresponding charge by one. Its chiral $R$-charges are therefore
\begin{equation}
    q_L^{\mathrm{R}} = -\frac{n}{2} + (n-q) = \frac{n}{2} - q \,, \qquad q_R^{\mathrm{R}} = -\frac{n}{2} + p = p - \frac{n}{2} \,.
    \label{eq:Ramond-R-charges-Hodge-bidegree}
\end{equation}
Performing half a unit of spectral flow \cite{Lerche:1989uy} relates these to NS-NS states with charges and conformal weights
\begin{equation}
    q^{\mathrm{NS}}_{L/R} = q^{R}_{L/R} + \frac{c_{L,R}}{6}  \,, \qquad h^{\mathrm{NS}}_{L/R} = h^{R}_{L/R} + \frac{q^R_{L/R}}{2} + \frac{c_{L/R}}{24} \,,
    \label{eq:NS-R-spectral-flow}
\end{equation}
By the state operator correspondence, these NS-NS states correspond to operators in the chiral ring $\mathcal R_{c,c}$. Ultimately, we find the following mapping between cohomologies and the chiral ring
\begin{equation}
    x_{\alpha} \in \mathcal{R}_{c,c} \text{ with }(q_L^{\mathrm{NS}}, q_R^{\mathrm{NS}}) = (n-q,p) \, \leftrightarrow \, \psi_{\alpha} \in H^{q,p} \, \leftrightarrow \, x_{\alpha} \in H^{0,p}(X,\wedge^{n-q} T^{1,0} X) \,.
    \label{chiral ring and bidegree charges}
\end{equation}
In particular, complex structure deformations (which we will keep referring to by the indices $I,J,...$) with $q=n-1$, $p=1$ are one-to-one with $(1,1)$ states whose conformal descendants give exactly marginal deformations. The resulting conformal manifold is the complex structure manifold $\mathcal{M}$. Primitive middle cohomology elements in $H^{n-p,p}$ have $R$-charge $(p,p)$.  

\paragraph{The $B$-model twist and the Poincaré pairing.}

The $(c,c)$ chiral sector of the theory can be studied using the B-twist, introduced in \cite{Witten:1991zz}: for a Calabi--Yau target, the axial $R$-symmetry is nonanomalous and can be used to twist the (worldsheet) Lorentz generator,
\begin{equation}
    J_{\mathrm{Lorentz}}^B \equiv J_{\mathrm{Lorentz}} + \frac{1}{2}J_A \,, \qquad Q_B \equiv (Q_+)^* + (Q_-)^* \,.
    \label{eq:B-twist-definition}
\end{equation}
With respect to the twisted Lorentz generator, $Q_B$ is a scalar nilpotent supercharge. The twisted stress tensor is $Q_B$-exact, so correlation functions of $Q_B$-closed local operators depend neither on their insertion points nor on the worldsheet metric. The $Q_B$-cohomology of local operators is precisely the $(c,c)$ chiral ring, and the OPE turns it into a finite-dimensional topological algebra on this cohomology. 

Through the $B$-twist and the ensuing statistics of the fermionic scalars, one can also recover the result that
\begin{equation}
    \mathcal{R}_{(c,c)} \cong \bigoplus_{p,q=0}^n H^p(X,\wedge^{n-q} T^{1,0} X). 
\end{equation}
In terms of this Hodge bidegree, the vector and axial $R$-charges are
\begin{equation}
    q_V =  n - q + p \,, \qquad q_A = n - p - q \,.
    \label{eq:R-charge-grading-coho}
\end{equation}
\Kahler deformations enter the B-twisted action through $Q_B$-exact terms. Consequently, away from singular loci, the chiral ring and its topological correlation functions can depend on the complex structure but not on the \Kahler moduli.

Sphere correlators in the $B$-twisted theory then encode the Poincaré pairing of the cohomologies. As the sphere is a curved worldsheet, one needs to implement the B-twist by coupling the axial $R$-current to a nondynamical background gauge field proportional to the spin connection. In a correlation function of operators with left- and right-moving $R$-charges $(q_{L,k},q_{R,k})$, there are then Ward identities which give the selection rule \cite{Witten:1991zz}
\begin{equation}
    \sum_k q_{L,k} = c/3 \,, \qquad \sum_k q_{R,k} = c/3 \,.
    \label{eq:B-model-genus-g-selection-rule}
\end{equation}
For the Calabi--Yau $n$-fold models considered here, $c = 3n$, so a sphere correlator can be nonzero only when the insertions have total left- and right-moving $R$-charges $(n,n)$. In particular, for a two-point function between two chiral ring operators $x_{\alpha}$, $x_{\beta}$, we get a nondegenerate bilinear pairing \cite{Witten:1991zz}
\begin{equation}
    \tilde{\eta}_{\alpha\beta} \equiv \left\langle x_\beta(\infty)x_\alpha(0)\right\rangle^{\mathrm{un}}_{S^2,B} \,,
    \label{eq:B-model-topological-pairing}
\end{equation}
with the superscript ``un'' meaning that this is an unnormalized correlator. This pairing has the selection rule
\begin{equation}
    q_{L,\alpha} + q_{L,\beta} = q_{R,\alpha} + q_{R,\beta} = n.
\end{equation}
The geometric counterpart is the following: by \eqref{chiral ring and bidegree charges}, $x_\alpha \in H^{q_\alpha^R}(X,\wedge^{q_\alpha^L}T^{1,0}X)$ and $x_\beta \in H^{q_\beta^R}(X,\wedge^{q_\beta^L}T^{1,0}X)$. Up to degree-dependent signs and a fixed overall normalization, the pairing is given by the expression 
\begin{equation}
    \tilde{\eta}_{\alpha\beta} = \int_X \Omega \wedge \bigl((x_\alpha \wedge x_\beta)\lrcorner\Omega\bigr) = \pm \int_X x_{\alpha}\lrcorner \Omega \wedge x_{\beta} \lrcorner \Omega = \pm \int_X \psi_{\alpha} \wedge \psi_{\beta} \,.
    \label{eq:B-model-topological-pairing-geometric}
\end{equation}
This expression is the Poincaré pairing and also vanishes unless $q_\alpha^R + q_\beta^R = n$ and $q_\alpha^L + q_\beta^L = n$. 

The pairing has a simple algebraic interpretation: the chiral ring contains a unique top form of left/right charge $(n,n)$ \cite{Lerche:1989uy} which geometrically is the complex conjugate of the holomorphic volume form. The nondegenerate pairing can be viewed as ring multiplication with subsequent projection onto this top form. This turns the ring into a Frobenius algebra. The holomorphic Yukawa couplings $C_{I\alpha \beta}$ are in this language topological three-point functions of one ring element with charge $(1,1)$ and two other ring elements such that the total charge adds up to $(n,n)$.

\paragraph{2d localization and the Hodge pairing.}

The 2d nonlinear sigma model also contains non-holomorphic data, such as the Hermitian metric defined by the physical flat-space two-point function of a chiral primary and its conjugate. We use vacuum-normalized expectation values on $\mathbb{R}^2$, so that
\begin{equation}
    g_{\alpha\bar\beta} \equiv \left\langle (\mathcal{O}_\beta)^*(\infty)\mathcal{O}_\alpha(0)\right\rangle_{\mathbb{R}^2} \,, 
    \qquad
    (\mathcal{O}_\beta)^*(\infty) \equiv \lim_{|x|\rightarrow\infty}|x|^{2\Delta_\beta}(\mathcal{O}_\beta)^*(x) \,.
    \label{eq:CFT-Hermitian-metric}
\end{equation}
This is now the physical correlator in the untwisted theory, rather than a correlator in the B-twisted theory. The path integral defining the expectation value is divided by the flat-space vacuum functional, and the identity has unit norm. When $\alpha$ and $\beta$ label marginal directions, supersymmetry relates $g_{\alpha\bar\beta}$ to the Zamolodchikov metric of their exactly marginal descendants.

For a Lagrangian description, the flat-space metric can be computed by supersymmetric localization \cite{Gerchkovitz:2014gta,Chen:2017fvl,Ishtiaque:2017trm}. One places the theory on the $SU(2|1)_B$ supersymmetric sphere\footnote{Note that this is not the topological B-twist, but a rigid supergravity background that preserves the supercharges used for localization.} and computes the correlator of a chiral and an antichiral operator at opposite poles. After including the Weyl factors relating the sphere to the plane and, when necessary, removing curvature-induced mixing with lower-dimensional operators \cite{Gerchkovitz:2014gta}, one obtains the flat-space metric in \eqref{eq:CFT-Hermitian-metric}. When the 2d CFT is a nonlinear sigma model, as in the geometric phase of a 2d GLSM described below, the supersymmetric sphere partition function is proportional to $\mathcal{V}_{\Omega}$, \eqref{eq:holomorphic-volume-form-Hodge-norm}, up to a moduli-independent proportionality constant. We can therefore also express the normalized metric $g$ and the proportionality constant $\mathcal V_\Omega$ relating it to the unnormalized metric $G_{\alpha\bar\beta} = \mathcal V_\Omega g_{\alpha\bar\beta}$ defined in \eqref{eq:Hodge-metric} as
\begin{equation}
    g_{\alpha\bar\beta} = (2R_{S^2})^{2\Delta_\alpha} \left\langle (\mathcal{O}_\beta)^*(\infty)\mathcal{O}_\alpha(0)\right\rangle_{S^2_{SU(2|1)_B}} \,, 
    \qquad 
    \mathcal V_\Omega = Z_{S^2_{SU(2|1)_B}} \,,
    \label{eq:hermitian-metric-localization-correlator}
\end{equation}
where $R_{S^2}$ is the radius of the sphere.

\paragraph{The Jacobian ring and the Calabi--Yau/Landau--Ginzburg correspondence.}

We now give an example where one can derive explicit expressions for all of these quantities. The example is the hyperquintic, given by
\begin{equation}
    X_{\zeta} = \mathbb{CP}^{n+1}|_{\mathcal{W}_0=0} \,,
\end{equation}
with $\mathcal{W}_0$ a degree-$(n+2)$ polynomial whose derivatives vanish simultaneously only at the origin and whose coefficients encode the complex structure $\zeta$. It is well known in the mathematical literature \cite{ffab916b-4326-37f5-aaf1-66df46a74bac,cb71e49b-28cb-3d4d-a302-c5ff3b25456b} that the primitive middle cohomology of such a hypersurface admits a description in terms of the Jacobian ring of $\mathcal{W}_0$, i.e. the set of all polynomials in the $x$'s modulo the relations generated by $\{\partial_i \mathcal W_0=0\}_{i=1}^{n+2}$,
\begin{equation}
     H_{\text{prim}}^{p}(X,\wedge^p T^{1,0} X) \cong \text{Jac}[\mathcal{W}_0]_{p(n+2)} \,, \quad \text{Jac}[\mathcal{W}_0] \equiv \frac{\mathbb{C}[x_1,...,x_{n+2}]}{\langle \partial_i \mathcal{W}_0 \rangle} \,. 
\end{equation}
The subscript of Jac denotes restriction to polynomials of degree $p(n+2)$. The primitive middle cohomology is therefore described by specific degree slices in the homogeneous coordinate ring of the ambient $\mathbb{CP}^{n+1}$, modulo relations induced by derivatives of $\mathcal{W}_0$. The relation respects the ring structure (after fixing the correct normalization) \cite{Carlson:1980}. There exists a linear functional called the Grothendieck residue \cite{griffiths1978principles} on the ring $\text{Jac}[\mathcal{W}_0]$ given by
\begin{equation}\label{Grothendieck residue}
    \tau_G: P \mapsto \left(\frac{1}{2\pi i}\right)^{n+2} \oint_{\{|\partial_i \mathcal{W}_0|=\epsilon\}} \frac{P\,  dx_1\wedge ...\wedge dx_{n+2}}{\prod_i \partial_i \mathcal{W}_0} \,.
\end{equation}
In particular, it induces a nondegenerate bilinear pairing $\eta(A,B) = \tau_G(AB)$. The residue vanishes on every homogeneous component except the component of degree $n(n+2)$, which is the top component. As the latter pairs with the degree-zero piece, it is one-dimensional.  Formulas for the Hodge metric had also been explored in the mathematical literature, albeit more recently \cite{hertling2002ttgeometryfrobeniusmanifolds,hertling2008nilpotentorbitsgeneralizationhodge,Tang_2025}.

These facts have an elegant physical derivation through the Calabi--Yau/Landau--Ginzburg correspondence \cite{Witten:1993yc}. The Jacobian ring is the chiral ring of the IR SCFT of a Landau--Ginzburg theory, the Grothendieck residue computes topological correlators of  ring using the B-twist \cite{Cecotti:1990kz,Vafa:1990mu}, and the Hodge metric is given by physical correlation function computed using localization \cite{Cecotti:1991me,Gerchkovitz:2014gta,Chen:2017fvl,Ishtiaque:2017trm}. We now review using the hyperquintic example.

Consider an $\mathcal{N} = (2,2)$ gauged linear sigma model (GLSM) with gauge group $U(1)$, one vector multiplet, $n + 2$ chiral multiplets $x_i$ of charge $1$, and one chiral multiplet $P$ of charge $-(n + 2)$ \cite{Witten:1993yc}. It has a Fayet--Iliopoulos parameter $r$, and the gauge-invariant superpotential is
\begin{equation}
    \mathcal{W}_{\mathrm{GLSM}} = P \, \mathcal{W}_0(x)\,.
    \label{eq:quintic-GLSM-superpotential}
\end{equation}
The F-term equations are
\begin{equation}
    \frac{\partial\mathcal{W}_{\mathrm{GLSM}}}{\partial P} = \mathcal{W}_0(x) = 0 \,, \qquad \frac{\partial\mathcal{W}_{\mathrm{GLSM}}}{\partial x_i} = P\partial_i \mathcal{W}_0(x) = 0 \,, \qquad i = 1,\ldots,n + 2\,,
    \label{eq:hypersurface-GLSM-F-terms}
\end{equation}
while the D-term equation is
\begin{equation}
    \sum_{i = 1}^{n + 2}|x_i|^2 - (n + 2)|P|^2 = r\,.
    \label{eq:hypersurface-GLSM-D-term}
\end{equation}
For $r > 0$, the D-term equation requires that the $x_i$ do not all vanish, and their $U(1)$ quotient gives $\mathbb{CP}^{n + 1}$. The F-term equations then restrict the vacuum to $\mathcal{W}_0(x) = 0$ and, for a smooth hypersurface, imply $P = 0$. The vacuum moduli space is therefore $X_{\zeta}$, and the infrared theory is the nonlinear sigma model on $X_{\zeta}$. For $r < 0$, the D-term equation requires $P \neq 0$ and its expectation value Higgses $U(1)$ to $\mathbb{Z}_{n + 2}$, leaving a Landau--Ginzburg orbifold with superpotential $\mathcal{W}_0$. These are two limiting descriptions in the same (complexified) \Kahler moduli space.

The geometric and Landau--Ginzburg descriptions are two phases of the same GLSM, reached by varying the FI parameter. In the B-twisted theory this variation is $Q_B$-exact \cite{Witten:1993yc}, so the chiral primaries, their ring product, and their topological pairing can be continued between the two phases, provided that no singular locus is crossed. We may therefore compute these quantities in the Landau--Ginzburg phase and interpret the resulting states and operators in the nonlinear sigma model on $X_{\zeta}$. We will also compute in the LG description the physical Hermitian metric \eqref{eq:CFT-Hermitian-metric} on the protected states. This metric is not a B-twisted observable and need not be \Kahler-independent on the full chiral ring. However, on the primitive\footnote{For the ambient cohomologies, we already argued that the unnormalized curvature $F$ vanishes. The normalized curvature $F$ is then only the universal scalar term in \eqref{eq:polyvector-Chern-curvature}, so no separate ring computation is needed.} middle cohomology sector $\alpha \in H^{n-p,p}(X)$ relevant below, the Hodge star acts as $*\alpha = (-1)^{n(n - 1)/2}i^{2p - n}\alpha$ without introducing powers of the \Kahler form, so the metric is independent of the \Kahler parameters. Its restriction to this sector can therefore also be continued between the geometric and LG phases.

The LG phase now implements the Jacobian ring description. For the hypersurface LG phase, the light fields are the $n + 2$ chiral multiplets $x_i$, the superpotential is $\mathcal{W}_{\mathrm{LG}} = \mathcal{W}_0(x)$, and the diagonal $\mathbb{Z}_{n + 2}$ generated by $\mathsf{j}:x_i \mapsto e^{2\pi i/(n + 2)}x_i$ is gauged. Before imposing the F-term relations, polynomial representatives of the untwisted chiral operators are linear combinations of the monomials $x^{\boldsymbol{a}} \equiv \prod_i x_i^{a_i}$, with $a_i \geq 0$. In the B-twisted theory, the F-term operators $\partial_i \mathcal{W}_0(x)$ are $Q_B$-exact, so polynomials that differ by an element of the ideal generated by the $\partial_i \mathcal{W}_0(x)$ represent the same chiral operator. The untwisted chiral ring is therefore the Jacobian ring
\begin{equation}
    R(\mathcal{W}_0) \equiv \operatorname{Jac}(\mathcal{W}_0) \equiv \frac{\mathbb{C}[x_1,\ldots,x_{n + 2}]}{J(\mathcal{W}_0)} \,, \qquad J(\mathcal{W}_0) \equiv \left(\partial_1 \mathcal{W}_0,\ldots,\partial_{n + 2}\mathcal{W}_0\right) \,.
    \label{eq:hypersurface-Jacobian-ring}
\end{equation}

We normalize the $R$-charges so that $q_L(x_i) = q_R(x_i) = q_i = \frac{1}{n + 2}$. A homogeneous polynomial $P_k$ of degree $k$ is then a $(c,c)$ chiral operator with charges $(\frac{k}{n + 2},\frac{k}{n + 2})$. By the state--operator correspondence it defines an NS sector state with the same charges,
\begin{equation}
    [P_k] \longleftrightarrow |P_k\rangle_{\mathrm{NS}} \equiv [P_k](0)|0\rangle_{\mathrm{NS}} \,, \qquad (q_L,q_R)(|P_k\rangle_{\mathrm{NS}}) = \left(\frac{k}{n + 2},\frac{k}{n + 2}\right)\,.
    \label{eq:hypersurface-LG-polynomial-R-charges}
\end{equation}
Under $\mathsf{j}$, this state acquires the phase $e^{2\pi i k/(n + 2)}$, so it survives the orbifold projection when $k = p(n + 2)$. Its charges are then $(q_L,q_R) = (p,p)$, which corresponds through \eqref{chiral ring and bidegree charges} to Hodge bidegree $(n - p,p)$. The LG/nl$\sigma$m correspondence identifies these invariant polynomial states with the primitive middle cohomology,
\begin{equation}
    \mathcal{H}_{\mathrm{untw}}^{(c,c)} \cong \bigoplus_{p = 0}^n R_{p(n + 2)}(\mathcal{W}_0) \cong \bigoplus_{p = 0}^n H_{\mathrm{prim}}^{n - p,p}(X_{\zeta})\,.
    \label{eq:hypersurface-untwisted-primitive-cohomology}
\end{equation}
The first two spaces describe the local $(c,c)$ chiral operators, or equivalently their NS chiral primary states under the operator--state correspondence.

One can also determine the rank of each primitive bundle, equivalently the degeneracy of the $(c,c)$ chiral primaries of charges $(p,p)$. These operators are represented by the degree-$p(n + 2)$ component of the Jacobian ring. The polynomial ring $\mathbb C[x_1,\dots,x_{n+2}]$ has Hilbert series (the generating function for its graded dimensions) $(1 - t)^{-(n + 2)}$, while each relation contributes a factor $1 - t^{n + 1}$. Therefore
\begin{equation}
    \sum_{m \geq 0}\bigl(\dim R_m(\mathcal{W}_0)\bigr) t^m = \left(\frac{1 - t^{n + 1}}{1 - t}\right)^{n + 2} = \left(1 + t + \cdots + t^n\right)^{n + 2}\,.
    \label{eq:hypersurface-Jacobian-Hilbert-series}
\end{equation}
For a generic polynomial $\mathcal{W}_0$ a basis consists of monomials $x_\alpha = \prod_i x_i^{\alpha_i}$ with $0 \leq \alpha_i \leq n$, so each variable contributes $1 + t + \cdots + t^n$. Extracting the coefficient of the required degree gives
\begin{equation}
    \begin{aligned}
        \dim H_{\mathrm{prim}}^{n - p,p}(X_{\zeta}) & = \dim R_{p(n + 2)}(\mathcal{W}_0) \\
        & = [t^{p(n + 2)}]\left(1 + t + \cdots + t^n\right)^{n + 2} \\
        & = \sum_{k = 0}^{p}(-1)^k\binom{n + 2}{k}\binom{p(n + 2) - (k - 1)(n + 1)}{n + 1}\,.
    \end{aligned}
    \label{eq:hypersurface-primitive-Hodge-degeneracy}
\end{equation}
As a consistency check, this formula can be shown to match \eqref{HRR hypersurface primitive}.
The number of states in the untwisted sector grows rapidly with $n$: the total number is proportional to $n^{n+1}$, centered around the ``middle-middle cohomology'' $p \approx \frac{n}{2}$ as a Gaussian of width $\sqrt{n}$. Near the edges ($p$ finite as $n\rightarrow \infty$), the number of cohomologies is exponentially growing with a coefficient scaling with $p$. In particular, this implies that ratios of dimensions of neighboring $p$ are exponentially growing. The first few primitive Hodge numbers are also listed in Table~\ref{tab:hodge-numbers-quintic}. 

The orbifolding also introduces the $\mathsf{j}^\ell$-twisted sectors, in which the fields obey the boundary conditions
\begin{equation}
    x_i(\sigma + 2\pi) = e^{2\pi i\ell/(n + 2)}x_i(\sigma) \,, \qquad \ell = 1,\ldots,n + 1\,.
    \label{eq:hypersurface-LG-twisted-boundary-condition}
\end{equation} 
Every $x_i$ is twisted in these sectors, so there are no $x_i$ zero modes and each sector contributes one protected ground state $|\ell\rangle$ \cite{Intriligator:1990ua}. There is thus a total of $n+1$ states in all the twisted sectors. Here $|\ell\rangle$ denotes the lowest NS sector state with the boundary condition \eqref{eq:hypersurface-LG-twisted-boundary-condition}, created by the corresponding twist operator. A short calculation shows that the $R$-charges are
\begin{equation}
    q_{L,\ell} = \ell-1, \quad q_{R,\ell} = n+1 - \ell .
\end{equation}
Consequently, we can use \eqref{chiral ring and bidegree charges} and match the twisted states to monotone cohomologies:
\begin{equation}
    |\ell\rangle \longleftrightarrow H^{n + 1 - \ell,n + 1 - \ell}(X_{\zeta}) \,, \qquad \ell = 1,\ldots,n + 1\,.
    \label{eq:hypersurface-twisted-sector-Hodge-bidegrees}
\end{equation}
As each (ambient) diagonal cohomology $H^{p,p}$ of $X_{\zeta}$ contains exactly one state, the twisted states recover the entire ambient cohomology. On those the commutator curvature $F$ along the complex structure moduli vanishes.\footnote{There is a slight subtlety related to the curvature, related to the fact that the isometries \eqref{eq:hypersurface-untwisted-primitive-cohomology} apply at fixed coupling, but not as bundles. The $H^{n-p,p}$ cohomologies map to the RR ground states (and the pure-Higgs states in the quiver QM) have the Chern curvature $F$. The cohomologies $H^p(X,\wedge^q T^{(1,0)}X)$ are mapped to the $B$-twisted chiral ring, and their curvature is $\mathcal F$. The curvature of the NS states in the untwisted 2d theory, $\mathbf{F}^{2d}$, is then shifted by the curvature of the $R$-charge line bundle times $(p+q)$, giving
\begin{equation}
    (\mathbf{F}^{\mathrm{2d}}_{I\bar J})_\alpha{}^\beta = (F_{I\bar J})_\alpha{}^\beta + \left(1 - \frac{p+q}{n}\right)g^{\mathrm{WP}}_{I\bar J}\delta_\alpha{}^\beta 
    = (\mathcal{F}_{I\bar J})_\alpha{}^\beta - \frac{p+q}{n}g^{\mathrm{WP}}_{I\bar J}\delta_\alpha{}^\beta \,.
    \label{eq:NS-Berry-versus-Hodge-Chern-curvature}
\end{equation}
In the derivation of \cite{Baggio:2017aww}, the $(p+q)$-shift arises from an $R$-current contact term. See also \cite{deBoer:2008ss}.}
Its normalized curvature $\mathcal{F}$ contains only the universal scalar term in \eqref{eq:polyvector-Chern-curvature}, and the curvature is therefore nontrivial only for the untwisted Jacobian ring.

In the Landau--Ginzburg phase, for the primitive middle cohomology, computing the objects $C_{I \alpha}{}^{\lambda}$, $\tilde{\eta}_{\beta \gamma}$ and $g_{\alpha \bar{\beta}}$ is then straightforward. Using the same notation as before, we pick the basis $\{\phi_{\alpha_{(p)}}(x)\}$ for $R_{p(n + 2)}(\mathcal{W}_0) \cong H^{p}(X,\wedge^p T^{1,0} X)$, such that $\deg(\phi_{\alpha_{(p)}}) = p(n+2)$ for any $\alpha_{(p)}$. Complex structure deformations correspond to the case $p=1$, whose indices we denote by $I \equiv \alpha_{(1)}$. 

Using the B-twist, the topological correlation functions can be computed using localization \cite{Vafa:1990mu,Herbst:2004ax}. The result is
\begin{multline}
    \left\langle \phi_{\alpha_{(i_1)}} \cdots \phi_{\alpha_{(i_k)}} \right\rangle_B = \operatorname{Res}_{\mathcal{W}_0}[\phi_{\alpha_{(i_1)}} \cdots \phi_{\alpha_{(i_k)}}] \\ 
    \equiv \oint_{\{|\partial_i \mathcal{W}_0|=\epsilon\}}  \frac{(\phi_{\alpha_{(i_1)}} \cdots \phi_{\alpha_{(i_k)}})(x)\,d x_1\wedge\cdots\wedge d x_{n + 2}}{(2\pi i)^{n + 2} \prod_{i = 1}^{n + 2}\partial_i\mathcal{W}_0(x)} \,,
    \label{eq:LG residue}
\end{multline}
where in the right hand side the $O$'s are treated as polynomials. This is precisely the Grothendieck residue in \eqref{Grothendieck residue}. Picking just three operators gives us the holomorphic structure constants $C_{\alpha\beta\gamma}$, and one index can be raised using the inverse of the topological two-point function, $\eta_{\alpha\beta} = \langle \phi_{\alpha} \phi_\beta \rangle _B$, which is also computed using the same localization formula.

Finally, for $\mathcal{N}=(2,2)$ Landau--Ginzburg models there are explicit expressions for the aforementioned localization formulas to compute the Hodge metric, via the sphere two-point function \cite{Cecotti:1991me} (see also \cite{Gerchkovitz:2014gta,Chen:2017fvl,Ishtiaque:2017trm,Chang:2023gow}). More precisely, one puts an operator $\phi_{\alpha_{(p)}}$ of the $(c,c)$-ring at the south pole of the sphere of radius $R_{S^2}$ and an operator $\bar{\phi}_{\beta_{(p)}}$ of the $(a,a)$-ring (i.e. a primary with $q_L=-2h_L$, $q_R=-2h_R$) at the north pole. The unnormalized two-point function is
\begin{equation}
    \widetilde G_{\alpha_{(p)}\bar{\beta}_{(p)}} = (2R_{S^2})^{2p} \int_{\mathbb{C}^{n+2}} \Bigl(\prod_{i=1}^{n+2} dx_i d\bar{x}_i\Bigr) \, \phi_{\alpha_{(p)}}(x) \bar{\phi}^{\beta_{(p)}}(\bar x) e^{-4\pi i R_{S^2} \mathcal{W}_0(x)-4\pi i R_{S^2} \overline{\mathcal{W}}_0(\bar{x})} \,.
\end{equation}
By this formal expression, we mean the following (see also Appendix C of \cite{Gaiotto:2013rk}): split the integral into holomorphic and antiholomorphic parts. As $\mathcal{W}_0$ is assumed to be a generic degree-$(n+2)$ polynomial of $n+2$ variables, there are $(n+1)^{n+2}$ steepest descent contours along which the integrand $e^{-4\pi i R_{S^2} \mathcal{W}_0}$ decays (as well as for the antiholomorphic part). Denote by $I_{ab}$ the intersection number (with some orientation of cycles absorbed) between the complex conjugate of a holomorphic thimble $\Gamma_a$ and an antiholomorphic thimble $\tilde{\Gamma}_b$. Then
\begin{multline}
    \widetilde G_{\alpha_{(p)} \bar{\beta}_{(p)}} = (2R_{S^2})^{2p} \sum_{a,b} \int_{\Gamma_a} \prod_{i=1}^{n+2} dx_i \,  \phi_{\alpha_{(p)}}(x) e^{-4\pi i R_{S^2} \mathcal{W}_0(x)} (I^{-1})_{ba} \\
    \times \int_{\tilde{\Gamma}_b} \prod_{i=1}^{n+2}  d\bar{x}_i \bar \phi_{\beta_{(p)}}(\bar x) e^{-4\pi i R_{S^2} \overline{\mathcal{W}}_0(\bar{x})} \,.
\end{multline}
The normalized correlator is, in the same sense defined via cycles,
\begin{equation}
        \tilde g_{\alpha_{(p)}\bar\beta_{(p)}} = \frac{(2R_{S^2})^{2p}}{Z_0} \widetilde G_{\alpha_{(p)}\bar\beta_{(p)}}  \,, \quad
        Z_0 = \int_{\mathbb{C}^{n + 2}} \prod_{i = 1}^{n + 2}d x_i \, d \bar x_i \, e^{-4\pi i R_{S^2}\mathcal{W}_0(x) - 4\pi i R_{S^2}\overline{\mathcal{W}}_0(\bar{x})} \,.
    \label{eq:Gepner-physical-metric-localization-tilde}
\end{equation}
This is in general not yet the metric. When placing the (anti-)chiral primaries on the sphere, one can have operator mixing induced by the background fields. There exists an algorithm which systematically subtracts the admixing operators \cite{Gerchkovitz:2016gxx,Ishtiaque:2017trm, Chen:2017fvl}. In this way, the CY/LG correspondence allows one to compute all the quantities necessary for computing the Berry curvature of the CY sigma model.

Things simplify significantly at a special point in the moduli space with enhanced symmetry, which is called the Gepner point. The superpotential there is given by
\begin{equation}\label{eqn:Gpt}
    \mathcal{W}_0(x) = \frac{1}{n+2} \sum_{i=1}^{n+2} x_i^{n+2} \,,
\end{equation}
and the constraints factorize, so the Jacobian ring becomes
\begin{equation}
    R(\mathcal{W}_0) = \frac{\mathbb{C}[x_1,\ldots,x_{n + 2}]}{(x_1^{n + 1},\ldots,x_{n + 2}^{n + 1})} \,.
    \label{eq:Jacobian-ring-Gepner}
\end{equation}
We can explicitly specify the basis $\phi_{\alpha(p)} \equiv x_{\alpha_{(p)}}$, where
\begin{equation}\label{basis of cohomologies in monomials}
    x_{\alpha_{(p)}} = \prod_{i=1}^{n+2} x_i^{\alpha^i_{(p)}} \,, \qquad 0 \leq \alpha^i_{(p)} \leq n \,, \qquad \sum_{i=1}^{n+2} \alpha^i_{(p)} = p(n+2) \,,
\end{equation}
which for the complex structure deformations becomes 
\begin{equation}
    x_I = \prod_{i=1}^{n+2} x_i^{I_i}\,, \qquad 0 \leq I_i \leq n \,, \qquad \sum_{i=1}^{n+2} I_i = n+2 \,.
\end{equation}
This basis is in fact a basis for a generic superpotential $\mathcal{W}_0$ (see the discussion around \eqref{eq:hypersurface-primitive-Hodge-degeneracy}).

At the Gepner point (due to the large symmetry), the structure constants of the ring are trivial in this basis. Multiplication of operators amounts to addition of multi-indices, as long as they fulfill the consistency conditions in \eqref{basis of cohomologies in monomials}; otherwise, the product vanishes. The structure constants are
\begin{equation}\label{monomial structure constants}
    C_{\alpha I}{}^{\lambda} = \delta_{\alpha + I}^{\lambda} \,.
\end{equation}
The metric also simplifies. At the Gepner point there is no mixing with lower dimensional operators, and the localization equations directly give the (unnormalized) metric 
\begin{multline}
    G_{\alpha_{(p)} \bar{\beta}_{(p)}} = (2R_{S^2})^{2p} \sum_{a,b} \int_{\Gamma_a} \prod_{i=1}^{n+2} dx_i \,  x_i^{\alpha^i_{(p)}} e^{-4\pi i R_{S^2} \mathcal{W}_0(x)} (I^{-1})_{ba} \\
    \times \int_{\tilde{\Gamma}_b} \prod_{i=1}^{n+2}  d\bar{x}_i \bar{x}_i^{\beta^i_{(p)}} e^{-4\pi i R_{S^2} \overline{\mathcal{W}}_0(\bar{x})} \,,
\end{multline}
and its normalized counterpart
\begin{equation}
        g_{\alpha_{(p)}\bar\beta_{(p)}} = \frac{(2R_{S^2})^{2p}}{Z_0} G_{\alpha_{(p)}\bar\beta_{(p)}}  \,, \quad
        Z_0 = \int_{\mathbb{C}^{n + 2}} \prod_{i = 1}^{n + 2}d x_i \, d \bar x_i \, e^{-4\pi i R_{S^2}\mathcal{W}_0(x) - 4\pi i R_{S^2}\overline{\mathcal{W}}_0(\bar{x})} \,.
    \label{eq:Gepner-physical-metric-localization}
\end{equation}

\section{BPS chaos at the Gepner point}\label{sec: the (hyper)quintic at the Gepner point}

The formalism introduced in the previous section allows us to write down an expression for the Berry curvature of the sigma models with hyperquintic target spaces. In this section, we study the Berry curvature of the primitive middle cohomology of these models, at a specific point in moduli space with a large amount of discrete symmetry: the Gepner point \cite{Gepner:1987qi,Gepner:1987vz}. This anchors our discussion in a concrete example and functions as a case study for the black hole model described in Section~\ref{sec:motivation}, which will be addressed in future work \cite{future-black-hole-paper}.

We summarize the results of this section:
\begin{itemize}
    \item \textbf{Chern curvature at the Gepner point.} At the special point \eqref{eqn:Gpt} in the complex structure moduli space, the localization integrals for the Hodge metric are exactly solvable, leading to analytic expressions for the curvature matrix. We then study its properties when averaging over generic directions in the complex structure moduli space (see \eqref{Curvature full expression}).

    \item \textbf{Level spacing ratio, SFF and number variance.} We numerically find the spectrum and analyze spectral diagnostics of chaos. In the spectral form factor, we find eigenvalue repulsion and a ramp whose length increases with the system size in units where the plateau time is fixed (see e.g. Figure~\ref{fig:ramp time scaling}), but also structures that point to physics beyond RMT. The number variance of the spectrum behaves similarly to that of SYK-like systems (see Figure~\ref{fig:number variance}).
    
    \item \textbf{Spectral density.} For the spectral density itself, we make a comparison to Wishart ensembles. We find a semi-analytic model based on sparse non-uniform Wishart ensembles, which describes the nontrivial features in the spectral density. We refer to these features as shoulders and tails (see e.g. Figure~\ref{fig:p=1 deformed Wishart}).
\end{itemize}

\subsection{Gepner point statics}

For the hypersurface Calabi--Yau, we specialize to the Gepner point\footnote{Chaos properties of a model with deformations around this point have also been studied in \cite{Berkooz:2021ehv,Berkooz:2022dfr}. The relation between the discussion there and the measures of chaos that we discuss here is not clear.}:
\begin{equation}\label{eqn:Gpt}
    \mathcal{W}_0(x) = \frac{1}{n+2} \sum_{i=1}^{n+2} x_i^{n+2} \,. 
\end{equation}
The infrared CFT is then the diagonal orbifold of $n + 2$ identical $\mathcal{N} = 2$ minimal models of level $k = n$:
\begin{equation}
    \mathcal{C}_{\mathrm{Gepner}} \equiv \frac{\left(\mathcal{M}_{k = n}\right)^{\otimes(n + 2)}}{\mathbb{Z}_{n + 2}^{\mathrm{diag}}} \,.
    \label{eq:hypersurface-Gepner-model}
\end{equation}
This point has a large discrete symmetry with a $\mathbb{Z}_{n+2}$ acting on each $x_i$ separately, as well as permutations of the $x$'s. This makes the protected ring coefficients, the topological pairing, and the physical metric exactly computable. By the continuation described above, the data computed in this orbifold CFT gives the corresponding data of the nonlinear sigma model on the Fermat hypersurface.

\paragraph{Cohomological data at the Gepner point.} As discussed in Section~\ref{subsec: 2D CFT calculations}, the Jacobian ring \eqref{eq:hypersurface-Jacobian-ring} reduces to:
\begin{equation}
    R(\mathcal{W}_0) = \frac{\mathbb{C}[x_1,\ldots,x_{n + 2}]}{(x_1^{n + 1},\ldots,x_{n + 2}^{n + 1})} \,,
    \label{eq:Jacobian-ring-Gepner}
\end{equation}
with the basis
\begin{equation}
    x_{\alpha_{(p)}} \equiv \prod_{i = 1}^{n + 2}x_i^{\alpha_{(p)}^i} \,, 
    \qquad 
    0 \leq \alpha_{(p)}^i \leq n \,, 
    \qquad 
    \sum_{i = 1}^{n + 2}\alpha_{(p)}^i = p(n + 2) \,.
    \label{eq:Gepner-invariant-monomial-basis}
\end{equation}
Recall that $p = 0,\ldots,n$ corresponds to the antiholomorphic degree of the middle cohomology. For multi-indices corresponding to complex structure deformations, we reserve capital Latin letters, $\alpha_{(1)}=I$. At times, we will suppress the subscript $p$.
The Poincaré pairing for two cohomologies $\alpha_{(p)}$, $\beta_{(n-p)} $ in this basis becomes:
\begin{equation}
    \tilde{\eta}_{\alpha \beta} = \oint_{\{|x_i^{n+1}|=\epsilon\}} \frac{x_{\alpha} x_{\beta}\;d x_1\wedge\cdots\wedge d x_{n + 2}}{(2\pi i)^{n+2} \prod_{i = 1}^{n + 2} x_i^{n+1}} = \begin{cases}
        1 &\text{ if } \forall i: \; \alpha^i_{(p)} +\beta^i_{(n-p)} = n \,, \\
        0 &\text{ otherwise. }
    \end{cases}
    \label{Poincare duality}
\end{equation}
The unique top component is $x^n \equiv \prod_{i=1}^{n+2} x_i^n$. The Poincaré dual of $x_{\alpha_{(p)}}$ is then simply $x_{n-\alpha_{(p)}}$. 

Finally, we can compute the Hodge metric. At the Gepner point, each basis monomial $x_{\alpha_{(p)}}$ has a unique charge with respect to the discrete symmetry group, so that the $x_{\alpha_{(p)}}$ are orthogonal and there is no sphere-induced mixing. The sphere two-point function is therefore already the unnormalized metric.
With the normalization factor included, the integration cycle in \eqref{eq:Gepner-physical-metric-localization} separates into one-variable cycles. The latter are easy to compute\footnote{Let $d \equiv n + 2$ and $\lambda \equiv 4\pi R_{S^2}/d$. Up to an $a$-independent normalization of the measure, which cancels from $\mathcal I_a/\mathcal I_0$, setting $x = r e^{i\vartheta}$ gives
\begin{equation}
    \mathcal I_a \propto 2\pi\int_0^\infty d r\,r^{2a + 1}J_0(2\lambda r^d) = \frac{2\pi}{d}(2\lambda)^{-\frac{2(a + 1)}{d}}\int_0^\infty d t\,t^{\frac{2(a + 1)}{d} - 1}J_0(t)\,.
    \label{eq:Gepner-one-variable-Bessel-integral}
\end{equation}
Using $\int_0^\infty d t\,t^{\mu - 1}J_0(t) = 2^{\mu - 1}\frac{\Gamma(\mu/2)}{\Gamma(1 - \mu/2)}$ with $\mu = 2(a + 1)/d$ gives
\begin{equation}
    \frac{\mathcal I_a}{\mathcal I_0} = \lambda^{-\frac{2a}{d}}\frac{\Gamma\!\left(\frac{a + 1}{d}\right)\Gamma\!\left(1 - \frac{1}{d}\right)}{\Gamma\!\left(1 - \frac{a + 1}{d}\right)\Gamma\!\left(\frac{1}{d}\right)}\,.
    \label{eq:Gepner-one-variable-integral-ratio}
\end{equation}}, 
giving

\begin{equation}
    \begin{aligned}
        \mathcal I_a(R_{S^2}) & \equiv \int_{\Gamma_1}d x\,d\bar x\,(x\bar x)^a\exp\!\left[-\frac{4\pi iR_{S^2}}{n + 2}\left(x^{n + 2} + \bar x^{n + 2}\right)\right] \,, \\
        \mathfrak n(a) &\equiv (2R_{S^2})^{\frac{2a}{n + 2}}\frac{\mathcal I_a(R_{S^2})}{\mathcal I_0(R_{S^2})} = \left(\frac{n + 2}{2\pi}\right)^{\frac{2a}{n + 2}}\frac{\Gamma\!\left(\frac{a + 1}{n + 2}\right)\Gamma\!\left(1 - \frac{1}{n + 2}\right)}{\Gamma\!\left(1 - \frac{a + 1}{n + 2}\right)\Gamma\!\left(\frac{1}{n + 2}\right)} \,,
    \end{aligned}
    \label{eq:Gepner-one-variable-metric-moment}
\end{equation}
for $a = 0,\ldots,n$. Using $\sum_i\alpha_{(p)}^i = p(n + 2)$, \eqref{eq:Gepner-physical-metric-localization} then becomes
\begin{equation}
    \begin{aligned}
        g_{\alpha_{(p)}\bar\beta_{(p)}} & = \mathcal{N}(\alpha_{(p)})\delta_{\alpha_{(p)},\beta_{(p)}} \,, 
        \\
        \mathcal{N}(\alpha_{(p)}) & \equiv \prod_{i = 1}^{n + 2}\mathfrak{n}(\alpha_{(p)}^i) = \left(\frac{n + 2}{2\pi}\right)^{2p} \prod_{i = 1}^{n + 2} \frac{\Gamma\Bigl(\frac{\alpha_{(p)}^i + 1}{n + 2}\Bigr) \Gamma\Bigl(1 - \frac{1}{n + 2}\Bigr)}{\Gamma\Bigl(1 - \frac{\alpha_{(p)}^i + 1}{n + 2}\Bigr)\Gamma\Bigl(\frac{1}{n + 2}\Bigr)} \,.
    \end{aligned}
    \label{eq:Gepner-physical-metric}
\end{equation}
Substituting the structure constants \eqref{monomial structure constants} and \eqref{eq:Gepner-physical-metric} into \eqref{eq:Chern-curvature-components} gives the curvature directly. We may use $g$ in place of $G = \mathcal V_{\Omega}g$ because the common factor cancels between the metric and its inverse. Using \eqref{eq:Higgs-field-components},
\begin{equation}
    \begin{aligned}
        (\theta_I\theta_{\bar J}^\dagger)_{\alpha_{(p)}}{}^{\epsilon_{(p)}} & = g^{\rho\bar\mu}g_{\alpha_{(p)}\bar\nu}(C_{J\mu}{}^\nu)^*C_{I\rho}{}^{\epsilon_{(p)}} = \delta_{I + \alpha_{(p)},{\bar J} + \epsilon_{(p)}}\frac{\mathcal{N}\left(\alpha_{(p)}\right)}{\mathcal{N}\left(\epsilon_{(p)} - I\right)} \,, \\
        (\theta_{\bar J}^\dagger\theta_I)_{\alpha_{(p)}}{}^{\epsilon_{(p)}} & = C_{I\alpha_{(p)}}{}^\lambda g^{\epsilon_{(p)}\bar\gamma}g_{\lambda\bar\delta}(C_{J\gamma}{}^\delta)^* = \delta_{I + \alpha_{(p)}, {\bar J} + \epsilon_{(p)}}\frac{\mathcal{N}\left(\alpha_{(p)} + I\right)}{\mathcal{N}\left(\epsilon_{(p)}\right)} \,.
    \end{aligned}
    \label{eq:Gepner-Chern-curvature-two-terms}
\end{equation}
Their difference, which computes $F_{I\bar J} = [\theta_I,\theta^\dagger_{\bar J}]$, therefore gives\footnote{
For the normalized curvature, the curvature is shifted by a diagonal term $\mathcal{N}(I) \delta_{I,J} \delta^{\epsilon_{(p)}}_{\alpha_{(p)}}$, as discussed in Section~\ref{subsec: 2D CFT calculations}.
}
\begin{equation}
    (F_{I\bar J})_{\alpha_{(p)}}{}^{\epsilon_{(p)}} = \delta_{I + \alpha_{(p)},{\bar J} + \epsilon_{(p)}}\left(\frac{\mathcal{N}\left(\alpha_{(p)}\right)}{\mathcal{N}\left(\epsilon_{(p)} - I\right)} - \frac{\mathcal{N}\left(\alpha_{(p)} + I\right)}{\mathcal{N}\left(\epsilon_{(p)}\right)}\right) \,.
    \label{eq:Gepner-Chern-curvature-explicit}
\end{equation}
Each term is understood to vanish if one of its multi-indices does not label a monomial in the Jacobian ring.

\paragraph{A random direction and its curvature.}
 
We want to study the statistical properties of the Berry curvature when deforming in a generic direction in moduli space $v^I$, drawn from a Gaussian distribution with zero mean and variance
\begin{equation}
\mathbb{E}\bigl[v^I(v^J)^*\bigr] = \frac{\mathcal{J}^2}{h^{n-1,1}} g^{I\bar J} \,,
    \label{eq:random-direction-variance}
\end{equation} 
where $h^{n-1,1} = \dim H^{n-1,1}$ is the dimension of the complex structure moduli space. In the following, we will set the variance to $\mathcal{J}=1$. The curvature 
\begin{equation}
    F_{v\bar v} \equiv v^{I}(v^J)^* F_{I\bar J} 
    \label{eq:curvature-v-direction}
\end{equation}
is then the leading contribution to the plaquette operator of a short loop in the moduli space of complex structures in the (complex) direction $v^I$, around the Gepner point.

$F$ maps a middle cohomology class $H^{n-p,p}(X_\zeta)$ to itself, and is therefore block-diagonal. We will denote the block acting only on that class by $F^{(p)}$. Following \eqref{eq:Chern-curvature-commutator} we have 
\begin{equation}
    F^{(p)}_{v\bar v} = \theta_v^{(p-1)}\theta_{\bar v}^{(p-1)\dagger} - \theta_{\bar v}^{(p)\dagger}\theta_v^{(p)} \,,
    \label{eq:curvature-block}
\end{equation}
where we have defined (in the obvious way) the rectangular matrices 
\begin{equation}
    \!\theta_{v}^{(p)}:H^{n-p,p}(X_{\zeta}) \longrightarrow H^{n-p-1,p+1}(X_{\zeta})\,,\quad \theta^{(p)\dagger}_{\bar{v}}:H^{n-p-1,p+1}(X_{\zeta}) \longrightarrow H^{n-p,p}(X_{\zeta}) \,.
    \label{eq:Higgs-bidegree-labeled}
\end{equation}
At times, we will suppress the subscript $v$. In the orthonormal monomial basis $e_{\lambda_{(p)}} \equiv \mathcal N(\lambda)^{-1/2}x_{\lambda_{(p)}}$, we have
\begin{equation}
    \theta_v^{(p)}e_{\lambda_{(p)}} = \sum_I v^I \sqrt{\frac{\mathcal{N}(\lambda_{_{(p)}}+I)}{\mathcal{N}(\lambda_{_{(p)}})}} e_{\lambda_{(p)}+I}.
\end{equation}
As $v^I$ comes with standard deviation proportional to $\mathcal{N}(I)^{-\frac{1}{2}}$ due to the inverse metric, the curvature is only dependent on the ratio $\sqrt{\frac{\mathcal{N}(\alpha+I)}{\mathcal{N}(\alpha) \mathcal{N}(I)}}$.

Before we write down the curvature, it is interesting to analyze the two terms in \eqref{eq:curvature-block} separately. By singular value decomposition, the nonzero eigenvalues of $\theta^{(p)\dagger}_{\bar v} \theta^{(p)}_v$ are the same as those of $\theta^{(p)}_{v} \theta^{(p)\dagger}_{\bar v}$. In practical calculations, it is therefore enough to find the spectrum of $\theta_v^{(p)}\theta_{\bar{v}}^{(p)\dagger}$, which can be written explicitly using \eqref{eq:Gepner-Chern-curvature-two-terms} and \eqref{eq:Gepner-physical-metric}. The matrix $\theta_v^{(p)}\theta_{\bar{v}}^{(p)\dagger}$ acts on $H^{n-p - 1,p + 1}(X_{\zeta})$, which corresponds to Jacobian-ring degree $(p + 1)(n + 2)$. In the orthonormal monomial basis $e_{\lambda_{(p + 1)}}$, its components are
\begin{equation}
    \begin{aligned}
        (\theta_v^{(p)}\theta_{\bar{v}}^{(p)\dagger})_{\lambda_{(p + 1)}}{}^{\mu_{(p + 1)}} & = \sum_{I,J} v^I (v^J)^* \delta_{I + \lambda_{(p + 1)},J + \mu_{(p + 1)}} \frac{\sqrt{\mathcal N(\lambda) \mathcal N(\mu)}}{\mathcal N(\mu - I)} \\
        & = \sum_{\alpha_{(p)}} \frac{\sqrt{\mathcal N(\lambda) \mathcal N(\mu)}}{\mathcal N(\alpha)} v^{\mu - \alpha} \left(v^{\lambda - \alpha}\right)^* \,.
    \end{aligned}
    \label{eq:A-explicit-expression}
\end{equation}
In the first line, a term contributes only when $\alpha \equiv \mu - I = \lambda - J$ labels a valid source monomial. Equivalently, $\sum_{\alpha_{(p)}}$ in the second line runs over the monomial basis of the degree-$p(n + 2)$ component of the Jacobian ring. 

The full curvature in the orthonormal frame is then assembled to be\footnote{With the curvature viewed as the Hamiltonian, the system is not $k$-local at large $n$ in the usual sense, but it is not all-to-all mixing: consider two indices $\alpha_{(p)}$, $\beta_{(p)}$. We can then define their ``distance'' 
\begin{equation}
    d(\alpha_{(p)},\beta_{(p)}) = \frac{1}{n+2}\sum_{i=1}^{n+2} |\alpha_{(p)}^i-\beta_{(p)}^i| \,.
\end{equation}
For $p\approx n/2$, where most of the cohomologies sit, the average distance is $\sim n$, while each application of the curvature changes the distance by at most $2$. Perhaps there is a basis where this structure is more apparent. }
\begin{equation}\label{Curvature full expression}
    (F_{v\bar{v}})_{\lambda_{(p)}}{}^{\mu_{(p)}} =  \sum_{I,J}v^I (v^J)^* \left(\sqrt{\frac{\mathcal{N}(\lambda+I)^2}{\mathcal{N}(\lambda)\mathcal{N}(\lambda + I -J)}} - \sqrt{\frac{\mathcal{N}(\lambda-J+I)\mathcal{N}(\lambda)}{\mathcal{N}(\lambda-J)^2}}\right)\delta_{\mu_{(p)+J},\lambda_{(p)}+I} \,.
\end{equation}
As in \eqref{eq:Gepner-Chern-curvature-explicit}, each term in the brackets is understood to vanish if one of its multi-indices does not label a monomial in the Jacobian ring.

\paragraph{Some numbers.}
The spectral statistics of the curvature ensemble \eqref{Curvature full expression} can now be studied using exact diagonalization, with the main numerical bottleneck being the dimension of the matrices, given by $h^{n-p,p}(X)$.

The first few Hodge numbers of the primitive middle cohomology $H^{n-p,p}_{\text{prim}}(X)$ of the hyperquintic with increasing $n$ are displayed in Table~\ref{tab:hodge-numbers-quintic}.\footnote{The displayed Hodge numbers $h^{n-p,p}$ count the number of the primitive cohomologies of the $H^{n-p,p}_{\text{prim}}(X)$ Hodge bundles. In cases where $n$ is even, in the full middle $H^{\frac{n}{2},\frac{n}{2}}(X)$ cohomology there is a single ambient cohomology coming from the restriction of an appropriate power of the \Kahler form to the complete intersection space. Therefore the full number of cohomologies in this slice is $h^{\frac{n}{2},\frac{n}{2}}=h_{\text{prim}}^{\frac{n}{2},\frac{n}{2}}+1$. This subtlety is inconsequential for the ensuing statistical analysis since even at small $n$, $h^{\frac{n}{2},\frac{n}{2}}\gg 1$. Note that the ambient cohomology does not have a representative in the polynomial ring.}
\begin{table}[htbp]
\centering
\setlength{\tabcolsep}{0pt}
\begin{tabular}{c@{\hspace{0.75em}}|@{\hspace{0.75em}}c}
$n$ & \\[4pt]
1 & \hodgecell{1}\hodgecell{1}\\[4pt]
2 & \hodgecell{1}\hodgecell{19}\hodgecell{1}\\[4pt]
3 & \hodgecell{1}\hodgecell{101}\hodgecell{101}\hodgecell{1}\\[4pt]
4 & \hodgecell{1}\hodgecell{426}\hodgecell{1751}\hodgecell{426}\hodgecell{1}\\[4pt]
5 & \hodgecell{1}\hodgecell{1667}\hodgecell{18327}\hodgecell{18327}\hodgecell{1667}\hodgecell{1}\\[4pt]
6 & \hodgecell{1}\hodgecell{6371}\hodgecell{154645}\hodgecell{398567}\hodgecell{154645}\hodgecell{6371}\hodgecell{1}\\[4pt]
7 & \hodgecell{1}\hodgecell{24229}\hodgecell{1170073}\hodgecell{6262237}\hodgecell{6262237}\hodgecell{1170073}\hodgecell{24229}\hodgecell{1}
\end{tabular}
\caption{Some Hodge numbers of the hyperquintic. The first row is the 1-fold (the torus), and the last row is the 7-fold. $p$ increases from left to right. We can compute these numbers using the Hirzebruch--Riemann--Roch theorem \cite{Hirzebruch1966} or by the monomial counting of the previous section. The dimension of the central cohomology $h^{\lceil\frac{n}{2}\rceil,\lfloor \frac{n}{2}\rfloor}$ grows much more rapidly than $h^{n-1,1}$, which encodes the number of directions in moduli space. The large-$n$ model is therefore sparse for generic middle cohomologies.}
\label{tab:hodge-numbers-quintic}
\end{table}
In practice, the complex structure slice ($p=1$) of $n=4,5,6,7$ and the $p=2$ slice of $n=4,5$ are both of sufficiently large dimension and numerically accessible on a computer. The $p$-slices differ in sparsity (the dimension of $H^{n-p,p}$ quickly becomes much larger than the number of parameters as $p$ grows), and we analyze them separately.\footnote{In the $p=1$ case, the $\theta_{v}^{(0)} \theta_{\bar{v}}^{(0)\dagger}$ part is rank 1, and so the full spectrum of $F_{v\bar{v}}^{(1)}$ and the spectrum of $\theta^{(1)\dagger}_{\bar{v}} \theta_v^{(1)}$ are the same up to a rank-one deformation (and a minus sign).  In this case, we thus focus on the curvature itself, and not on the individual pieces. }

\subsection{Spectral diagnostics of chaos}\label{subsec:BPS-chaos-moduli-space}

A standard quantitative measurement of chaos is the chaos exponent. Since it is not accessible in the BPS setting, we instead first diagnose the nearest-neighbor eigenvalue repulsion (which is believed to be nearly universal \cite{Bohigas1984CharacterizationOC}) using the level spacing ratio. A possible way of quantifying BPS chaos \cite{Chen:2024oqv} is to study the ratio between plateau and Thouless time in the connected spectral form factor, and observe its scaling with the system size. Effectively, it can be thought of as a measure of what fraction of the Hilbert space participates in the chaotic dynamics. In our case, the ratio between plateau and Thouless time does grow with system size, although it remains unclear if there is strong chaos in the sense of \cite{Chen:2024oqv}. A related measure is the number variance, which we also determine. We find deviations from RMT-like behavior that were also observed in SYK-like systems \cite{Altland:2017eao,Jia:2019orl}.

In the following analysis, we always remove the zero eigenvalues from the spectra, where they appear due to rank deficiency.

\subsubsection{The level spacing ratio}

The level spacing ratio of the spectrum $\{\lambda_i\}$ of a Hermitian matrix is defined as follows. We order the eigenvalues $\lambda_i$ by their size and form the ratios:\footnote{The reason why we use here a definition of the level spacing ratio that is by construction between $0$ and $1$ is that for computing moments of $r$ and comparing with other distributions, it is the more reliable quantity \cite{Oganesyan_2007} (for example, the variance of the ordinary GOE level spacing distribution does not converge). }
\begin{equation}
    r_i = \frac{\min (s_i,s_{i-1})}{\max (s_i,s_{i-1})}, \qquad \text{ where }
    s_i = \lambda_{i+1}-\lambda_i \,.
\end{equation}
We then consider a histogram of these ratios and extract a probability distribution $P(r)$.  For integrable systems, one usually obtains a Poisson distribution:
\begin{flalign}
    &P(r) = \frac{2}{(1+r)^2} \,, \qquad
    \langle r \rangle_{\text{Poisson}} = 2 \log 2 -1 \approx 0.386 \,.
\end{flalign}
For a GUE random matrix ensemble, one finds the curve\footnote{This curve and the GOE curve below are the curves one finds for the level spacing ratios of three levels. It is known that the (numerically found) large-dimension curves are very close to that.} \cite{Atas_2013}
\begin{equation}
    \begin{split}
        P_{\text{GUE}}(r) =\frac{81\sqrt{3}}{2\pi } \frac{(r+r^2)^2}{(1+r+r^2)^4} \,, \qquad
        \langle r \rangle_{\text{GUE}} = 0.5996 \,.
    \end{split}
\end{equation}
For a GOE random matrix ensemble, the probability distribution reads:
\begin{equation}
    \begin{split}
        P_{\text{GOE}}(r) =\frac{27}{4 } \frac{(r+r^2)}{(1+r+r^2)^{\frac{5}{2}}} \, \qquad
        \langle r \rangle_{\text{GOE}} = 0.5307 \,.
    \end{split}
\end{equation}
\vspace{1pt}
\paragraph{The complex structure slice.} When $p=1$, $H^{n-1,1}$ describes the complex structure deformations, and the Berry curvature is the Riemannian curvature of the conformal manifold. The level spacing ratios of the curvature for $n=4,5,6,7$ and $p=1$ follow the GUE curve to very high accuracy. An example for $n=4$ and $n=7$ is plotted in Figure~\ref{fig:d=6,9 complex structure LSR}. 
\begin{figure}[h]
    \centering
    \includegraphics[width=0.95\linewidth]{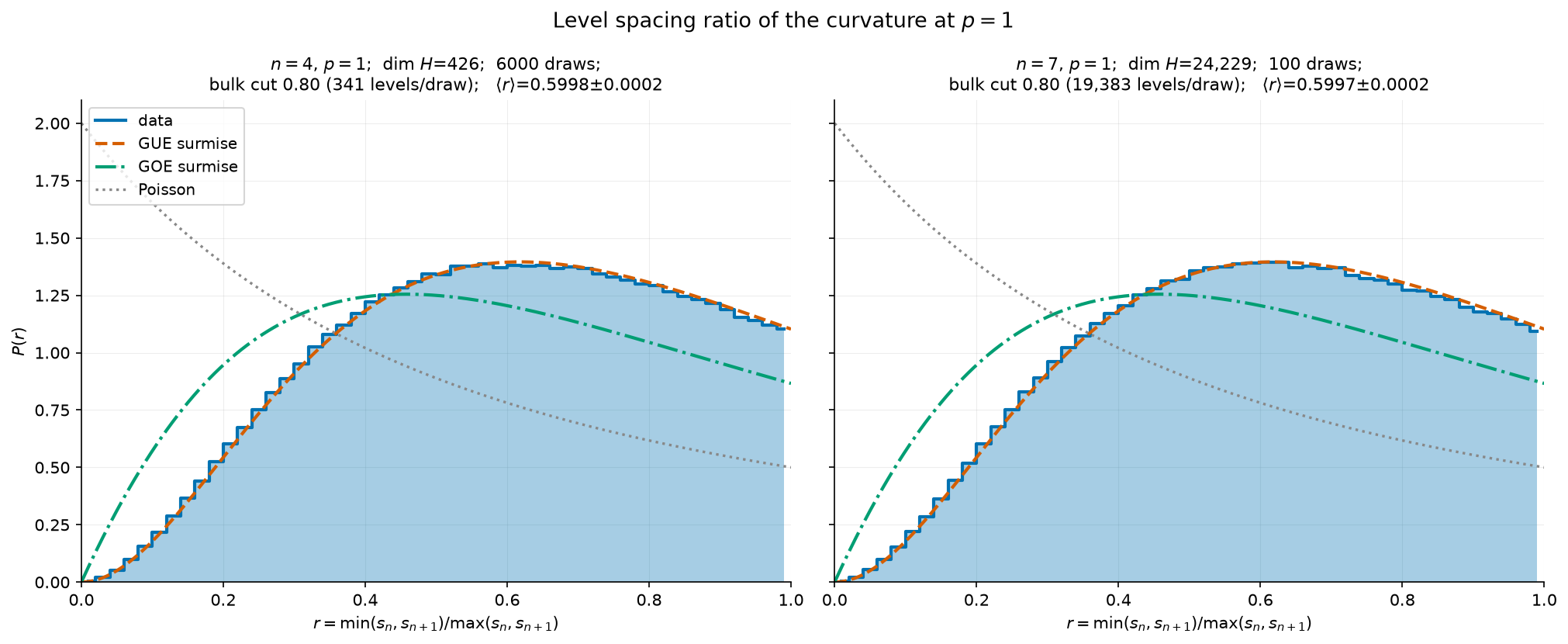}
    \caption{The level spacing ratio histograms for the curvature on the complex structure slice of the 4-fold (left) and the 7-fold (right) for $p=1$. The histograms follow the GUE expectation closely. Errors are the standard error of the mean over the draws. There are many more draws for $n=4$, as the curvature matrix is smaller.}
    \label{fig:d=6,9 complex structure LSR}
\end{figure}

\begin{figure}
    \centering
    \includegraphics[width=0.95\linewidth]{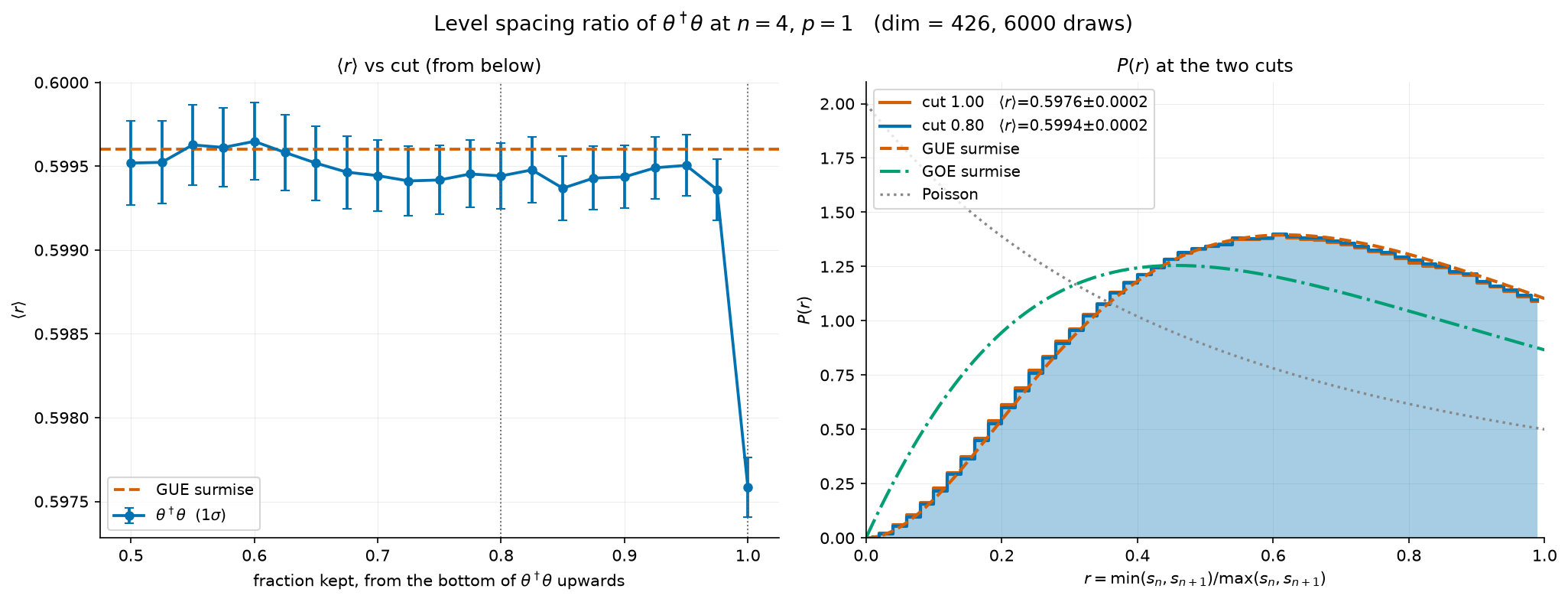}
    \caption{Right: Level spacing ratio histograms of $\theta_{\bar{v}}^{(1)\dagger}\theta_v^{(1)}$ for $n=4, p=1$. Left: the average level spacing ratio compared to the choice of bulk cut, starting from the bottom upwards. Similar results apply for a symmetric cut. It is close to GUE value except for the very top edge. We plotted $\theta_{\bar{v}}^{(1)\dagger}\theta_v^{(1)}$ instead of the curvature to avoid contamination of $F_{v\bar{v}}^{(1)}$ from $\theta_v^{(0)}\theta_{\bar{v}}^{(0)\dagger}$ at the very edge of the spectrum.}
    \label{fig:lsr with different cuts.png}
\end{figure}

In Figure~\ref{fig:d=6,9 complex structure LSR}, we have chosen a bulk cut: we have removed the tail ends of the spectrum and only kept the middle 80\%. One can check that the results for different bulk cuts are almost identical, except when one includes everything, including the very top of the spectrum (see Figure~\ref{fig:lsr with different cuts.png}). A GUE matrix of the same size shows no such behavior. The relative size of the effect decreases with increasing $n$. A possible origin of that will be discussed later. 

For completeness, we also state the results of the $n=5,6$ level spacing ratios, together with $n=4$ and $n=7$, in Table~\ref{tab:lsrp=1}. All of them are within $1\sigma$ of the large-$N$ GUE result.

\begin{table}[h]
  \centering
  \begin{tabular}{lrrcl}
    \toprule
    $n$ & $\dim\mathcal{H}_{p=1}$ & $N_{\mathrm{draws}}$ & $\bar{r}$ \\
    \midrule
    4 & 426   & 6000 & 0.5998(2) \\
    5 & 1667  & 1000 & 0.5995(3) \\
    6 & 6371  & 300  & 0.5995(3) \\
    7 & 24229 & 100  & 0.5997(2) \\
    \bottomrule
  \end{tabular}
  \caption{Mean level spacing ratio $ r_i=\min(s_i,s_{i-1})/\max(s_i,s_{i-1})$ of the complex structure slice $(p=1)$. A bulk cut of $0.80$ has been taken for all of the spectra. All of them are within $1\sigma$ of the large-$N$ GUE value.}
  \label{tab:lsrp=1}
\end{table}

\paragraph{The central slice ($p=2$ for $n=4,5$).}

For $n=4$, we sit exactly in the slice $H^{2,2}$ and $\theta^{{(2)} \dagger}_{\bar{v}}\theta^{(2)}_v$ and $\theta_{v}^{(1)} \theta^{(1)\dagger}_{\bar{v}}$ are related to each other via Poincar\'{e} duality \eqref{Poincare duality}. The spectrum comes exactly in $\pm$ eigenvalue pairs.

The two pieces in $F_{v\bar{v}}^{(2)}$ do not factorize: their images have nonzero overlap, which causes the total curvature to have a narrower spectrum than the individual components. Each individual piece has rank $426$ (exactly the dimension of the $p=1$ and $p=3$ slices), and the full curvature has rank $852=2\times 426$, so there are no cancellations. 

\begin{figure}[H]
    \centering
    \includegraphics[width=0.95\linewidth]{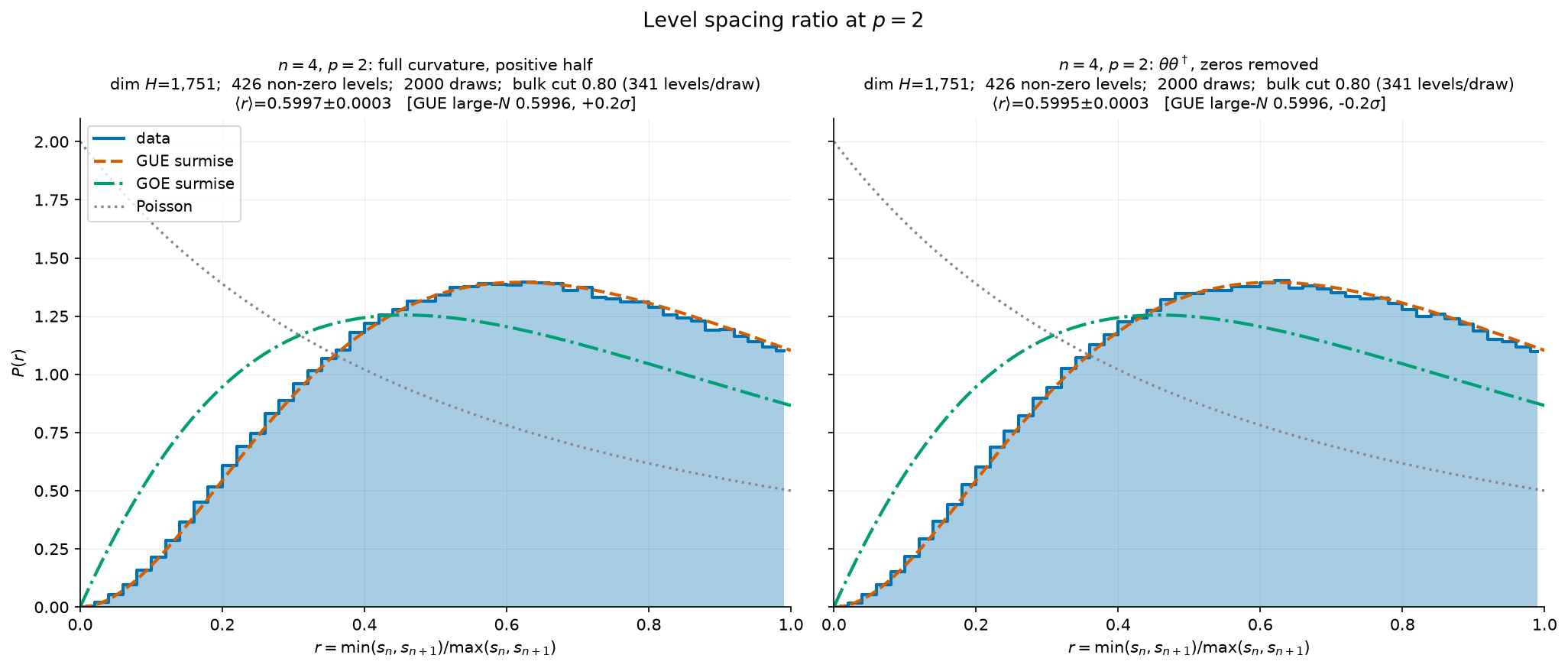}
    \caption{The level spacing ratio histograms for the curvature $F_{v\bar v}$ (left) and $\theta_v^{(1)} \theta^{(1)\dagger}_{\bar{v}}$ (right) on the central slice of the 4-fold at $p=2$. The histograms follow the GUE expectation closely. Errors are the standard error of the mean over the draws.}
    \label{fig:lsr 4-fold middle}
\end{figure}

We can study the level spacing ratio of both $\theta_{v}^{(1)} \theta^{(1)\dagger}_{\bar{v}}$ (individually) and the full  curvature $F_{v\bar{v}}^{(2)}$. 
In both cases, we disregard the zero modes. For the full curvature, we also have to project to the positive spectrum. In both cases, the result closely matches the GUE curve (see Figure~\ref{fig:lsr 4-fold middle}).

Similarly to the complex structure slice, one can observe that the average level spacing ratio of $\theta_{\bar{v}}^{(1)\dagger}\theta_v^{(1)}$ stays essentially constant across bulk cuts, except when one includes the very top of the spectrum. 

The case $n=5$ (the 5-fold) is more interesting. As $\theta_{v}^{(2)}$ maps from $H^{3,2}(X_\zeta)$ to $H^{2,3}(X_\zeta)$, the combination $\theta^{(2)\dagger}_{\bar{v}} \theta_v^{(2)}$ is full rank, while the combination $\theta_v^{(1)} \theta^{(1)\dagger}_{\bar{v}}$ has rank $1667$ (the complex structure slice of the 5-fold). In Figure~\ref{fig:d=7, p=2 lsr}, we plot the level spacing ratio for both individual pieces and for the entire $F_{v\bar{v}}^{(2)}$.

\begin{figure}[h]
    \centering
    \includegraphics[width=0.9\linewidth]{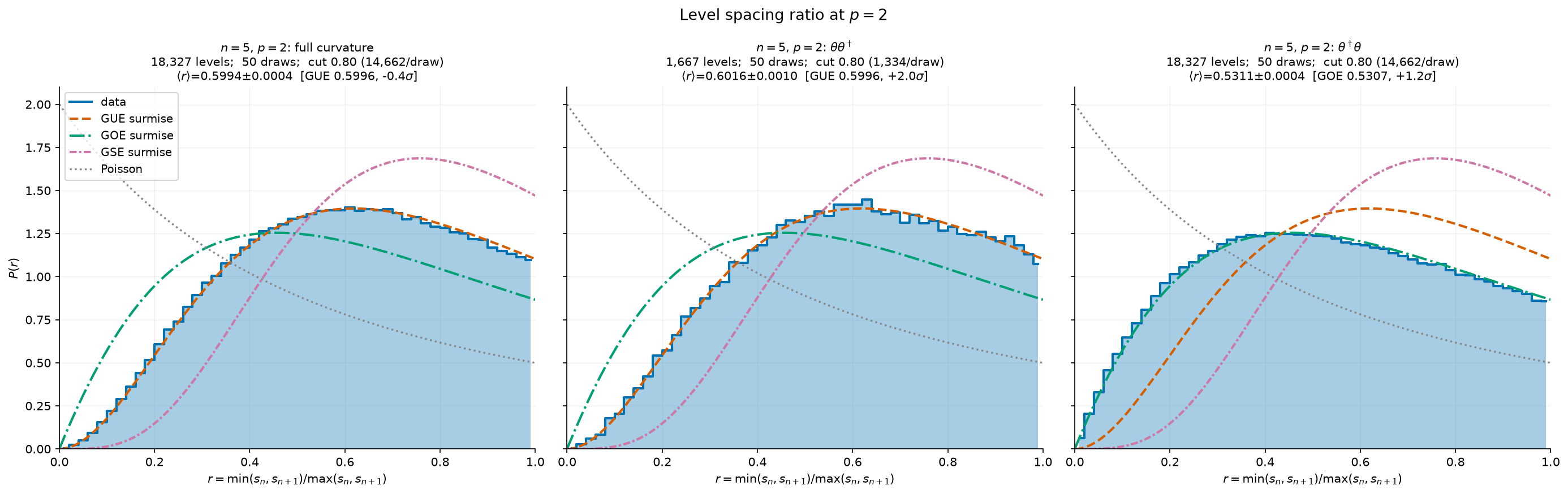} 
    \caption{The level spacing ratio histograms for the curvature (left), $\theta_v^{(1)} \theta^{(1)\dagger}_{\bar{v}}$ (middle), and $\theta^{(2)\dagger}_{\bar{v}} \theta^{(2)}_v$ (right) on the $p=2$ slice of the 5-fold. The first two follow GUE closely, while the third mimics a GOE distribution. Errors are the standard error of the mean over the draws. They are larger here as the number of draws is comparatively small.}
    \label{fig:d=7, p=2 lsr}
\end{figure}
We see that while the rank-deficient $\theta_v^{(1)} \theta^{(1)\dagger}_{\bar{v}}$ and the curvature are close to GUE, the full-rank $\theta_{\bar{v}}^{(2)\dagger} \theta_v^{(2)}$ is GOE. This is because of an antiunitary symmetry: for the $H^{\frac{n+1}{2},\frac{n-1}{2}}(X_{\zeta})$ slice, one can use the fact that applying $\theta_v^{(2)}$ together with the Poincar\'{e} pairing \eqref{Poincare duality} gives a complex-symmetric matrix $S$ on $H^{\frac{n+1}{2},\frac{n-1}{2}}(X_{\zeta})$. Together with complex conjugation, that $S$ gives rise to an antiunitary operator that commutes with $\theta_{\bar{v}}^{(2)\dagger} \theta_v^{(2)}$, which causes the GOE behavior. The rank-deficient $\theta_v^{(1)} \theta^{(1)\dagger}_{\bar{v}}$ is enough to lift that symmetry and return the full curvature to the GUE ensemble.

In Table~\ref{tab:lsrp=2}, we summarize the number of draws and average level spacing ratio for $n=4$ and $n=5$ at their central slice. Both of them are within $1\sigma$ of the large-$N$ GUE result.

\begin{table}[h]
  \centering
  \begin{tabular}{lrrcl}
    \toprule
    $n$ & $\dim\mathcal{H}_{p=2}$ & $N_{\mathrm{draws}}$ & $\bar{r}$ \\
    \midrule
    4 & 1751 (rank 426)   & 2000 & 0.5997(4) \\
    5 & 18327  & 50 & 0.5994(4) \\
    \bottomrule
  \end{tabular}
  \caption{Mean level spacing ratio $ r_i=\min(s_i,s_{i-1})/\max(s_i,s_{i-1})$ of the central slice $(p=2)$ for the 4-fold and 5-fold. A bulk cut of $0.80$ has been taken for all of the spectra. Both of them are within $1\sigma$ of the large-$N$ GUE value.}
  \label{tab:lsrp=2}
\end{table}

\subsubsection{The spectral form factor}

The level spacing ratio only probes eigenvalue correlations between nearest neighbors. To probe longer-range correlations, we look at the spectral form factor. It is defined as
\begin{equation}
    \langle |Z(it)|^2 \rangle \sim \bigg\langle \sum_{n,m} e^{i(\lambda_n-\lambda_m)t} \bigg\rangle \,.
\end{equation}
Similarly to \cite{Gharibyan:2018jrp}, we consider a Gaussian filter of the spectral form factor. More precisely, per draw $a$ and associated eigenvalues $\{\lambda_n^{(a)}\}$, we define
\begin{equation}
    Y_a(\alpha,t)=\sum_n e^{-i \lambda_n^{(a)} t} e^{-\alpha (\lambda_n^{(a)}-\lambda_0)^2} \,.
\end{equation}
Here $\lambda_0$ is the median of the average distribution (not zero, as the spectrum is not symmetric around zero). At zero time, we have
\begin{equation}
    \langle |Y_a(\alpha,0)|^2 \rangle = \bigg\langle \left(\sum_n e^{-\alpha(\lambda_n^{(a)}-\lambda_0)^2} \right)^2\bigg\rangle \equiv \langle (Z_1^{(a)})^2 \rangle \,.
\end{equation}
At late times, all the non-diagonal contributions of the sum in $|Y_a(\alpha,t)|^2$ dephase and we have
\begin{equation}
    \lim_{t\rightarrow \infty}\langle |Y_a (\alpha,t)|^2 \rangle = \bigg\langle \sum_n e^{-2\alpha (\lambda_n^{(a)}-\lambda_0)^2} \bigg\rangle= \langle Z_2^{(a)} \rangle \,.
\end{equation}
The precise object we then calculate is
\begin{equation}
    h(\alpha,t)=\frac{\langle |Y_a (\alpha,t)|^2 \rangle}{\langle |Y_a (\alpha,0)|^2 \rangle} \,.
\end{equation}
This ratio starts at one. At very long times, it approaches the asymptotic value
\begin{equation}
    \lim_{t\rightarrow \infty} h(\alpha,t)=\frac{\langle Z_2^{(a)}\rangle }{\langle (Z_1^{(a)})^2 \rangle} = h_{\text{plateau}}= \frac{1}{L_{\text{eff}}} \,.
\end{equation}
This is the height of the plateau; its inverse is a measure of the effective degrees of freedom participating (which depends on $\alpha$). The Heisenberg time at which the plateau appears is given by 
\begin{equation}
    t_H = 2\pi \bar{\rho}_f = \sqrt{8\pi \alpha} \langle Z_2^{(a)} \rangle \,,
\end{equation}
where on the right-hand side we have used the fact that we have a Gaussian filter $f$. 

We can also separate out the connected piece, by defining
\begin{equation}
    \overline{Y}(t)=\frac{1}{N_s} \sum_a Y_a(t) \,, \qquad \hat{S}(t)=\frac{1}{N_s-1}\sum_a (Y_{a}(t)-\overline{Y}(t))^2 \,.
\end{equation}
Here $N_s$ is the number of draws. Then the disconnected and connected pieces are
\begin{equation}
    \begin{split}
        h_d(t)=\frac{|\overline{Y}(t)|^2-\frac{1}{N_s}\hat{S}(t)}{\langle (Z_1^{(a)})^2 \rangle} \,, \qquad 
        h_c(t) = \frac{\hat{S}(t)}{\langle (Z_1^{(a)})^2 \rangle} \,.
    \end{split}
\end{equation}
It is natural to compare the data to the ramp--plateau behavior of GUE and GOE, for which we have
\begin{equation}
    h_{\text{GUE/GOE}}(\alpha,t) = h_{\text{plateau}} K_{\text{GUE/GOE}}(t/t_H) \,, 
\end{equation}
with \cite{Meh2004}
\begin{equation}
    K_{GUE}(x) = 
        \begin{cases}
             x & x \leq 1 \,, \\
             1 & x \geq 1 \,,
        \end{cases}
     \qquad    
     K_{GOE}(x) = 
        \begin{cases}
            2 x- x\log(1+2x) & x \leq 1 \,, \\
            2-x\log \frac{2x+1}{2x-1} & x \geq 1 \,.
        \end{cases}
\end{equation}
The modeling choice is essentially in the value of the gaussian window $\alpha$, which we determine as follows (see \cite{Gharibyan:2018jrp}). Too small $\alpha$ implies that the Gaussian window is across the whole spectrum and the fluctuations in the density obscure the ramp. For too large $\alpha$, the number of states becomes small, and the Gaussian filter's Fourier transform $e^{-t^2/(4\alpha)}$ also obscures the ramp.\footnote{One would naively think that only the disconnected part would see the Fourier transform of the filter. However, with the density fluctuating from draw to draw (for example with its median fluctuating), the Fourier transform of the filter survives, with its amplitude suppressed by the relative size of the fluctuations.} Essentially, for the purposes of determining the ramp time, we want to pick as small an $\alpha$ as we can get away with. We measure it in units of the variance of the spectrum $\sigma^2_E$, which we compute as the mean of the per-draw variances. The dimensionless ratio is 
\begin{equation}
    r = \frac{1}{2\alpha \sigma_E^2}, 
\end{equation}
which we will determine by picking the largest $r$ that still suppresses the fluctuations.

\paragraph{The complex structure slice $(p=1)$.}

For the complex structure slice, we once again consider the data from the 4-fold, 5-fold, 6-fold and 7-fold. The fact that we have a series of $p=1$ slices will allow us to make some qualitative statements about the scaling of the ramp time with the system size. 

\begin{figure}
    \centering
    \includegraphics[width=0.9\linewidth]{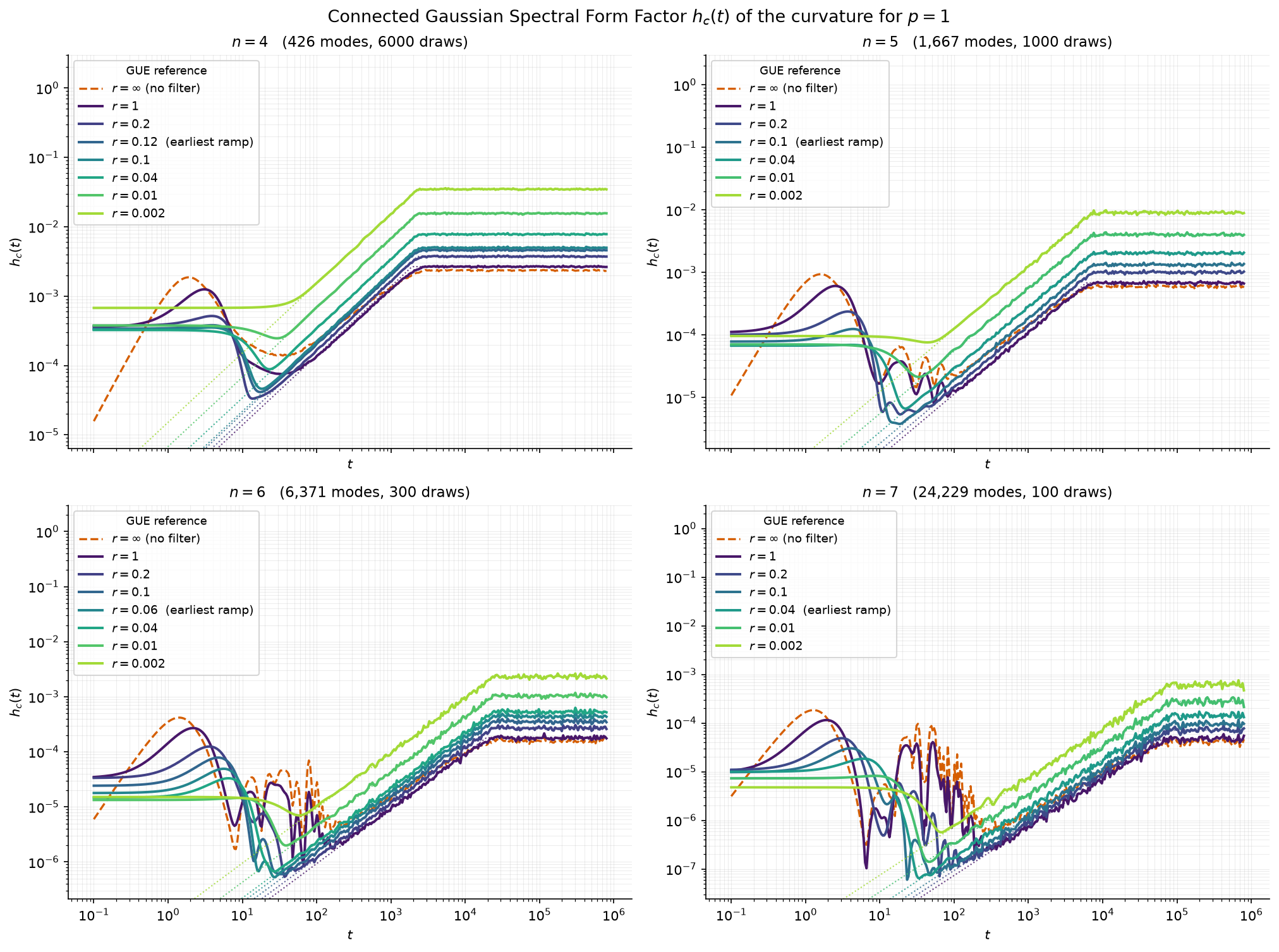}
    \caption{The (unnormalized) connected Gaussian spectral form factor of the complex structure slice with varying window size. Different colors represent different filter variance $r$. As the $r$ decreases, the fluctuations get suppressed and the ramp time decreases. For too small $r$, the Fourier transform of the Gaussian filter obscures the ramp. The ``optimal'' $r$ decreases with increasing system size. The $n=4$ curves look smoother because of the much larger number of draws.}
    \label{fig:Gaussian window}
\end{figure}

\begin{figure}
    \centering
    \includegraphics[width=0.7\linewidth]{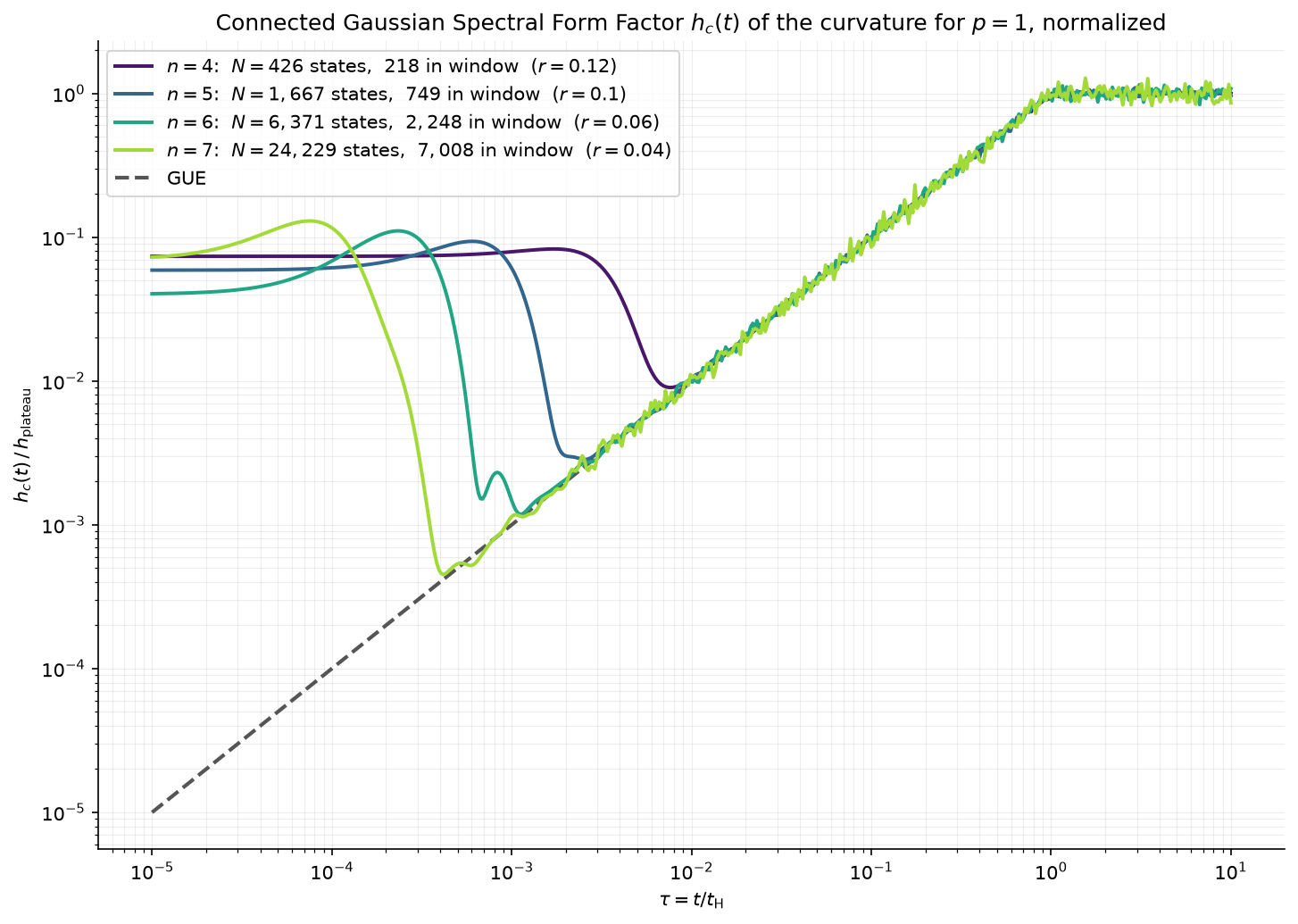}
    \caption{The (Gaussian) spectral form factor with a filter chosen such that the earliest ramp time is achieved for each $n$. Time is measured in units of the Heisenberg time and the plateau is normalized to one. We see that the ramp time scales with system size.}
    \label{fig:ramp time scaling}
\end{figure}

In Figure~\ref{fig:Gaussian window}, we show the behavior of the Gaussian spectral form factor with varying window size.
We see exactly the two effects described above: too small $r$ obscures the ramp because of the Fourier transform of the Gaussian window, while too large $r$ does not suppress the fluctuations that also obscure the ramp. The value of $r$ with the earliest ramp time is decreasing with increasing system size. Practically, the $2\sigma$-Gaussian window encompasses at least $30\%$ of the total number of states. Normalizing the plateau now to one and measuring time in units of the Heisenberg time, we find Figure~\ref{fig:ramp time scaling}: the ramp time scales with the system size. 
We can determine a proxy for the ramp time by asking about when $h$ is first ``close'' to the GUE ramp. That is, we determine the ramp time as
\begin{equation}
    t_{\text{ramp}}=\min\bigg\{t_*|\;\forall t \in [t_*,t_{\text{max}}] \quad \bigg|\log \frac{h(t)}{h_{GUE}}\bigg| < \epsilon \bigg\} \,.
\end{equation}
Here $t_{\text{max}}$ is some time close to the plateau (for example $t_H/4)$ and $\epsilon$ is an arbitrary parameter. The precise value of the ramp time depends on the parameter $\epsilon$, but it only fluctuates mildly with changing $\epsilon$. The resulting ramp times are summarized in Table~\ref{tab: ramp times}.

\begin{table}[h]
  \centering
  \begin{tabular}{cccccc}
    \toprule
    $n$ & $N$ & draws
        & \multicolumn{3}{c}{$\tau_{\text{ramp}} = t_{\text{ramp}}/t_{\text{H}}$} \\
    \cmidrule(lr){4-6}
    & & & $\epsilon = 0.3$ & $\epsilon = 0.25$ & $\epsilon = 0.2$ \\
    \midrule
    4 & 426  & 6000  & $9.64\times10^{-3}$ & $9.43\times10^{-3}$ & $9.32\times10^{-3}$ \\
    5 & 1667 & 1000   & $2.99\times10^{-3}$ & $2.92\times10^{-3}$ & $2.89\times10^{-3}$ \\
    6 & 6371 & 300 & $1.17\times10^{-3}$ & $1.16\times10^{-3}$ & $1.15\times10^{-3}$ \\
    7 & 24229 & 100 & $4.45\times10^{-4}$ & $4.40 \times 10^{-4}$ &  $4.35 \times 10^{-4}$ \\
    \bottomrule
  \end{tabular}
  \caption{The ramp time as a function of $\epsilon$, for $n=4,5,6,7$. One can see that it is not very sensitive to the choice of $\epsilon$, and that the ratio between Heisenberg time and ramp time increases with system size. As the earliest ramp $r$ decreases with system size, it is hard to read off a direct power law, but the inverse ramp time is increasing roughly with the number of degrees of freedom in the Gaussian window. We have employed a short-time averaging (average of $\log h_c$ over 0.02 decades) to remove fluctuations.}
  \label{tab: ramp times}
\end{table}

\paragraph{The central slice ($p=2$ for $n=4,5$).}

\begin{figure}[h]
    \centering
    \includegraphics[width=0.85\linewidth]{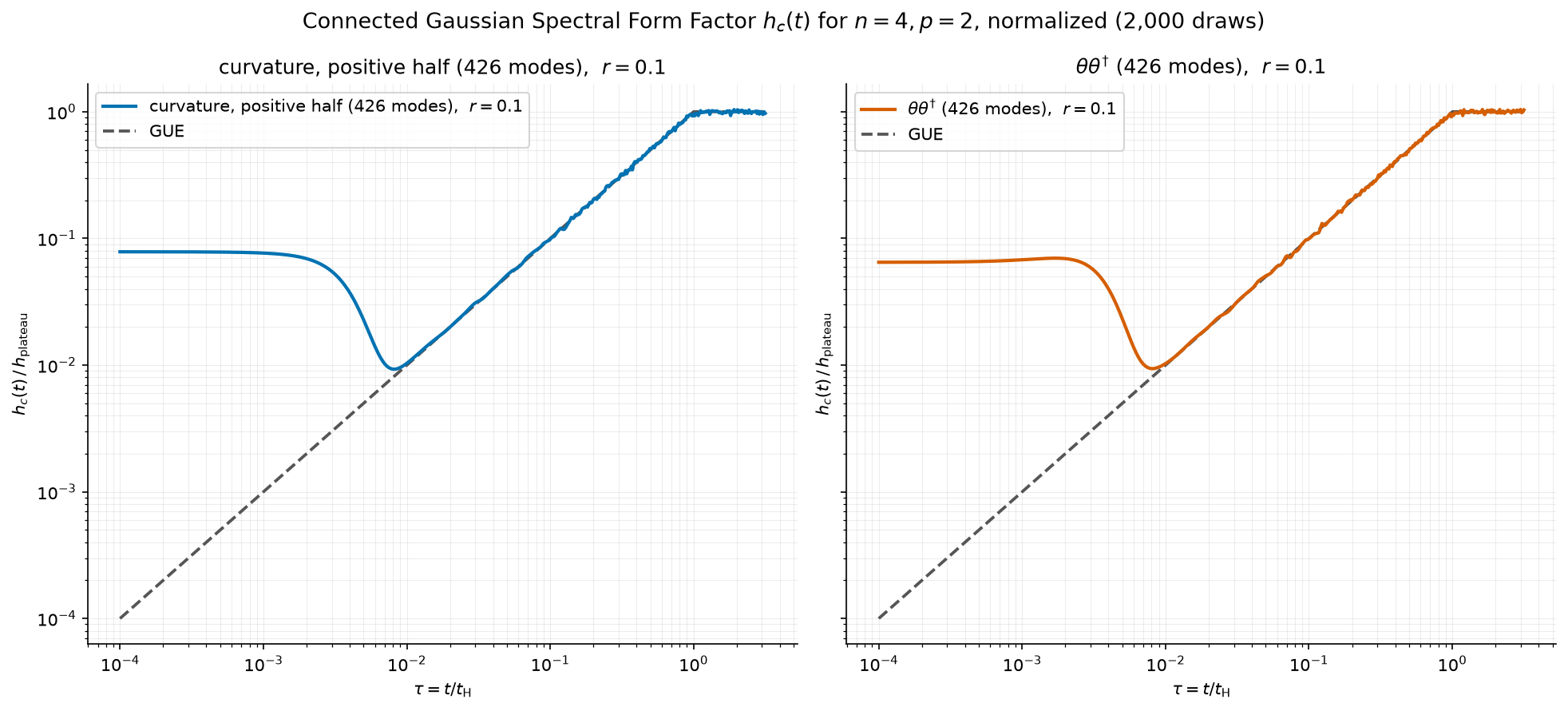}
    \caption{The Gaussian spectral form factor for the central slice of the 4-fold, at $r = 0.1$. The full curvature and the $\theta_v^{(1)} \theta^{(1)\dagger}_{\bar{v}}$ piece show almost the same behavior.}
    \label{fig:d=6 p=2 SFF}
\end{figure}

For the central slice of the 4-fold ($n=4$), we can compute the Gaussian SFF in the same way as we just sketched: vary the window size, pick the largest possible window which does not obscure the ramp onset through its Fourier transform, and compare to the GUE expectation. One can do this both for the positive part of the full curvature and for the nonzero part of $\theta_v^{(1)} \theta^{(1)\dagger}_{\bar{v}}$. The result is in Figure~\ref{fig:d=6 p=2 SFF}.

We do not extract a ramp time here as we do not have several cases of the central slice of an even-dimensional hyperquintic, so we cannot read off any $N$-scaling.

\begin{figure}[h]
    \centering
    \includegraphics[width=1.0\linewidth]{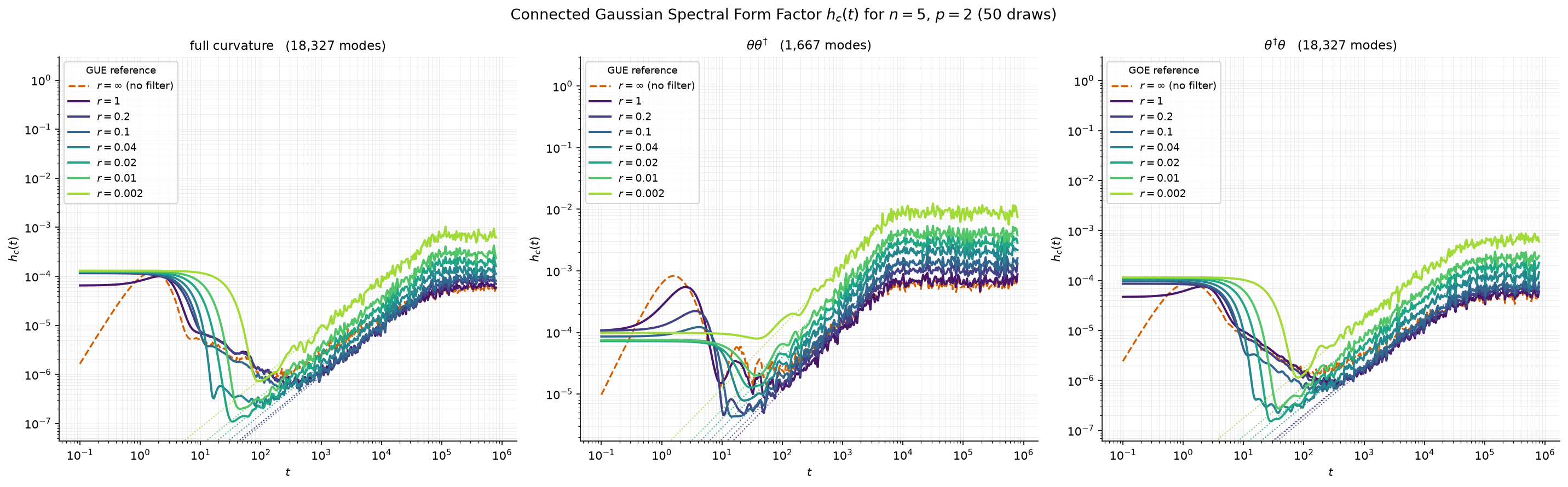}
    \caption{The (unnormalized) connected Gaussian spectral form factor with varying window size. Plotted are the full curvature, the full-rank piece $\theta^{(2)\dagger}_{\bar{v}} \theta_v^{(2)}$ and the rank-deficient $\theta_v^{(1)} \theta^{(1)\dagger}_{\bar{v}}$. For the two full-rank cases, the ramp is obscured at large window sizes due to a fluctuation in the spectral edge (see main text).}
    \label{fig:Gaussian-scan-d=7 central}
\end{figure}

\begin{figure}[h]
    \centering
        \includegraphics[width=1.0\linewidth]{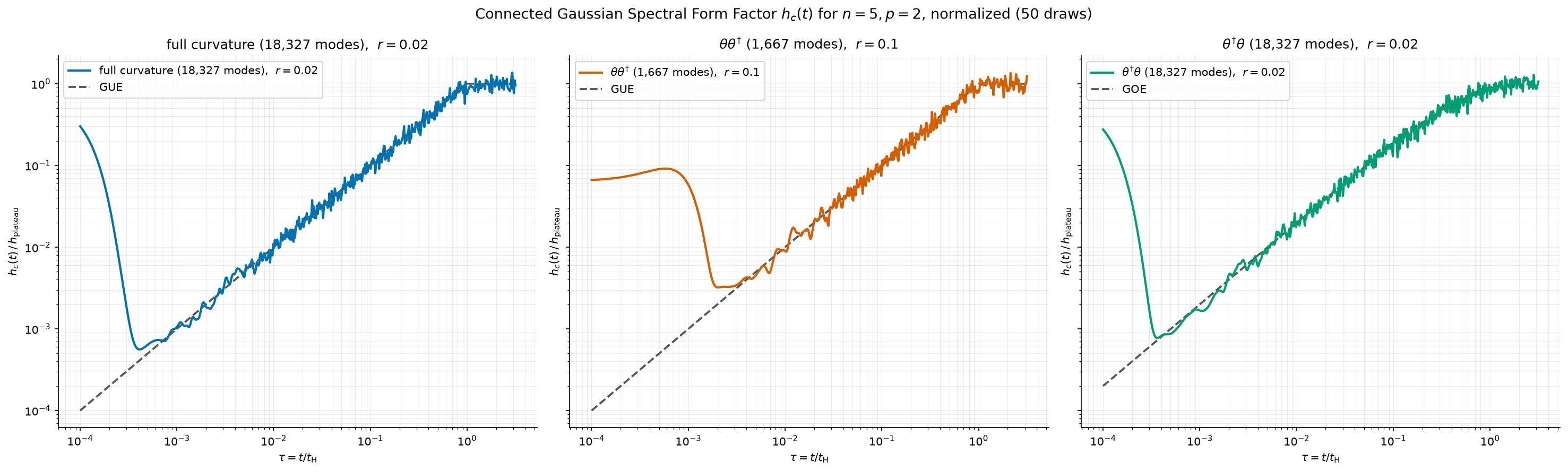}
    \caption{The (normalized) connected Gaussian spectral form factor of the full curvature and individual $\theta_v^{(p)} \theta_{\bar{v}}^{(p)\dagger}$, $\theta_{\bar{v}}^{(p)\dagger} \theta_v^{(p)}$ on the central slice of the 5-fold. For the two full-rank pieces, the optimal $r$ is $0.02$, while it is $r=0.1$ for the rank-deficient piece. Notice that the ramp time is larger for the rank-deficient piece, as expected. 
    $\theta_{\bar{v}}^{(p)\dagger}\theta_{v}^{(p)}$ follows the GOE curve.}
    \label{fig:Gaussian SFF d=7 p=2}
\end{figure}

The $n=5, p=2$ case is a bit more interesting. Scanning through the Gaussian variance, we find Figure~\ref{fig:Gaussian-scan-d=7 central}.
In agreement with the discussion of the level spacing ratios, the piece $\theta^{(2)\dagger}_{\bar{v}} \theta_v^{(2)}$ follows the GOE curve because of the antiunitary symmetry there, while the other two mimic GUE. The two full-rank pieces show some slowly decaying connected piece for values of $r$ that are not too small. For $\theta^{(2)\dagger}_{\bar{v}} \theta_v^{(2)}$, that piece can be fit to have a $1/t$ decay. We interpret it as follows: for the special case that the intermediate dimension is the same as the exterior, a Wishart matrix has a $1/\sqrt{\lambda}$ spectral edge at zero \eqref{eqn:WishartRho}. The Fourier transform of the spectral edge is $1/\sqrt{t}$, so in the spectral form factor it would appear as $1/t$. In a pure Wishart case, such an effect would not be visible in the connected piece. Numerically, we find that each draw of $\theta^{(2)\dagger}_{\bar{v}} \theta_v^{(2)}$ has the spectral edge $\sim 1/\sqrt{\lambda}$, but with a coefficient that is slightly fluctuating (by about the inverse number of parameters, similar to what happens in SYK \cite{Berkooz:2020fvm,Altland:2017eao,Altland:2024ubs,Raz:2025wjw}). The variance of this coefficient causes the $1/t$ behavior to appear also in the connected piece. It is visible only for large enough $r$, as only then is the spectral edge inside the Gaussian window.
For better visibility, the Gaussian SFF at the earliest ramp values of $r$ for the curvature and for the separate terms is plotted in Figure~\ref{fig:Gaussian SFF d=7 p=2}.

\begin{figure}
    \centering
    \includegraphics[width=0.7\linewidth]{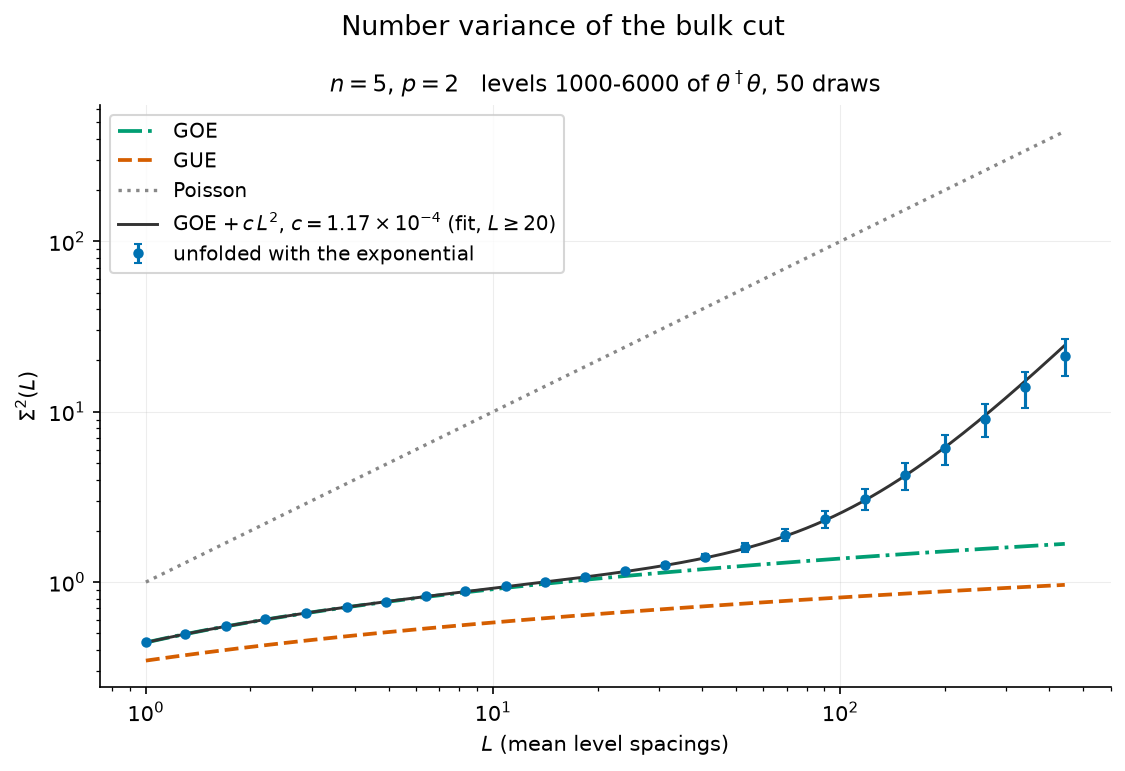}
    \caption{The number variance $\Sigma_2$ as a function of the number of levels $L$, for the $\theta^{(2)\dagger}_{\bar{v}}\theta^{(2)}_v$ term of the $p=2$ curvature of the 5-fold. We have taken a bulk cut of $5001$ eigenvalues away from the ``shoulders'' (see next subsection), and unfolded using a single exponential fit.}
    \label{fig:number variance}
\end{figure}

We see that in all cases, there is BPS chaos. There are some features which point towards structures beyond Wishart/RMT. Some symptoms of beyond-RMT physics also appear in the number variance $\Sigma_2(L)$ (see also \cite{Gharibyan:2018jrp}), with $L$ being the window size, which we have plotted in Figure~\ref{fig:number variance}.

The number variance initially follows the GOE curve (because of the Poincaré-induced symmetry), and then deviates with a $\sim L^2$ law. This is the behavior expected from SYK-like systems \cite{Altland:2017eao,Jia:2019orl}. The coefficient is expected to scale like the inverse number of fluctuating parameters. In our case, the coefficient is $c\sim 10^{-4}$, which roughly matches the inverse number of fluctuating parameters. 

\subsection{Some finer features of the spectrum}\label{sec:fine-features}

The observables studied in Section \ref{subsec:BPS-chaos-moduli-space} indicate that correlations between eigenvalues signal the expected chaotic behavior at longer time scales, while at shorter times, they are sensitive to other dynamics. In this section, we study the spectral density of the ensemble of curvature matrices both by direct diagonalization and by writing analytic equations for the resolvent of $\theta\theta^\dagger$. The spectral distribution is that of a non-uniform sparse Wishart ensemble, and its features can be explained by considering the product structure of the Jacobian ring.

\subsubsection{Relation to Wishart ensembles}

The Chern curvature \eqref{eq:curvature-block} on a fixed $(n,p)$ slice is reminiscent of a difference between two Wishart matrices. Naively, one may imagine $F_{v\bar{v}}^{(p)} \sim XX^\dagger-YY^\dagger$, where $X$ and $Y$ would be two random matrices of dimensions $h^{p,n-p}\times h^{p-1,n-p+1}$ and $h^{p,n-p}\times h^{p+1,n-p-1}$ respectively. When  $h^{p+1,n-p-1}\gg h^{p,n-p}$, as in $p=1$, one of the matrices becomes a low-rank correction to the other and thus one might expect that the spectrum of $F_{v\bar{v}}^{(p)}$ goes over to a Wishart distribution.

The main reason why this intuition fails is that the matrices $X$ and $Y$ in our case are not random, but highly ordered. Consider for example the entries of $\theta_v^{(p)}$:
\begin{flalign}
    (\theta_v^{(p)})_{\alpha_{(p)}}^{\beta_{(p+1)}} =
    v_{\beta_{(p+1)}-\alpha_{(p)}} 
        \sqrt{\frac{\mathcal{N}(\beta_{(p+1)})}{\mathcal{N}(\alpha_{(p)})}} \mathbf{1}_{\beta_{(p+1)}-\alpha_{(p)}}.
\end{flalign}
The projector $\mathbf{1}_{\beta_{(p+1)}-\alpha_{(p)}}$ is one if $\beta_{(p+1)}-\alpha_{(p)}$ is an allowed multi-index, but zero otherwise. This matrix is therefore very sparse. Furthermore, its nonzero entries on a fixed $\beta_{(p+1)}-\alpha_{(p)}$ diagonal\footnote{The term ``diagonal" here is used loosely, as matrix elements with fixed $\beta_{(p+1)}-\alpha_{(p)}$ do not lie on any kind of straight line.} are all correlated by the same $v_{\beta_{(p+1)}-\alpha_{(p)}}$.

To obtain an analytic model which captures the essential features of such a distribution, we will relax the diagonal constraint and assume that the main features of the spectrum come from the sparsity of $\theta_v^{(p)}$. Let us rescale the random variables to $\xi_I = v_I\sqrt{h^{n-1,1}\mathcal{N}(I)}$ so that $\text{Var}[\xi_I]=1$. The matrix elements can be approximated by:
\begin{flalign}
    &(\theta_{v}^{(p)})^{\beta_{(p+1)}}_{\alpha_{(p)}} \approx X^{\beta_{(p+1)}}_{\alpha_{(p)}} = \xi_{{\alpha_{(p)}} \beta_{(p+1)}} \sqrt{S_{\alpha_{(p)}\beta_{(p+1)}}} \,,
    \label{eqn:def-wishart-theta}
    \\[3pt]
    &
    S_{\alpha_{(p)}\beta_{(p+1)}} = \frac{1}{h^{n-1,1}}\frac{\mathcal{N}(\beta_{(p+1)})}{\mathcal{N}(\alpha_{(p)})\mathcal{N}(\beta_{(p+1)}-\alpha_{(p)})} \mathbf{1}_{\beta_{(p+1)}-\alpha_{(p)}}  \,.
    \label{eqn:def-wishart-S}
\end{flalign}
The problem thus reduces to studying the spectrum of a Wishart matrix, with a non-uniform variance \eqref{eqn:def-wishart-S}. Such problems have been studied before in the literature \cite{Alt_2017,hachem2005empiricaldistributioneigenvaluesgram,Hachem_2007}, with the upshot being that one can typically derive a deterministic set of equations describing the density of states. This is done by considering the resolvent equations and averaging over the random variables $\xi_{\alpha_{(p)}\beta_{(p+1)}}$. In Appendix \ref{app:def-wishart} we argue that such a set of equations can be derived for our case. Let us briefly describe the resulting picture. 

On the slice $H^{n-p,p}(X)$, the operator $\theta_v^{(p-1)}\theta_{\bar{v}}^{(p-1)\, \dagger}$ acts like a diffusive process, which takes some input multi-index $\mu_{(p)}$ and converts it into an output multi-index $\nu_{(p)}$. Note that the variance profile is symmetric under a simultaneous permutation of indices, $S_{\alpha\beta}=S_{\pi\alpha,\pi\beta}$ for any $\pi\in S_{n+2}$, and that this is in fact a symmetry of the entire ensemble.
The problem thus simplifies by splitting all input/output indices $\gamma$ into permutation orbits $[\gamma]=\{\pi \gamma \,|\, \pi\in S_{n+2}\}$. To each orbit $[\gamma]$ we associate a representative multi-index $\hat{\gamma}$, whose elements are given in descending order. We define the multiplicity $N_{[\gamma]}$ of the orbit $[\gamma]$ to be the total number of multi-indices that are related to $\hat{\gamma}$ by a permutation of indices. If $\hat{\gamma}$ has $n_l$ indices of value $l$, one has $N_{[\gamma]} = \tfrac{(n+2)!}{\prod_{l} n_l !}$.
\begin{table}[h]
  \centering
  \begin{tabular}{lcc}
    \toprule
    $(n,p)$ & number of orbits $[\gamma]$ & representatives and multiplicities $\hat{\gamma}(N_{[\gamma]})$ \\
    \midrule 
    (2,1) & 3   & \scalebox{1}{\{2200(6),\ 2110(12),\ 1111(1)\}} \\
    (3,1) & 5   & \scalebox{1}{\{32000(20),\ 31100(30),\ 22100(30),\ 21110(20),\ 11111(1)\}} \\
    (5,5) & 1  & \{5555555(1)\}\\
    \bottomrule
  \end{tabular}
  \caption{Examples of multi-index orbit representatives for some simple $(n,p)$ cases. The listed representatives and their multiplicities correspond to the primitive cohomologies of the appropriate Hodge slice.}
  \label{tab:shapes}
\end{table}

\noindent As discussed in Appendix \ref{app:def-wishart}, we obtain a deterministic coupled system of equations for functions $g_{\hat{\beta}}(E+i\epsilon)$,\footnote{There exists an equivalent formulation using column instead of row vectors. Which one of them is more economical depends on whether $p\leq n/2$ or $p\geq n/2$.}
\begin{flalign}\label{eqn:shape-eq1}
    &-\frac{1}{g_{\hat{\beta}}(z)} = z - \sum_{[\alpha]}\sum_{\alpha\in[\alpha]} \frac{S_{\alpha \hat{\beta}}}{1 +\sum_{[\gamma]} g_{\hat{\gamma}}(z)\big(\sum_{\gamma\in[\gamma]}S_{\alpha\gamma}\big)} \,.
\end{flalign}
The density of states of $\theta_v^{(p-1)}\theta_{\bar{v}}^{(p-1)\, \dagger}$ is then given by:
\begin{flalign}\label{eqn:rho-ttd}
    \rho_{\theta\theta^\dagger}(E) =
    \frac{1}{\pi} \text{Im}\bigg(\sum_{[\beta]} \frac{N_{[\beta]}}{h^{n-p,p}} \, g_{\hat{\beta}}(E+i\epsilon)\bigg)_{\text{at fixed }0<\epsilon\ll 1} \,.
\end{flalign}
The unknown functions $g_{\hat{\beta}}(E+i\epsilon)$ are diagonal elements of the resolvent of $\theta_{\hat{\beta}}^{(p-1)}\theta_{\bar{{\hat{\beta}}}}^{(p-1)\, \dagger}$. Intuitively from \eqref{eqn:rho-ttd} we may understand (the imaginary parts of) these functions to describe the degree to which the monomials $x^{\pi\hat{\beta}}$ ($\pi\in S_{n+2}$) participate in the diffusion process at some fixed eigenvalue $E$. The small parameter $\epsilon$ effectively controls the resolution of neighboring eigenvalues in $\rho_{\theta\theta^\dagger}(E)$ (see \eqref{eqn:eps-rho-m}). In practice we keep it at some small nonzero value, when solving the system of equations \eqref{eqn:shape-eq1}.

If one sets a uniform variance $S_{\alpha\beta} = \frac{1}{h^{n-p+1,p-1}}$ in \eqref{eqn:shape-eq1}, this leads to the Marchenko--Pastur law for $\rho_{\theta\theta^\dagger}(E)$, which describes the ordinary Wishart distribution \cite{Pastur:1967zca}. In contrast, we will refer to \eqref{eqn:rho-ttd} as the deformed Wishart distribution. This is a relatively small system of equations, compared to the dimensions of the matrix $\theta\theta^\dagger$, which we can solve numerically at some fixed $0<\epsilon\ll 1$. We now present a comparison of these distributions against data obtained by direct diagonalization.

\begin{figure}[h]
    \centering
    \includegraphics[width=1.0\linewidth]{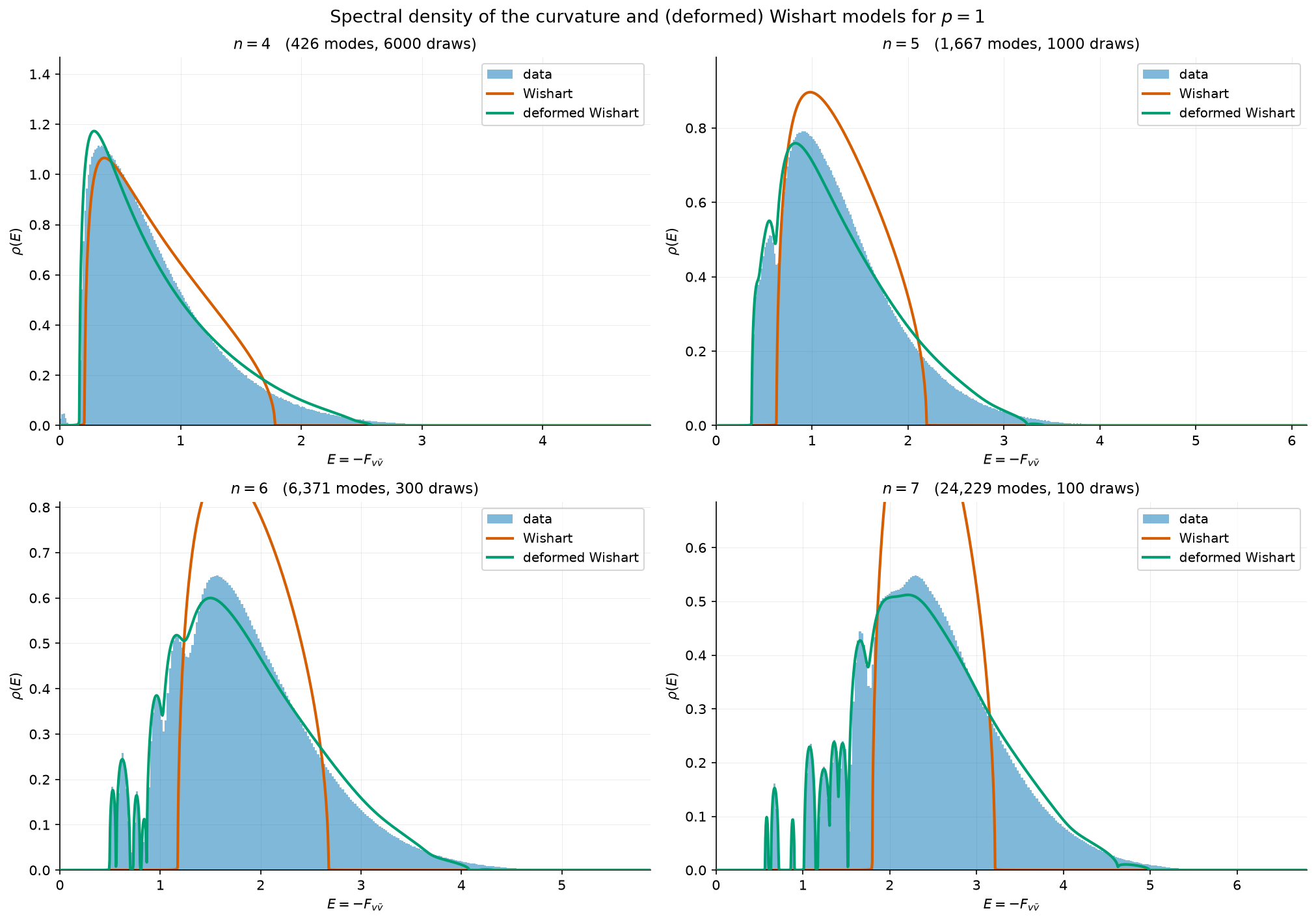}
    \caption{The spectral density of the curvature for $n=4,5,6,7$ and $p=1$ in comparison to (deformed) Wishart models (with zero eigenvalues removed). The blue shaded histogram is the data obtained by direct diagonalization. The orange curve is the distribution of pure Wishart, while the solid green curve is the prediction for a deformed Wishart matrix, obtained from the resolvent equations. For all plots we use $\epsilon=3\cdot 10^{-4}$.}
    \label{fig:p=1 deformed Wishart}
\end{figure}

In Figure~\ref{fig:p=1 deformed Wishart}, we plot the spectral density of the curvature (more precisely $-F_{v\bar{v}}^{(p)}$) on the complex structure slice and compare to the Wishart prediction. $F^{(p)}_{v\bar{v}}$ on the complex structure slice is equal to $ \theta^{(1)\dagger}_{\bar{v}} \theta_v^{(1)}$ up to the rank-one deformation $\theta_v^{(0)}\theta^{(0)\dagger}_{\bar{v}}$, which does not distort the result significantly. We see that the deformed Wishart distribution captures especially the lower part of the spectrum very well, which has a series of peaks that we will refer to as shoulders. The ordinary Wishart distribution fails to capture them. What distinguishes the two distributions is sparsity and the normalization factors $\mathcal{N}$. In numerical tests, we find that both\footnote{Here, ``sparsity'' is enforced by a non-uniform variance $S_{\alpha\beta}$. One can check that if one enforces the sparsity explicitly per draw, the resulting spectral density behaves in the same way.} of these are needed to produce the correct strength and multiplicity of the shoulders. 
\begin{figure}[h]
    \centering
    \includegraphics[width=1.0\linewidth]{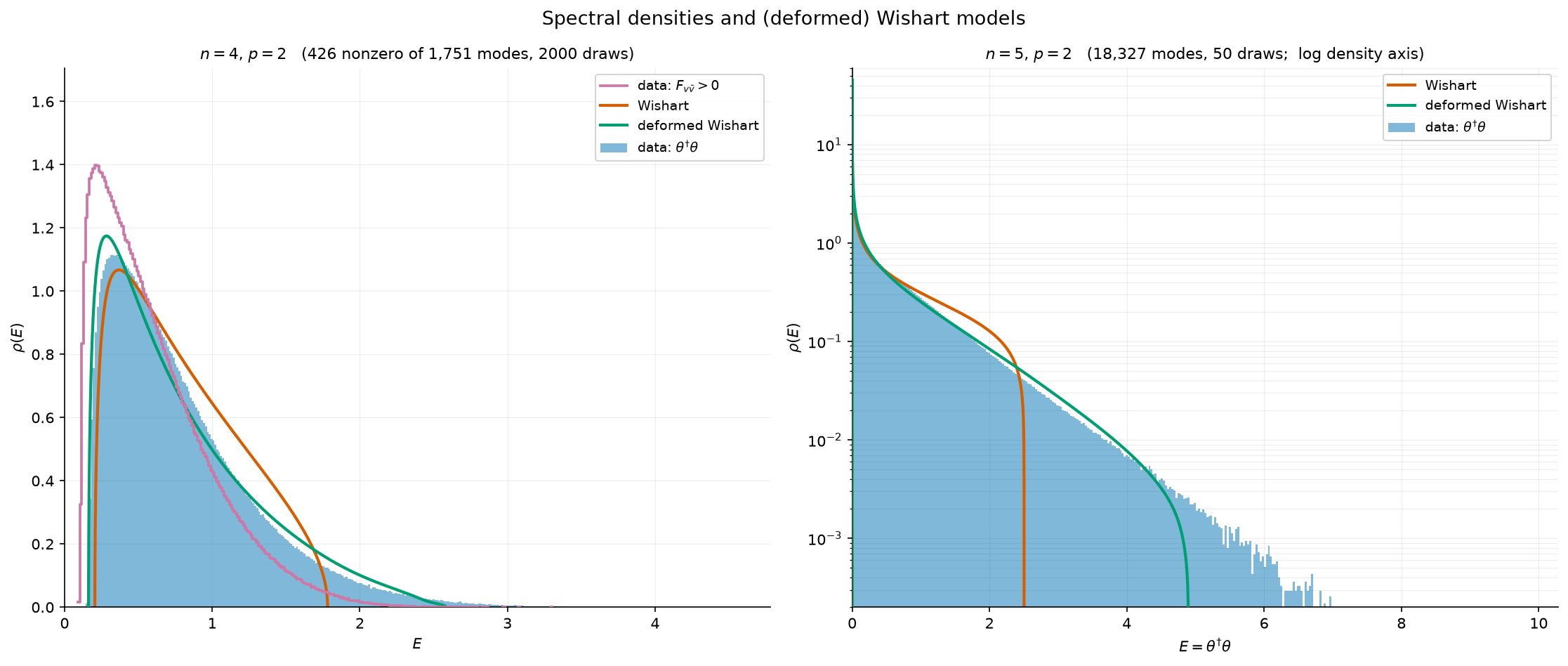}
    \caption{The spectral density of $\theta^{(2)\dagger}_{\bar{v}} \theta_v^{(2)}$ at $p=2$ for the 4-fold and 5-fold. For the 5-fold we plotted a $\log$-density as it better resolves the $c=1$ Wishart edge. Here we use $\epsilon=3\cdot 10^{-4}$ for $n=4$ and $\epsilon = 1\cdot 10^{-5}$ for $n=5$.}
    \label{fig:n=4,5 density}
\end{figure}
The discussion of the $p=2$ spectrum is once again presented for the $4$-fold and $5$-fold cases. 
In the $4$-fold case, the curvature is the difference of two equal-rank pieces, $\theta^{(2)\dagger}_{\bar{v}} \theta_v^{(2)}$ and $\theta_v^{(1)}\theta^{(1)\dagger}_{\bar{v}}$. Both of them are individually identical to the $n=4,p=1$ plot in Figure~\ref{fig:p=1 deformed Wishart}. Due to cancellations between them, the full curvature has a narrower spectrum. 

In the $5$-fold case, the piece $\theta^{(2)\dagger}_{\bar{v}} \theta_v^{(2)}$ is a product of two square matrices. In that case, the Wishart analogue develops a $1/\sqrt{\lambda}$ spectral edge at zero (see Figure~\ref{fig:n=4,5 density}). The deformed Wishart also reproduces that spectral edge correctly, and fits the upper parts of the spectrum significantly better. Similarly to the $n=4$, $p=2$ case, the spectrum of the full curvature turns out to be narrower than the $\theta^{(2)\dagger}_{\bar{v}} \theta_v^{(2)}$-part (see Figure~\ref{fig:n=4,5 density}). We expect similar behavior for the middle slices of odd $n$ in general.

\begin{figure}
    \centering
    \includegraphics[width=1.0\linewidth]{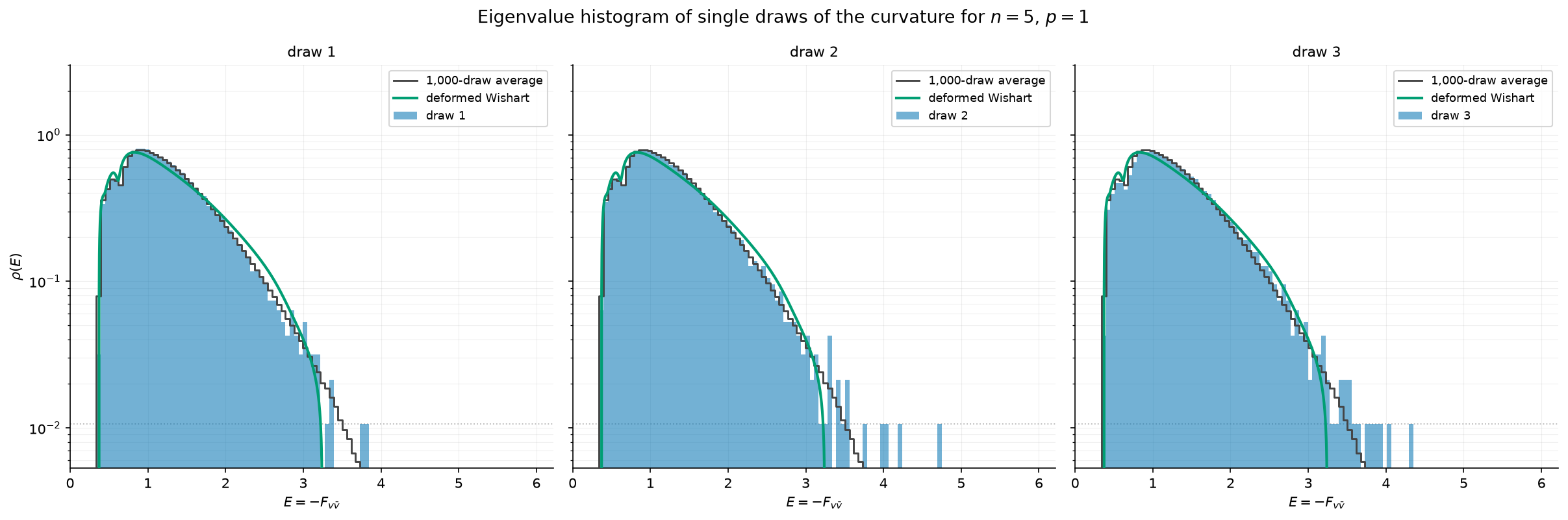}
    \caption{The spectral $\log$-density of individual draws of $F_{v\bar{v}}^{(1)}$ for the 5-fold. The spectrum remains qualitatively the same except for the very edge of the spectrum.}
    \label{fig:single-draws}
\end{figure}

It is important to note that, unlike the spectral form factor discussed in Section \ref{subsec:BPS-chaos-moduli-space}, the spectral density of $F_{v\bar{v}}^{(p)}$ or its Wishart-like components does not depend strongly on performing the ensemble average over the moduli space directions, as demonstrated in Figure~\ref{fig:single-draws}. The same qualitative features are observed for a single, but sufficiently generic, draw of $v^I,\bar{v}^{\bar{J}}$. This suggests that the structure observed in the curvature spectrum is a typical property of generic directions on moduli space, rather than an artifact of ensemble averaging. This observation also holds in other $(n,p)$ cases.

\subsubsection{Shoulders, tails and participation numbers}

The spectral densities of both the curvature and its individual Wishart-like components display two features which distinguish them from standard Wishart random ensembles. At the lower end of the spectrum, the sharp Wishart edge is replaced by the shoulders. At the upper end of the spectrum, the density develops an extended tail. We now present an intuitive combinatorial explanation for these features, which originates from the structure of polynomial multiplication in the Jacobian ring. The shoulders are probably related to the fact that we are working around the Gepner hypersurface.

\paragraph{The low shoulders.} The matrix $\theta_v^{(p)}$ represents multiplication by the polynomial $v(x)=v^I x_I$ in the Jacobian ring. A source monomial $x^\alpha$ can connect to a target monomial $x^\beta$ only when every component of $\beta - \alpha$ is nonnegative, in which case $x^{\beta - \alpha}$ is one of the degree-$(n + 2)$ monomials appearing in $v(x)$. Monomials with several powers near the cutoff therefore have fewer allowed products and a smaller average multiplication strength. Moreover, two monomials mix through $\theta_v^{(p)}\theta_{\bar{v}}^{(p)\dagger}$ only when they can be reached from a common source monomial. Different groups of target monomials can consequently have different typical eigenvalue scales while mixing only weakly.

We illustrate this in a concrete example. For $n = 5$ and $p = 1$, the matrix $\theta_{\bar{v}}^{(1)\dagger}\theta_v^{(1)}$ acts on $H^{4,1}(X_{\zeta})$, which has dimension $h^{4,1} = 1667$. The companion matrix $\theta_v^{(1)}\theta_{\bar v}^{(1)\dagger}$ acts on the $h^{3,2}=18327$-dimensional space $H^{3,2}(X_{\zeta})$ and has the same nonzero eigenvalues. We therefore use the smaller matrix $\theta_{\bar{v}}^{(1)\dagger}\theta_v^{(1)}$ to study the nonzero spectrum. Let $x^\lambda$ be a target monomial and let $\mathcal N(\lambda)$ denote its norm. The Gaussian average of $\theta_{\bar{v}}^{(1)\dagger}\theta_v^{(1)}$ is diagonal in the orthonormal monomial basis,
\begin{equation}
        \Sigma  \equiv \mathbb E_v[\theta_{\bar{v}}^{(1)\dagger}\theta_v^{(1)}] \,, \qquad \Sigma_\lambda{}^\mu = \delta_\lambda^\mu s_\lambda \,, \qquad 
        s_\lambda \equiv \frac{1}{h^{4,1}} \sum_{\substack{0 \leq I_i \leq 5\,,\ \sum_i I_i = 7\\ 0 \leq \lambda_i + I_i \leq 5}} \frac{\mathcal N(\lambda+I)}{\mathcal N(\lambda) \mathcal N(I)}  \,.
    \label{eq:CY5-p2-average-target-strength}
\end{equation}
The number $s_\lambda$ is the average squared norm of the row labeled by $\lambda$. It depends only on the permutation orbit of $\lambda$, whose entries sum to seven. With $\overline s \equiv (h^{4,1})^{-1}\operatorname{Tr}\Sigma$, the lowest orbits are:
\begin{equation}
    \begin{array}{c|c|c}
        \hat{\lambda} & N_{[\lambda]} & s_\lambda/\overline s \\ \hline
        5200000 \;&\; 42 \;&\; 0.383 \\
        5110000 \;&\; 105 \;&\; 0.461 \\
        4300000 \;&\; 42 \;&\; 0.525 \\
        4210000 \;&\; 210 \;&\; 0.710
    \end{array} \,.
    \label{eq:CY5-p2-low-strength-orbits}
\end{equation}
We call the $189$-dimensional span of the first three orbits $\mathcal H_{\mathrm{weak}}$, denote its projector by $P_{\mathrm{weak}}$, and call its $1478$-dimensional orthogonal complement $\mathcal H_{\mathrm{strong}}$, with $P_{\mathrm{strong}} \equiv \mathbf 1 - P_{\mathrm{weak}}$. These two spaces account for fractions $189/1667 \simeq 0.113$ and $1478/1667 \simeq 0.887$ of the nonzero spectrum. Averaging over $s_\lambda$ separately in each sector gives
\begin{equation}
    \frac{\overline s_{\mathrm{weak}}}{\overline s} = 0.458 \,, \qquad \frac{\overline s_{\mathrm{strong}}}{\overline s} = 1.069 \,.
    \label{eq:CY5-p2-weak-strong-average-strengths}
\end{equation}
The weak and strong spaces are not exactly decoupled, but their mixing is small. A direct Gaussian contraction gives
\begin{equation}
    \varepsilon_{\mathrm{mix}} \equiv \frac{2\mathbb E_v\!\left[\left\lVert P_{\mathrm{weak}}\,(\theta_{\bar{v}}^{(1)\dagger}\theta_v^{(1)})\,P_{\mathrm{strong}}\right\rVert^2\right]}{\mathbb E_v\!\left[\operatorname{Tr}(\theta_{\bar{v}}^{(1)\dagger}\theta_v^{(1)})^2\right]} = 0.00486 \,.
    \label{eq:CY5-p2-weak-strong-mixing}
\end{equation}
Thus the off-diagonal weak--strong blocks contribute only about $0.5\%$ of the second spectral moment. The two spaces are therefore approximately, but not exactly, independent sectors. 
The lower typical norm of the weak sector produces the small low-eigenvalue peak, while the strong sector produces the main peak.

\begin{figure}
    \centering
    \includegraphics[width=0.9\linewidth]{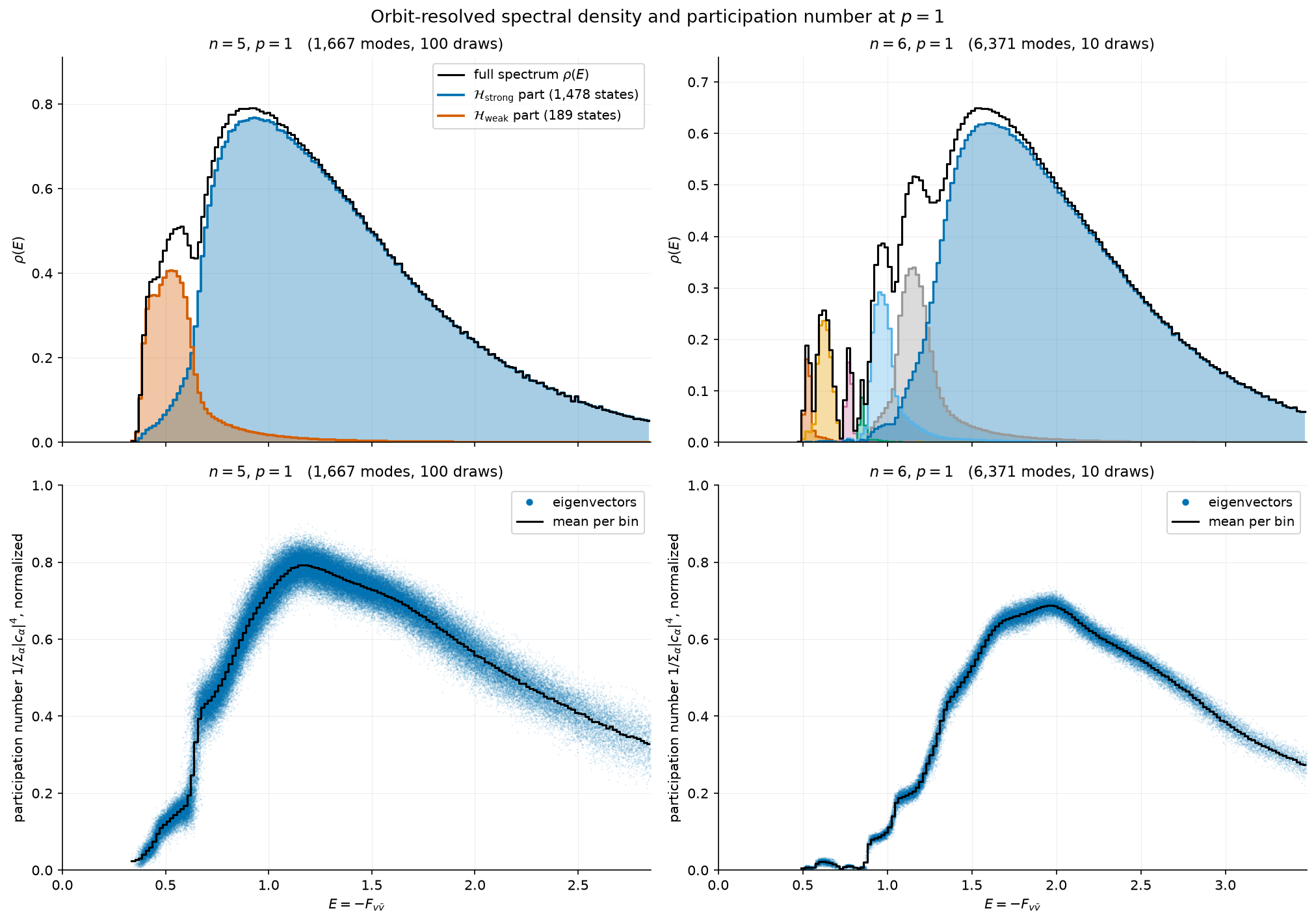}
    \caption{Top: the spectral density, weighted by the projection of the eigenvector onto different subspaces. Bottom: the (normalized) participation number. Left: $n=5$, $p=1$. Right: $n=6$, $p=1$. In the top left figure, we have split the states into $\mathcal{H}_{weak}$ containing orbit groups (excluding zero indices): $\{52,511,43\}$ and $\mathcal{H}_{strong}$ containing the rest. In the top right figure, we have split the states into several orbit groups (excluding zero indices): $\{62\}, \{611\}, \{53\}, \{44\}, \{521\}, \{5111,431\}$, and all others.}
    \label{fig:CY5-p1-participation-number}
\end{figure}

In order to see that there are indeed two sectors separated by their index, and that the eigenvectors are not uniformly spread in the index basis, in the top panels of Figure~\ref{fig:CY5-p1-participation-number} we overlay two histograms capturing the projection of an eigenvector onto the weak and strong subspaces. The left panel supports the analysis \eqref{eq:CY5-p2-weak-strong-mixing} of the $n=5$, $p=1$ case, while the right panel discusses a similar decomposition for the $n=6$, $p=1$ slice. The $x$-axis denotes the eigenvalue, and the $y$-axis counts the number of eigenvalues in this bin weighted by their projection to the relevant subspace $||P_{\mathrm{weak / strong}} \psi_\lambda ||$. We show this both for the $n=5$, $p=1$ case where the subspaces are those defined above and for the $n=6$, $p=1$ case where there are more distinct orbit subspaces, chosen in a similar way by noting their $s_\lambda$ differs from that of other orbits.
In the bottom panels of Figure~\ref{fig:CY5-p1-participation-number} we plot a related quantity, the participation number $N_p(\lambda) = \frac{2}{(\dim \mathcal H+1)} \frac{1}{\sum_\alpha |\langle \alpha | \psi_\lambda\rangle |^4}$ for the index basis $\{ |\alpha\rangle \}$. We have inserted a normalization factor such that the participation number approaches $1$ for the eigenvectors of a GUE matrix (i.e. for Haar random vectors), and decreases as the eigenvector gets more concentrated around a single multi-index. As can be seen there, both the shoulder and the tail come from a smaller part of the Hilbert space.

\begin{figure}[h]
    \centering
    \includegraphics[width=1.0\linewidth]{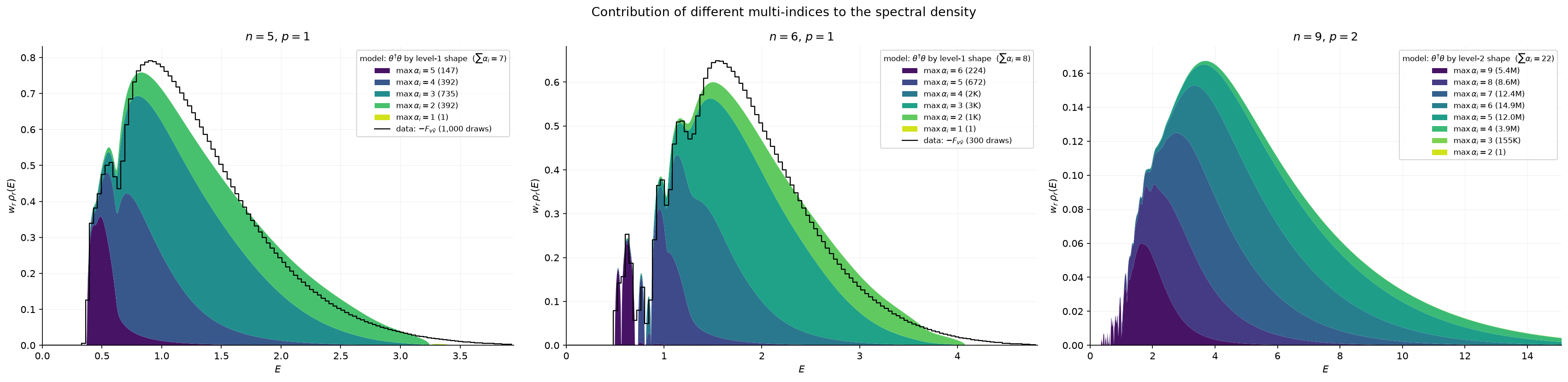}
    \caption{The spectral density of $\theta_{\bar{v}}^{(p) \dagger}\theta_{v}^{(p)}$ obtained from the deformed Wishart ensemble (shaded region), obtained by solving the system of resolvent equations. Each color represents the contribution of a group of multi-indices with a specific maximal element to the spectral density. $w_r$ is the relative weight of such a group. For $n=5, p=1$ (left) and $n=6,p=1$ (center) the prediction is compared against the direct diagonalization result (the solid black line shows $-F_{v\bar{v}}$, which deviates from $\theta_{\bar{v}}^{(p) \dagger}\theta_{v}^{(p)}$ only by the usual rank-one piece). In the $n=9, p=2$ case (right), $\theta_{\bar{v}}^{(p) \dagger}\theta_{v}^{(p)}$ contains roughly 57.4M input/output multi-indices, making direct diagonalization infeasible. }
    \label{fig:shape-stack-lightning}
\end{figure}

Another way to probe the same distribution of eigenvectors is by using the system of resolvent equations \eqref{eqn:shape-eq1}--\eqref{eqn:rho-ttd} (see Figure~\ref{fig:shape-stack-lightning}). After solving the system numerically, one obtains the density of states as a sum of resolvent functions $\text{Im}\big(g_{\hat{\beta}}(E+i\epsilon)\big)$. In a loose sense, each such function signifies which fraction of the spectral density at a fixed eigenvalue arises from multiplication of monomials $x^{\pi\hat{\beta}}$ ($\pi\in S_{n+2}$). The spectral densities per $\hat{\beta}$ distribute themselves differently according to $S_{\alpha\beta}$. 
The resulting distribution of eigenvalues is in good agreement with the results obtained by direct diagonalization. The general behavior which persists across all $n$ is that the distribution contains shoulders which are dominated by multi-indices that are concentrated at one index, while the more uniform multi-indices populate the bulk of the spectrum. At higher $p$, the shoulders become less pronounced although, interestingly, the stratification by features of $\hat\beta$ remains very clear. For example, the right panel in Figure~\ref{fig:shape-stack-lightning} displays only minor oscillations at the lower end of the spectrum.

\paragraph{The long tails.}
Very large eigenvalues are not captured by the resolvent equations. For these eigenvalues, it is important that the matrix elements of $\theta_v^{(p)}$ are correlated with each other. A very primitive model for those can be formulated as follows: assume for a moment that the indices had periodic boundary conditions. Then (ignoring the normalization factors) the combination $v^I \theta_I$ would be a shift operator and diagonalized (with unit norm eigenvalue) by plane waves.

Our actual boundary conditions prevent plane waves from being eigenvectors and also break the degeneracy between different momenta. We can still compute the expectation value of the plane waves. The momenta with the largest eigenvalues show the same qualitative behavior as the real eigenvalues, up to some universal prefactor which is expected to be related to the normalizations $\mathcal{N}$ that we neglected. Such expectation values depend sensitively on the precise values of the $v^I$, which is in agreement with the draw-to-draw fluctuations of the edge of the spectrum that we see (see Figure~\ref{fig:single-draws}).

\acknowledgments

It is a pleasure to thank Yiming Chen, Sean Colin-Ellerin, Jan de Boer, Luca Iliesiu, Juan Maldacena, Boris Pioline, Shlomo Razamat, Steve Shenker, Douglas Stanford, Erez Urbach and Edward Witten for useful discussions. OM is supported by the ERC-COG grant NP-QFT No. 864583, ``Non-perturbative dynamics of quantum fields: from new deconfined phases of matter to quantum black holes'', the MUR-FARE2020 grant No. R20E8NR3HX, ``The Emergence of Quantum Gravity from Strong Coupling Dynamics'' and the MUR-PRIN2022 grant No. 2022NY2MXY.
The work of MB, TK and JS is supported in part by the Israel Science Foundation grants No. 2159/22 and No. 3192/26, by the Minerva Foundation, and by a German-Israeli Project Cooperation (DIP) grant ``Holography and the Swampland". MB is an incumbent of the Charles and David Wolfson Professorial Chair of Theoretical Physics. The work of JS is supported by a research grant from the Chaim Mida Prize in Theoretical Physics at the Weizmann Institute of Science, and by a Dean Award of Excellence of the Weizmann School of Science.

The authors used the assistance of Claude and of ChatGPT for numerical simulations and proofreading. All derivations were carried out and written up by the authors, who take full responsibility.

\appendix

\section{Lefschetz $SU(2)$ and Chern connection}\label{app:mathematical minimum}

In the main text, we considered a complete intersection Calabi--Yau manifold $X$ and its moduli space of complex structures, $\mathcal{M}$. We employ the following conventions for coordinates on these two spaces: for the $CY_n$-fold, we denote the complex coordinates as
\begin{equation}
    z^i, \bar{z}^{\bar{j}}, \quad i \in \{1,...,n\}.
\end{equation}
The moduli space of complex structures is also (locally) a complex orbifold (even \Kahler \cite{1991NuPhB.355..455C}); we denote its coordinates as
\begin{equation}
    \zeta^I, \bar{\zeta}^{\bar{J}}, \quad I \in \{1,...,\dim \mathcal{M}\}.
\end{equation}
As described in the main text, there exist natural notions of connections on this space.

There is a vast literature on this subject; some classic references are \cite{griffiths1978principles,Voisin2002,Voisin2003,MorrowKodaira1971,Kodaira1986,Hori:2003ic} and, from a physics point of view, \cite{Greene:1996cy,Green:1987mn,Hubsch:1992nu,Bershadsky:1993cx}, to which we refer for proofs.

Here we only review two basic facts that we need in the main text.

\paragraph{Lefschetz $SU(2)$.}

For a compact \Kahler manifold, we have the \Kahler identities: there is the Lefschetz multiplication, which is just wedging with $\omega$,
\begin{equation}
    L(\alpha) = \omega \wedge \alpha,
\end{equation}
and its adjoint $\Lambda= L^{\dagger}$. Then one can prove (using a calculation in local coordinates) that
\begin{equation}
    \begin{split}
        & [L,\Lambda]=p+q-n \\
        & [\Lambda,\partial] = i \bar{\partial}^{\dagger} \\
        & [\Lambda,\bar{\partial}] = - i \partial^{\dagger}. 
    \end{split}
\end{equation}
$L,\Lambda$ and their commutator form an $\mathfrak{sl}_2$ algebra (``the Lefschetz $SU(2)$''). Using these algebra relations, one can prove the Hodge decomposition. The Laplacian is in particular a singlet under the Lefschetz $SU(2)$. The $\mathfrak{sl}_2$ structure then descends to harmonic forms $H^{q,p}(X)$. Each cohomology (represented by its unique harmonic representative) is an eigenvector of $[L,\Lambda]$ with eigenvalue $p+q-n$. The primitive cohomologies are cohomologies that are lowest-weight states:
\begin{equation}
    P^{q,p}(X) = \{ \alpha \in H^{q,p}(X)| \Lambda \alpha = 0\}.
\end{equation}
We then have the Lefschetz decomposition
\begin{equation}
    H^{q,p}(X)=\bigoplus_k L^k P^{q-k,p-k}(X).
\end{equation}
A local computation in a unitary frame shows that for a primitive cohomology $\alpha$ of degree $(q,p)$, with $p+q=k$, we have:
\begin{equation}\label{Hodge primitive}
    \ast \alpha = i^{p-q} (-1)^{\frac{k(k-1)}{2}} \frac{\omega^{n-k}}{(n-k)!} \wedge \alpha.
\end{equation}
For a primitive middle cohomology, $k=n$ and the action of the Hodge star on the harmonic representative is then trivial up to a phase.

\paragraph{Relation between Gauss--Manin and Chern connections.}

In Section~\ref{subsec:Chern connection on the moduli space}, we stated the relation
\begin{equation}
    \nabla^{\mathrm{Ch}}_I = \nabla^{\mathrm{GM}}_I - \theta_I, \quad \nabla^{\mathrm{Ch}}_{\bar{I}} = \nabla^{\mathrm{GM}}_{\bar{I}} - (\theta_I)^{\dagger}.
\end{equation}
Here we show that $\nabla_{\text{Ch}}$ defined in this way is indeed metric-compatible with \eqref{eq:Hodge-metric}.

In a GM-constant frame $\omega_{\alpha}$, all the dependence of the Hermitian Hodge metric comes from the change in the Hodge star of \eqref{eq:Hodge-metric}. For a primitive cohomology of degree $(q,p)$ with total degree $n$, \eqref{Hodge primitive} gives
\begin{equation}
    \ast_{\zeta_0} \omega_\beta = i^{p-q}  (-1)^{\frac{n(n-1)}{2}} \omega_\beta.
\end{equation}
To compute the change of the Hodge metric, we then compute the change of $\ast$ acting on a GM-flat section $\omega_\beta$ which is $(q,p)$ at $\zeta_0$:
\begin{equation}
    \ast_{\zeta_0+\delta \zeta} \overline{\omega_\beta} =\ast_{\zeta_0+\delta \zeta}\overline{\omega_\beta + (\delta \zeta^I\theta_I + \overline{\delta \zeta^I} \theta_I^{\dagger})\omega_\beta} - \ast_{\zeta_0+\delta \zeta} \overline{(\delta \zeta^I\theta_I + \overline{\delta \zeta^I} \theta_I^{\dagger})\omega_\beta}.
\end{equation}
The first term is $(q,p)$ in the new complex structure, so that the Hodge star acts via a phase; the second term is a linear combination of $(q-1,p+1)$ and $(q+1,p-1)$. Both of those are eigenvectors of the Hodge star with a negative sign relative to the first term, such that we get 
\begin{equation}
    \ast_{\zeta_0+\delta \zeta} \overline{\omega_\beta}-\ast_{\zeta_0} \overline{\omega_\beta}= -2\ast_{\zeta_0}\overline{(\delta \zeta^I\theta_I + \overline{\delta \zeta^I} \theta_I^{\dagger})\omega_\beta}.
\end{equation}
Then, for two GM-flat sections $\omega_\alpha,\omega_\beta$, we have that
\begin{equation}
    \begin{split}
        & G(\nabla^{\mathrm{Ch}}\omega_\alpha,\omega_\beta) + G(\omega_\alpha,\nabla^{\mathrm{Ch}}\omega_\beta) 
        \\[3pt]&
        = -G((\delta \zeta^I\theta_I + \overline{\delta \zeta^I} \theta_I^{\dagger})\omega_\alpha,\omega_\beta)-G(\omega_\alpha,(\delta \zeta^I\theta_I + \overline{\delta \zeta^I} \theta_I^{\dagger})\omega_\beta) 
        \\[3pt]&=  
        -G(\omega_\alpha,2(\delta \zeta^I\theta_I + \overline{\delta \zeta^I} \theta_I^{\dagger})\omega_\beta) 
        \\[1pt]&= 
        \int_{X_{\zeta_0}} \omega_\alpha \wedge (-2\ast_{\zeta_0}\overline{(\delta \zeta^I\theta_I + \overline{\delta \zeta^I} \theta_I^{\dagger})\omega_\beta})=\nabla^{\mathrm{Ch}}(G(\omega_\alpha,\omega_\beta)),
    \end{split}
\end{equation}
proving metric compatibility.
The facts that the GM connection is flat and that the Chern connection preserves the Hodge decomposition lead directly to\footnote{We use the convention that for a connection $\nabla$, its curvature is $F_{I\bar{J}}=-[\nabla_I,\nabla_{\bar{J}}]$.}:
\begin{equation}
    \begin{split}
        & F_{I\bar{J}}^{\mathrm{Ch}} =  [\theta_I, (\theta_J)^{\dagger}],
    \end{split}
\end{equation}
as well as $[\theta_I,\theta_J]=0$.

\section{Deformed Wishart ensembles} \label{app:def-wishart}

The goal of this appendix is to establish a simple analytic model which qualitatively captures the density of states of the ensemble of curvature matrices \eqref{eq:curvature-block}. For simplicity, we will focus on the spectrum of $\theta^{(p)}_v\theta^{(p)\,\dagger}_{\bar{v}}$. For cases where one of the terms in \eqref{eq:curvature-block} becomes highly rank-deficient, this model leads to a good approximation for the density of states of the full $F_{v\bar{v}}^{(p)}$ ensemble.

For a random matrix $X\in \mathbb{C}^{d_1,d_2}$ at large $d_{1}$ with $c=d_1/d_2$ fixed, the spectrum of $XX^\dagger$ tends to a universal distribution, known as the Marchenko--Pastur distribution \cite{Wishart1928THEGP,8a873437-53d6-348d-9b7a-19c7327502a1,Pastur:1967zca}:
\begin{equation}
    \rho(x)=\frac{1}{2\pi \sigma^2}\frac{\sqrt{(x_+-x)(x-x_-)}}{cx}, \qquad \text{with } x_{\pm}=\sigma^2 (1\pm \sqrt{c})^2.
\end{equation}
At $c\rightarrow 0$ (and also $c\rightarrow \infty$ after appropriate rescaling), the distribution approaches a semicircle, which is intuitive from a central-limit argument. For the special case $c=1$ (the product of two Gaussian random square matrices), the spectral density becomes
\begin{equation}\label{eqn:WishartRho}
    \rho(x)=\frac{1}{2\pi\sigma^2}\frac{\sqrt{4\sigma^2-x}}{\sqrt{x}}\underset{x\to 0^+}{\sim}
    \frac{1}{\pi\sigma\sqrt{x}}. 
\end{equation}
\paragraph{A deformed Wishart distribution.} The difference between this case and ours is that some of the entries of $\theta^{(p)}_v$ are zero and others are correlated. Consider the matrix element $(\theta_v^{(p)})_{\alpha_{(p)}}^{\beta_{(p+1)}}$. We can write it using an ordinary Gaussian variable $\xi_I = v_I \sqrt{h^{n-1,1}\mathcal{N}(I)}$, for which $\mathbb{E}[\xi_I]=0$ and $\text{Var}[\xi_I]=1$:
\begin{flalign}
    (\theta_v^{(p)})_{\alpha_{(p)}}^{\beta_{(p+1)}} =
    \xi_{\beta_{(p+1)}-\alpha_{(p)}} 
        \sqrt{\frac{1}{h^{n-1,1}}\frac{\mathcal{N}(\beta_{(p+1)})}{\mathcal{N}(\alpha_{(p)})\mathcal{N}(\beta_{(p+1)}-\alpha_{(p)})}} \mathbf{1}_{\beta_{(p+1)}-\alpha_{(p)}}.
\end{flalign}
For an allowed multi-index $\beta_{(p+1)}-\alpha_{(p)}$, elements on the same $\beta_{(p+1)}-\alpha_{(p)}$ diagonal are correlated. To make the problem tractable, we will approximate the matrix elements of $\theta_{v}^{(p)}$ by
\begin{flalign}
    &(\theta_{v}^{(p)})^{\beta_{(p+1)}}_{\alpha_{(p)}} \approx X^{\beta_{(p+1)}}_{\alpha_{(p)}} \equiv \xi_{{\alpha_{(p)}} \beta_{(p+1)}} \sqrt{S_{\alpha_{(p)}\beta_{(p+1)}}},\label{eqn:X-def}\\[3pt]&
    S_{\alpha_{(p)}\beta_{(p+1)}} = \frac{1}{h^{n-1,1}}\frac{\mathcal{N}(\beta_{(p+1)})}{\mathcal{N}(\alpha_{(p)})\mathcal{N}(\beta_{(p+1)}-\alpha_{(p)})} \mathbf{1}_{\beta_{(p+1)}-\alpha_{(p)}} .\label{eqn:Sab}
\end{flalign}
Here, $\xi_{{\alpha_{(p)}} \beta_{(p+1)}}$ are independent complex Gaussian variables, for each $\alpha_{(p)}$, $\beta_{(p+1)}$. We thus neglect correlations between nonzero matrix elements. The problem now reduces to studying the spectrum of a Wishart ensemble $XX^\dagger$ with a non-uniform variance profile $S_{\alpha\beta}$ (which we will call a deformed Wishart ensemble).\\[-5pt]

Such random matrix ensembles have been well studied in the literature \cite{Alt_2017,hachem2005empiricaldistributioneigenvaluesgram,Hachem_2007}: in the limit of large matrix dimensions, one can typically derive a deterministic system of equations for the resolvent of $XX^\dagger$, which only depends on the properties of $S_{\alpha\beta}$. In the ensuing discussion, we will argue that such a system of equations exists in our case, and that it becomes deterministic at large $n$ (meaning that the fluctuations associated with ensemble averaging go to zero as $n$ becomes large).

To keep the presentation concise, we will not show a complete proof that this set of resolvent equations becomes self-averaging at large matrix dimensions. Under our initial assumptions, this proof is straightforward, but mostly technical. We will only mention the main ingredients and sketch out a derivation in Section \ref{sec:shape-eq}, skipping several technical steps. As we discuss in Section \ref{sec:fine-features}, the model closely reproduces the spectrum obtained by direct diagonalization.

\paragraph{The resolvent and the density of states.} Let $A=XX^\dagger$ be a matrix of dimensions $d_1 \times d_1$. We will define the resolvent of $A$ and its trace:
\begin{flalign}
     &G(z) = (A-z\mathbf{1})^{-1}\\&
     m(z) = \frac{1}{d_1} \text{Tr}(G(z)).
\end{flalign}
Let us denote the eigenvalues of $A$ by $\lambda_\alpha$ (note that $\lambda_\alpha\in\mathbb{R}$). The trace of the resolvent and the density of states at large $d_1$ are related by:
\begin{flalign}
    &m(z) = \frac{1}{d_1} \text{Tr}(G(z)) = \frac{1}{d_1} \sum_{\alpha=1}^{d_1} \frac{1}{\lambda_\alpha-z} \xrightarrow{\scriptscriptstyle d_1\to\infty} \int d\lambda\frac{\rho(\lambda)}{\lambda-z}\\&
    \frac{1}{\pi} \text{Im}\bigg( m(E+i\epsilon)\bigg) = \frac{1}{d_1} \sum_{\alpha=1}^{d_1} \frac{1}{\pi} \frac{\epsilon}{(\lambda_\alpha-E)^2 + \epsilon^2} \xrightarrow{\scriptscriptstyle \epsilon\to 0^+} \rho(E). \label{eqn:eps-rho-m}
\end{flalign}
Let $z=E+i\epsilon$. The meaning of the parameter $\epsilon$ is clear from \eqref{eqn:eps-rho-m}. It controls the width of the Lorentzian profiles in $\rho(E)$ and thus the resolution of nearby eigenvalues. In practice, our plots are represented at some finite binning width; therefore we will not assume the exact $\epsilon \to 0^+$ limit, but keep $\epsilon$ at some small, finite value. This gives us a very useful finite upper bound:
\begin{flalign}\label{eqn:ebound1}
    |\lambda_\alpha - z|^{-1}\leq \epsilon^{-1},
\end{flalign}
which leads to establishing a finite upper bound on matrix elements of the resolvent.

\paragraph{Bounds on the variance profile $S_{\alpha\beta}$.} To show that the resolvent equations become self-averaging at large $n$, it is useful to first establish some constraints satisfied by the variance profile \eqref{eqn:Sab}. 

Consider a general nonzero $S_{\alpha\beta}$, where $\beta_{(p+1)}=I+\alpha_{(p)}$ and $I$ is an allowed complex structure multi-index. By expressing the ratio of $\Gamma$-functions as an integral over digamma functions, and using some elementary inequalities, one can show that:
\begin{flalign}\label{eqn:Sbound1}
    S_{\alpha,\alpha+I} = \frac{1}{h^{n-1,1}}\frac{\mathcal{N}(\alpha+I)}{\mathcal{N}(\alpha)\mathcal{N}(I)} \leq \frac{1}{h^{n-1,1}} \prod_{i=1 \atop \sum_i I_i=(n+2)}^{n+2}(I_i+1)\leq \frac{2^{n+2}}{h^{n-1,1}} .
\end{flalign}
This bound also applies to the maximal element $\max_{\alpha\beta}S_{\alpha\beta}$. A similar procedure also leads to a bound on the sum of the row or column elements $S$:
\begin{flalign}
    &\sum_{\alpha}S_{\alpha\beta} \leq \frac{1}{h^{n-1,1}}\sum_{n\geq I_i\geq0 \atop \sum I_i=n+2} \prod_{i=1}^{n+2}(I_i+1)\leq \frac{1}{h^{n-1,1}}\sum_{I_i\geq0 \atop \sum I_i=n+2}\prod_{i=1}^{n+2}(I_i+1)\\&
    \sum_{\beta}S_{\alpha\beta} \leq \frac{1}{h^{n-1,1}}\sum_{n\geq I_i\geq0 \atop \sum I_i=n+2} \prod_{i=1}^{n+2}(I_i+1)\leq \frac{1}{h^{n-1,1}}\sum_{I_i\geq0 \atop \sum I_i=n+2}\prod_{i=1}^{n+2}(I_i+1).
\end{flalign}
The sums on the right-hand side can be evaluated explicitly, leading to identical bounds for the sums of row or column elements of $S_{\alpha\beta}$:
\begin{flalign}\label{eqn:Sbound2}
    \sum_{\alpha}S_{\alpha\beta} \leq \frac{1}{h^{n-1,1}}\binom{3n+5}{n+2},
    \qquad\qquad
    \sum_{\beta}S_{\alpha\beta} \leq \frac{1}{h^{n-1,1}}\binom{3n+5}{n+2}.
\end{flalign}

\subsection{A deterministic system of equations}\label{sec:shape-eq}
Let us split the matrix $X$, defined in \eqref{eqn:X-def}, into column vectors $y_\alpha$:
\begin{flalign}
    &X = \bigg[y_1 \,\bigg|\, y_2 \,\bigg|\, ... \,\bigg|\, y_{d_2}\bigg]\\&
    A = XX^\dagger = \sum_{\alpha=1}^{d_2} y_{\alpha}y_\alpha^\dagger\\&
    G(z) = \big(\sum_{\alpha=1}^{d_2} y_{\alpha}y_\alpha^\dagger-z\mathbf{1}\big)^{-1}.
\end{flalign}
We also define a cavity resolvent $G^{(\alpha)}(z)$ as follows:
\begin{flalign}
    &A^{(\alpha)} = A - y_\alpha y^{\dagger}_\alpha = \sum_{\beta\neq\alpha} y_{\beta}y_{\beta}^\dagger\\&
    G^{(\alpha)}(z) = (A^{(\alpha)}-z\mathbf{1})^{-1}.
\end{flalign}
We will use $\mathbb{E}[\cdot]$ to denote the average over all column vectors, and $\mathbb{E}_\alpha[\cdot]$ to denote averaging over a specific column vector $y_\alpha$. Note that $\mathbb{E}[\cdot]=\mathbb{E}[\mathbb{E}_\alpha[\cdot]]$. The matrix entries of $X$ have a non-uniform variance, described by:
\begin{flalign}\label{eqn:yavg-S}
    \mathbb{E}_{\alpha}[y_{\alpha,\beta}^\dagger y_{\alpha,\gamma}] = S_{\alpha\beta}\delta_{\beta\gamma}.
\end{flalign}
By definition, the resolvent $G(z)$ satisfies $1 + zG(z) = AG(z)$. Let us consider the $\beta\beta$ component of this equation:
\begin{flalign}\label{eqn:MPx1}
   1+zG_{\beta\beta}(z) = (AG)_{\beta\beta} = \sum_{\alpha}y_{\alpha,\beta}(y_\alpha^\dagger G(z))_\beta.
\end{flalign}
The right-hand side can be expressed using the Sherman--Morrison identity:
\begin{flalign}\label{eqn:SM-identity}
    (M+uu^\dagger)^{-1} = M^{-1} - \frac{M^{-1}uu^\dagger M^{-1}}{1+u^\dagger M^{-1}u}.
\end{flalign}
For $M=A^{(\alpha)}-z\mathbf{1}$ and $u=y_\alpha$ we find:
\begin{flalign}
    &G(z) = G^{(\alpha)}(z)-\frac{G^{(\alpha)}(z)y_\alpha y_\alpha^\dagger G^{(\alpha)}(z)}{1+y_\alpha^\dagger G^{(\alpha)}(z) y_\alpha}. \label{eqn:GGa}
\end{flalign}
Using this identity, we can express $(y_\alpha^\dagger G(z))_\beta$ as:
\begin{flalign}\label{eqn:sMPx2}
    (y_\alpha^\dagger G(z))_\beta = (y_\alpha^\dagger G^{(\alpha)}(z))_\beta - \frac{y_\alpha^\dagger G^{(\alpha)}(z)y_\alpha (y_\alpha^\dagger G^{(\alpha)}(z))_\beta}{1+ y_\alpha^\dagger G^{(\alpha)}(z)y_\alpha} = \frac{(y_\alpha^\dagger G^{(\alpha)}(z))_\beta}{1+ y_\alpha^\dagger G^{(\alpha)}(z)y_\alpha}.
\end{flalign}
Substituting this into the right-hand side of \eqref{eqn:MPx1}, we obtain:
\begin{flalign}\label{eqn:MPx4}
    1+z G_{\beta\beta}(z) = \sum_{\alpha} \frac{y_{\alpha,\beta} (y_\alpha^\dagger G^{(\alpha)}(z))_\beta }{1+y_\alpha^\dagger G^{(\alpha)}(z) y_\alpha}= \sum_{\alpha} \frac{y_{\alpha,\beta}y_{\alpha,\gamma}^\dagger G^{(\alpha)}_{\gamma\beta}(z)}{1+ y_\alpha^\dagger G^{(\alpha)}(z)y_\alpha}.
\end{flalign}
We now average this equation over all column vectors:
\begin{flalign}\label{eqn:MPx3}
    1+z \mathbb{E}[G_{\beta\beta}(z)] = \sum_{\alpha}\mathbb{E}\bigg[\mathbb{E}_\alpha\bigg[\frac{y_{\alpha,\beta}y_{\alpha,\gamma}^\dagger G_{\gamma\beta}^{(\alpha)}(z)}{1+ y_\alpha^\dagger G^{(\alpha)}(z)y_\alpha}\bigg]\bigg] .
\end{flalign}
Let us define $q_{\alpha}(z) =  y_\alpha^\dagger G^{(\alpha)}(z)y_\alpha$. The average over $y_\alpha$ on the right-hand side is made difficult by the fact that the denominator explicitly depends on $y_\alpha$. However, it is straightforward to establish:
\begin{flalign}%\label{eqn:EqaS}
    &\mathbb{E}_\alpha[q_\alpha(z)] = \sum_{\beta} S_{\alpha\beta} G^{(\alpha)}_{\beta\beta}(z)  \label{eqn:E-q} 
    \\&
    \text{Var}_\alpha[q_\alpha(z)] \leq (\max_{\beta} S_{\alpha\beta})\sum_{\mu}S_{\alpha\mu} \sum_{\rho} |G_{\mu\rho}^{(\alpha)}(z)|^2. 
\end{flalign}
Using the inequalities \eqref{eqn:ebound1}, \eqref{eqn:Sbound1} and \eqref{eqn:Sbound2}, it is simple to verify that $\text{Var}_{\alpha}[q_{\alpha}(z)]\to 0$ exponentially as $n\to \infty$.  Therefore, up to exponential corrections we can replace the denominator on the right-hand side of \eqref{eqn:MPx3} using \eqref{eqn:E-q} and perform the ensemble average using \eqref{eqn:yavg-S}. The end result is:
\begin{flalign}\label{eqn:MPx6}
    1+z\mathbb{E}[G_{\beta\beta}(z)] \approx \mathbb{E}\bigg[\sum_{\alpha}\frac{S_{\alpha\beta}G_{\beta\beta}^{(\alpha)}(z)}{1+\sum_{\gamma}S_{\alpha\gamma}G_{\gamma\gamma}^{(\alpha)
    }(z)}\bigg].
\end{flalign}
From here, we will describe the calculation in two main steps. 

\paragraph{Step 1. Replacing $\mathbf{G_{\beta\beta}^{(\alpha)}}$ by $\mathbf{G_{\beta\beta}}$.} First, we argue that within the expectation value on the right-hand side of \eqref{eqn:MPx6}, in the large-$n$ limit, one can replace $G_{\beta\beta}^{(\alpha)}(z)$ by $G_{\beta\beta}(z)$. The following identity follows from some straightforward algebra:
\begin{flalign}
\begin{aligned}\label{eqn:MPx7}
    &\frac{G_{\beta\beta}^{(\alpha)}(z)}{1+\sum_{\gamma}S_{\alpha\gamma}G_{\gamma\gamma}^{(\alpha)}(z)} -  \frac{G_{\beta\beta}(z)}{1+\sum_{\gamma}S_{\alpha\gamma}G_{\gamma\gamma}(z)} \\&= 
    \frac{G_{\beta\beta}^{(\alpha)}-G_{\beta\beta}}{1+\sum_{\gamma}S_{\alpha\gamma}G_{\gamma\gamma}^{(\alpha)}(z)} + G_{\beta\beta}
    \frac{\sum_\gamma S_{\alpha\gamma}(G_{\gamma\gamma}-G_{\gamma\gamma}^{(\alpha)})}{(1+\sum_\delta S_{\alpha\delta} G_{\delta\delta}^{(\alpha)})(1+\sum_{\rho}S_{\alpha\rho}G_{\rho\rho})}.
\end{aligned}
\end{flalign}
By taking the expectation value of this equation and using the bounds \eqref{eqn:ebound1}, \eqref{eqn:Sbound1}, and \eqref{eqn:Sbound2} one can show that within any finite spectral window:
\begin{flalign}
    \mathbb{E}\bigg[
    \frac{G_{\beta\beta}^{(\alpha)}(z)}{1+\sum_{\gamma}S_{\alpha\gamma}G_{\gamma\gamma}^{(\alpha)}(z)}\bigg] - \mathbb{E}\bigg[
    \frac{G_{\beta\beta}(z)}{1+\sum_{\gamma}S_{\alpha\gamma}G_{\gamma\gamma}(z)}\bigg] \to 0, \text{ as }n\to\infty,
\end{flalign}
with corrections being exponentially suppressed in $n$. From \eqref{eqn:MPx6} it then follows:
\begin{flalign}\label{eqn:MPx7}
    1+z\mathbb{E}[G_{\beta\beta}(z)] \approx \mathbb{E}\bigg[\sum_{\alpha}\frac{S_{\alpha\beta}G_{\beta\beta}(z)}{1+\sum_{\gamma}S_{\alpha\gamma}G_{\gamma\gamma}(z)}\bigg].
\end{flalign}

\paragraph{Step 2. Replacing $G_{\beta \beta}(z)$ by $\mathbf{\mathbb{E}[G_{\beta\beta}(z)]}$.} We can evaluate the following difference:
\begin{flalign}\begin{aligned}\label{eqn:MPx8}
    &\sum_{\alpha} S_{\alpha\beta}\bigg(\mathbb{E}\bigg[\frac{G_{\beta\beta}}{1+\sum_\gamma S_{\alpha\gamma}G_{\gamma\gamma}}\bigg] - \frac{\mathbb{E}[G_{\beta\beta}]}{1+\sum_{\gamma}S_{\alpha\gamma}\mathbb{E}[G_{\gamma\gamma}]}\bigg) =\\& 
    \sum_{\alpha} S_{\alpha\beta}\bigg(\mathbb{E}\bigg[\frac{G_{\beta\beta}-\mathbb{E}[G_{\beta\beta}]}{1+\sum_\gamma S_{\alpha\gamma}G_{\gamma\gamma}}\bigg] +\mathbb{E}[G_{\beta\beta}] \bigg(
    \mathbb{E}\bigg[\frac{1}{1+\sum_\gamma S_{\alpha\gamma} G_{\gamma\gamma}}\bigg]
    -\frac{1}{1+\sum_{\gamma}S_{\alpha\gamma} \mathbb{E}[G_{\gamma\gamma}]}\bigg)\bigg).
\end{aligned}
\end{flalign}
Using the bounds \eqref{eqn:ebound1}, \eqref{eqn:Sbound1} and \eqref{eqn:Sbound2}, it is straightforward but tedious to show that the sum of terms on the right-hand side becomes exponentially small over any finite spectral window, as $n\to \infty$. Hence from \eqref{eqn:MPx7}, it follows that:
\begin{flalign}\label{eqn:MPx8}
    1+z\mathbb{E}[G_{\beta\beta}(z)] \approx \sum_{\alpha} \frac{S_{\alpha\beta}\,\mathbb{E}[G_{\beta\beta}]}{1+\sum_{\gamma}S_{\alpha\gamma}\mathbb{E}[G_{\gamma\gamma}]}.
\end{flalign}
In this context $\approx$ means up to exponentially small corrections in $n$. At large $n$, this becomes a deterministic equation for $\mathbb{E}[G_{\beta\beta}(z)]$.

\subsection{A set of equations for multi-index shapes}
We will use the notation $g_\beta(z) = \mathbb{E}[G_{\beta\beta}(z)]$. We now rewrite \eqref{eqn:MPx8}, and restore the summation indices as:
\begin{flalign}\label{eqn:MPx9}
    -\frac{1}{g_{\beta}(z)} = z - \sum_{\alpha \in \text{columns}} \frac{S_{\alpha\beta}}{1 +\sum_{\gamma \in \text{rows}} S_{\alpha\gamma}g_{\gamma}(z)}.
\end{flalign}
By ``rows'' and ``columns'' here we mean the corresponding sets of indices of $\theta_v^{(p)}$. The density of states follows from the trace of the resolvent \eqref{eqn:eps-rho-m}:
\begin{flalign}
    \rho(E) = \frac{1}{\pi}\text{Im}\, m(E+i\epsilon) = 
    \frac{1}{\pi d_1} \text{Im}\bigg(\sum_{\beta \in \text{rows}}g_{\beta}(E+i\epsilon)\bigg)_{\text{at fixed }0<\epsilon\ll 1}.
\end{flalign}
The system still appears intractable. We have a large number of coupled equations for equally many unknown functions $g_{\beta}$.  

A key simplification is due to the symmetry of our variance profile $S_{\alpha\beta}$. In particular, it is invariant under the same permutation acting on the input and output multi-index. Given any $\pi\in S_{n+2}$, it is straightforward to see that $S_{\alpha\beta} = S_{\pi\alpha,\pi\beta}$. Hence, the entire ensemble \eqref{eqn:X-def} is invariant under such simultaneous permutations, implying:
\begin{flalign}
    g_{\beta}(z) = \mathbb{E}[G_{\beta\beta}(z)] = \mathbb{E}[G_{\pi\beta,\pi\beta}(z)] = g_{\pi\beta}(z), \qquad \pi \in S_{n+2} .
\end{flalign}
Let us split indices into permutation orbits $[\gamma]=\{\pi \gamma \,|\, \pi\in S_{n+2}\}$. To each orbit $[\gamma]$ we uniquely associate a representative multi-index $\hat{\gamma}$, whose elements are given in descending order. We refer to this representative as the shape of the orbit. We define the multiplicity $N_{[\gamma]}$ of the orbit $[\gamma]$ to be the total number of multi-indices that are related to $\hat{\gamma}$ by a permutation of indices. If $\hat{\gamma}$ has $n_l$ indices of value $l$, then
\begin{flalign}
    N_{[\gamma]} = \frac{(n+2)!}{\prod_{l} n_l !}. 
\end{flalign}
We can now write the reduced set of equations:
\begin{flalign}
    &-\frac{1}{g_{\hat{\beta}}(z)} = z - \sum_{[\alpha]}\sum_{\alpha\in[\alpha]} \frac{S_{\alpha \hat{\beta}}}{1 +\sum_{[\gamma]} g_{\hat{\gamma}}(z)\big(\sum_{\gamma\in[\gamma]}S_{\alpha\gamma}\big)}
    \label{eqn:reduced-1}\\&
    \rho_{\theta\theta^\dagger}(E) = \frac{1}{\pi}\text{Im}\, m(E+i\epsilon) = 
    \frac{1}{\pi d_1} \text{Im}\bigg(\sum_{[\beta]} N_{[\beta]} g_{\hat{\beta}}(E+i\epsilon)\bigg)_{\text{at fixed }0<\epsilon\ll 1}.\label{eqn:reduced-2}
\end{flalign}
Hence, the problem is reduced to solving only a small number of equations. Compared to the total number of rows of $\theta_{v}^{(p)}$, the number of row orbit representatives of $\theta_{v}^{(p)}$ grows very slowly. In the main text, we study some explicit examples, and show that this analytic system closely reproduces the spectra obtained by direct diagonalization.

\bibliographystyle{JHEP}
\bibliography{Draft/quiversbib}

\providecommand{\href}[2]{#2}\begingroup\raggedright\begin{thebibliography}{100}

\bibitem{Strominger:1996sh}
A.~Strominger and C.~Vafa, \emph{{Microscopic origin of the Bekenstein-Hawking entropy}}, \href{https://doi.org/10.1016/0370-2693(96)00345-0}{\emph{Phys. Lett. B} {\bfseries 379} (1996) 99} [\href{https://arxiv.org/abs/hep-th/9601029}{{\ttfamily hep-th/9601029}}].

\bibitem{Maldacena:1997de}
J.M.~Maldacena, A.~Strominger and E.~Witten, \emph{{Black hole entropy in M theory}}, \href{https://doi.org/10.1088/1126-6708/1997/12/002}{\emph{JHEP} {\bfseries 12} (1997) 002} [\href{https://arxiv.org/abs/hep-th/9711053}{{\ttfamily hep-th/9711053}}].

\bibitem{Peet:2000hn}
A.W.~Peet, \emph{{TASI lectures on black holes in string theory}},  in \emph{{Theoretical Advanced Study Institute in Elementary Particle Physics (TASI 99): Strings, Branes, and Gravity}}, pp.~353--433, 8, 2000, \href{https://doi.org/10.1142/9789812799630_0003}{DOI} [\href{https://arxiv.org/abs/hep-th/0008241}{{\ttfamily hep-th/0008241}}].

\bibitem{Cotler:2016fpe}
J.S.~Cotler, G.~Gur-Ari, M.~Hanada, J.~Polchinski, P.~Saad, S.H.~Shenker et~al., \emph{{Black Holes and Random Matrices}}, \href{https://doi.org/10.1007/JHEP05(2017)118}{\emph{JHEP} {\bfseries 05} (2017) 118} [\href{https://arxiv.org/abs/1611.04650}{{\ttfamily 1611.04650}}].

\bibitem{Bohigas1984CharacterizationOC}
O.~Bohigas, M.J.~Giannoni and C.~Schmit, \emph{Characterization of chaotic quantum spectra and universality of level fluctuation laws}, {\emph{Physical Review Letters} {\bfseries 52} (1984) 1}.

\bibitem{Jafferis:2022uhu}
D.L.~Jafferis, D.K.~Kolchmeyer, B.~Mukhametzhanov and J.~Sonner, \emph{{Matrix Models for Eigenstate Thermalization}}, \href{https://doi.org/10.1103/PhysRevX.13.031033}{\emph{Phys. Rev. X} {\bfseries 13} (2023) 031033} [\href{https://arxiv.org/abs/2209.02130}{{\ttfamily 2209.02130}}].

\bibitem{Jafferis:2022wez}
D.L.~Jafferis, D.K.~Kolchmeyer, B.~Mukhametzhanov and J.~Sonner, \emph{{Jackiw-Teitelboim gravity with matter, generalized eigenstate thermalization hypothesis, and random matrices}}, \href{https://doi.org/10.1103/PhysRevD.108.066015}{\emph{Phys. Rev. D} {\bfseries 108} (2023) 066015} [\href{https://arxiv.org/abs/2209.02131}{{\ttfamily 2209.02131}}].

\bibitem{Maldacena:2015waa}
J.~Maldacena, S.H.~Shenker and D.~Stanford, \emph{{A bound on chaos}}, \href{https://doi.org/10.1007/JHEP08(2016)106}{\emph{JHEP} {\bfseries 08} (2016) 106} [\href{https://arxiv.org/abs/1503.01409}{{\ttfamily 1503.01409}}].

\bibitem{Lin:2022rzw}
H.W.~Lin, J.~Maldacena, L.~Rozenberg and J.~Shan, \emph{{Holography for people with no time}}, \href{https://doi.org/10.21468/SciPostPhys.14.6.150}{\emph{SciPost Phys.} {\bfseries 14} (2023) 150} [\href{https://arxiv.org/abs/2207.00407}{{\ttfamily 2207.00407}}].

\bibitem{Lin:2022zxd}
H.W.~Lin, J.~Maldacena, L.~Rozenberg and J.~Shan, \emph{{Looking at supersymmetric black holes for a very long time}}, \href{https://doi.org/10.21468/SciPostPhys.14.5.128}{\emph{SciPost Phys.} {\bfseries 14} (2023) 128} [\href{https://arxiv.org/abs/2207.00408}{{\ttfamily 2207.00408}}].

\bibitem{Chen:2024oqv}
Y.~Chen, H.W.~Lin and S.H.~Shenker, \emph{{BPS chaos}}, \href{https://doi.org/10.21468/SciPostPhys.18.2.072}{\emph{SciPost Phys.} {\bfseries 18} (2025) 072} [\href{https://arxiv.org/abs/2407.19387}{{\ttfamily 2407.19387}}].

\bibitem{Chen:2026vml}
Y.~Chen, S.~Colin-Ellerin, O.~Mamroud and K.~Papadodimas, \emph{{Chaos of Berry curvature for BPS microstates}},  \href{https://arxiv.org/abs/2604.23287}{{\ttfamily 2604.23287}}.

\bibitem{Chang:2024zqi}
C.-M.~Chang and Y.-H.~Lin, \emph{{Holographic covering and the fortuity of black holes}},  \href{https://arxiv.org/abs/2402.10129}{{\ttfamily 2402.10129}}.

\bibitem{Chang:2022mjp}
C.-M.~Chang and Y.-H.~Lin, \emph{{Words to describe a black hole}}, \href{https://doi.org/10.1007/JHEP02(2023)109}{\emph{JHEP} {\bfseries 02} (2023) 109} [\href{https://arxiv.org/abs/2209.06728}{{\ttfamily 2209.06728}}].

\bibitem{Denef:2002ru}
F.~Denef, \emph{{Quantum quivers and Hall / hole halos}}, \href{https://doi.org/10.1088/1126-6708/2002/10/023}{\emph{JHEP} {\bfseries 10} (2002) 023} [\href{https://arxiv.org/abs/hep-th/0206072}{{\ttfamily hep-th/0206072}}].

\bibitem{Bena:2006kb}
I.~Bena, C.-W.~Wang and N.P.~Warner, \emph{{Mergers and typical black hole microstates}}, \href{https://doi.org/10.1088/1126-6708/2006/11/042}{\emph{JHEP} {\bfseries 11} (2006) 042} [\href{https://arxiv.org/abs/hep-th/0608217}{{\ttfamily hep-th/0608217}}].

\bibitem{Denef:2007vg}
F.~Denef and G.W.~Moore, \emph{{Split states, entropy enigmas, holes and halos}}, \href{https://doi.org/10.1007/JHEP11(2011)129}{\emph{JHEP} {\bfseries 11} (2011) 129} [\href{https://arxiv.org/abs/hep-th/0702146}{{\ttfamily hep-th/0702146}}].

\bibitem{Bena:2012hf}
I.~Bena, M.~Berkooz, J.~de~Boer, S.~El-Showk and D.~Van~den Bleeken, \emph{{Scaling BPS Solutions and pure-Higgs States}}, \href{https://doi.org/10.1007/JHEP11(2012)171}{\emph{JHEP} {\bfseries 11} (2012) 171} [\href{https://arxiv.org/abs/1205.5023}{{\ttfamily 1205.5023}}].

\bibitem{Manschot:2012rx}
J.~Manschot, B.~Pioline and A.~Sen, \emph{{From Black Holes to Quivers}}, \href{https://doi.org/10.1007/JHEP11(2012)023}{\emph{JHEP} {\bfseries 11} (2012) 023} [\href{https://arxiv.org/abs/1207.2230}{{\ttfamily 1207.2230}}].

\bibitem{Lee:2012sc}
S.-J.~Lee, Z.-L.~Wang and P.~Yi, \emph{{Quiver Invariants from Intrinsic Higgs States}}, \href{https://doi.org/10.1007/JHEP07(2012)169}{\emph{JHEP} {\bfseries 07} (2012) 169} [\href{https://arxiv.org/abs/1205.6511}{{\ttfamily 1205.6511}}].

\bibitem{deBoer:2008zn}
J.~de~Boer, S.~El-Showk, I.~Messamah and D.~Van~den Bleeken, \emph{{Quantizing N=2 Multicenter Solutions}}, \href{https://doi.org/10.1088/1126-6708/2009/05/002}{\emph{JHEP} {\bfseries 05} (2009) 002} [\href{https://arxiv.org/abs/0807.4556}{{\ttfamily 0807.4556}}].

\bibitem{deBoer:2009un}
J.~de~Boer, S.~El-Showk, I.~Messamah and D.~Van~den Bleeken, \emph{{A Bound on the entropy of supergravity?}}, \href{https://doi.org/10.1007/JHEP02(2010)062}{\emph{JHEP} {\bfseries 02} (2010) 062} [\href{https://arxiv.org/abs/0906.0011}{{\ttfamily 0906.0011}}].

\bibitem{Martinec:2015pfa}
E.J.~Martinec and B.E.~Niehoff, \emph{{Hair-brane Ideas on the Horizon}}, \href{https://doi.org/10.1007/JHEP11(2015)195}{\emph{JHEP} {\bfseries 11} (2015) 195} [\href{https://arxiv.org/abs/1509.00044}{{\ttfamily 1509.00044}}].

\bibitem{Balasubramanian:2006gi}
V.~Balasubramanian, E.G.~Gimon and T.S.~Levi, \emph{{Four Dimensional Black Hole Microstates: From D-branes to Spacetime Foam}}, \href{https://doi.org/10.1088/1126-6708/2008/01/056}{\emph{JHEP} {\bfseries 01} (2008) 056} [\href{https://arxiv.org/abs/hep-th/0606118}{{\ttfamily hep-th/0606118}}].

\bibitem{Gimon:2009gk}
E.G.~Gimon, F.~Larsen and J.~Simon, \emph{{Constituent Model of Extremal non-BPS Black Holes}}, \href{https://doi.org/10.1088/1126-6708/2009/07/052}{\emph{JHEP} {\bfseries 07} (2009) 052} [\href{https://arxiv.org/abs/0903.0719}{{\ttfamily 0903.0719}}].

\bibitem{deBoer:2008ss}
J.~de~Boer, J.~Manschot, K.~Papadodimas and E.~Verlinde, \emph{{The Chiral ring of AdS(3)/CFT(2) and the attractor mechanism}}, \href{https://doi.org/10.1088/1126-6708/2009/03/030}{\emph{JHEP} {\bfseries 03} (2009) 030} [\href{https://arxiv.org/abs/0809.0507}{{\ttfamily 0809.0507}}].

\bibitem{Bena:2011zw}
I.~Bena, B.D.~Chowdhury, J.~de~Boer, S.~El-Showk and M.~Shigemori, \emph{{Moulting Black Holes}}, \href{https://doi.org/10.1007/JHEP03(2012)094}{\emph{JHEP} {\bfseries 03} (2012) 094} [\href{https://arxiv.org/abs/1108.0411}{{\ttfamily 1108.0411}}].

\bibitem{Douglas:1996sw}
M.R.~Douglas and G.W.~Moore, \emph{{D-branes, quivers, and ALE instantons}},  \href{https://arxiv.org/abs/hep-th/9603167}{{\ttfamily hep-th/9603167}}.

\bibitem{Shi:2026rbr}
X.~Shi and G.J.~Turiaci, \emph{{BPS black holes in flat space from quantum mechanics}},  \href{https://arxiv.org/abs/2610.18951}{{\ttfamily 2610.18951}}.

\bibitem{Anninos:2016szt}
D.~Anninos, T.~Anous and F.~Denef, \emph{{Disordered Quivers and Cold Horizons}}, \href{https://doi.org/10.1007/JHEP12(2016)071}{\emph{JHEP} {\bfseries 12} (2016) 071} [\href{https://arxiv.org/abs/1603.00453}{{\ttfamily 1603.00453}}].

\bibitem{Chryssanthacopoulos:2026mni}
J.~Chryssanthacopoulos and D.~Vegh, \emph{{Fortuity and fragility in supersymmetric SYK}},  \href{https://arxiv.org/abs/2608.12160}{{\ttfamily 2608.12160}}.

\bibitem{Choi:2022caq}
S.~Choi, S.~Kim, E.~Lee and J.~Park, \emph{{The shape of non-graviton operators for SU(2)}}, \href{https://doi.org/10.1007/JHEP09(2024)029}{\emph{JHEP} {\bfseries 09} (2024) 029} [\href{https://arxiv.org/abs/2209.12696}{{\ttfamily 2209.12696}}].

\bibitem{Kim:2026mch}
S.~Kim, J.~Lee, S.~Lee and J.~Song, \emph{{Black hole cohomologies in N=4 SYM}},  \href{https://arxiv.org/abs/2610.20340}{{\ttfamily 2610.20340}}.

\bibitem{Kibe:2026qid}
T.~Kibe, S.~van Leuven and K.~Mathieson, \emph{{Non-Coulomb $Q$-Cohomology and Quantum Lifting}},  \href{https://arxiv.org/abs/2610.22332}{{\ttfamily 2610.22332}}.

\bibitem{Benjamin:2026fortuity}
N.~Benjamin, A.~Guerrieri, V.A.~Rodriguez, D.~Simmons-Duffin, Y.~Wang and Z.~Zheng, ``{Fortuity: fortuitous BPS states of $\mathcal{N}=4$ SYM at weak coupling}.'' \url{https://github.com/davidsd/fortuity-ai}, Oct., 2026.

\bibitem{Belin:2025hsg}
A.~Belin, P.~Singh, R.~Vadala and A.~Zaffaroni, \emph{{Fortuity in ABJM}}, \href{https://doi.org/10.1007/JHEP08(2026)028}{\emph{JHEP} {\bfseries 08} (2026) 028} [\href{https://arxiv.org/abs/2512.04146}{{\ttfamily 2512.04146}}].

\bibitem{Behan:2025hbx}
C.~Behan and L.P.~de~Gioia, \emph{{Two roads to fortuity in ABJM theory}}, \href{https://doi.org/10.1007/JHEP06(2026)217}{\emph{JHEP} {\bfseries 06} (2026) 217} [\href{https://arxiv.org/abs/2512.23603}{{\ttfamily 2512.23603}}].

\bibitem{Choi:2025pqr}
J.~Choi and S.~Kim, \emph{{Fortuity and relevant deformation}}, \href{https://doi.org/10.1007/JHEP07(2026)097}{\emph{JHEP} {\bfseries 07} (2026) 097} [\href{https://arxiv.org/abs/2512.12674}{{\ttfamily 2512.12674}}].

\bibitem{Choi:2026faq}
J.~Choi, S.~Choi and S.~Kim, \emph{{Fortuity on the conifold}},  \href{https://arxiv.org/abs/2609.19297}{{\ttfamily 2609.19297}}.

\bibitem{deMelloKoch:2025cec}
R.~de~Mello~Koch, A.~Ghosh and H.J.R.~Van~Zyl, \emph{{Bosonic fortuity in vector models}}, \href{https://doi.org/10.1007/JHEP06(2025)246}{\emph{JHEP} {\bfseries 06} (2025) 246} [\href{https://arxiv.org/abs/2504.14181}{{\ttfamily 2504.14181}}].

\bibitem{Fliss:2026rad}
J.R.~Fliss, V.~Jejjala and O.~Parrikar, \emph{{Fortuity and complexity in a simple quark model}}, \href{https://doi.org/10.1007/JHEP09(2026)040}{\emph{JHEP} {\bfseries 09} (2026) 040} [\href{https://arxiv.org/abs/2605.16254}{{\ttfamily 2605.16254}}].

\bibitem{Kim:2025vup}
S.~Kim, J.~Lee, S.~Lee and H.~Oh, \emph{{BPS phases and fortuity in higher spin holography}}, \href{https://doi.org/10.1007/JHEP07(2026)218}{\emph{JHEP} {\bfseries 07} (2026) 218} [\href{https://arxiv.org/abs/2511.03105}{{\ttfamily 2511.03105}}].

\bibitem{Chen:2025sum}
Y.~Chen, \emph{{Fortuity with a single matrix}},  \href{https://arxiv.org/abs/2511.00790}{{\ttfamily 2511.00790}}.

\bibitem{Chae:2026vgn}
J.-S.~Chae, \emph{{Microscopic Organization of BPS Black-Hole States in the D1--D5--P CFT}},  \href{https://arxiv.org/abs/2609.35688}{{\ttfamily 2609.35688}}.

\bibitem{Chang:2025rqy}
C.-M.~Chang, Y.-H.~Lin and H.~Zhang, \emph{{Fortuity in the D1-D5 system}},  \href{https://arxiv.org/abs/2501.05448}{{\ttfamily 2501.05448}}.

\bibitem{Hughes:2025tdy}
M.R.R.~Hughes and M.~Shigemori, \emph{{Fortuity and supergravity}}, \href{https://doi.org/10.1007/JHEP03(2026)130}{\emph{JHEP} {\bfseries 03} (2026) 130} [\href{https://arxiv.org/abs/2505.14888}{{\ttfamily 2505.14888}}].

\bibitem{Djukic:2026ejk}
V.~Djuki{\'c}, M.~Stepanovi{\'c} and M.~{\v{C}}ubrovi{\'c}, \emph{{Berry Picking: Random Wave Chaos Hierarchy for BPS Microstate Geometries}},  \href{https://arxiv.org/abs/2607.03434}{{\ttfamily 2607.03434}}.

\bibitem{Djukic:2026phb}
V.~Djuki{\'c} and M.~{\v{C}}ubrovi{\'c}, \emph{{Random wave chaos on momentum-fractionated microstate geometries}},  \href{https://arxiv.org/abs/2610.15918}{{\ttfamily 2610.15918}}.

\bibitem{Cecotti:1991me}
S.~Cecotti and C.~Vafa, \emph{{Topological antitopological fusion}}, \href{https://doi.org/10.1016/0550-3213(91)90021-O}{\emph{Nucl. Phys. B} {\bfseries 367} (1991) 359}.

\bibitem{Papadodimas:2009eu}
K.~Papadodimas, \emph{{Topological Anti-Topological Fusion in Four-Dimensional Superconformal Field Theories}}, \href{https://doi.org/10.1007/JHEP08(2010)118}{\emph{JHEP} {\bfseries 08} (2010) 118} [\href{https://arxiv.org/abs/0910.4963}{{\ttfamily 0910.4963}}].

\bibitem{Baggio:2017aww}
M.~Baggio, V.~Niarchos and K.~Papadodimas, \emph{{Aspects of Berry phase in QFT}}, \href{https://doi.org/10.1007/JHEP04(2017)062}{\emph{JHEP} {\bfseries 04} (2017) 062} [\href{https://arxiv.org/abs/1701.05587}{{\ttfamily 1701.05587}}].

\bibitem{future-black-hole-paper}
M.~Berkooz, T.~Kukolj, O.~Mamroud and J.~Seitz. To appear.

\bibitem{Alt_2017}
J.~Alt, L.~Erdős and T.~Krüger, \emph{Local law for random gram matrices}, \href{https://doi.org/10.1214/17-ejp42}{\emph{Electronic Journal of Probability} {\bfseries 22} (2017) }.

\bibitem{hachem2005empiricaldistributioneigenvaluesgram}
W.~Hachem, P.~Loubaton and J.~Najim, \emph{The empirical distribution of the eigenvalues of a gram matrix with a given variance profile},  2005.

\bibitem{Hachem_2007}
W.~Hachem, P.~Loubaton and J.~Najim, \emph{Deterministic equivalents for certain functionals of large random matrices}, \href{https://doi.org/10.1214/105051606000000925}{\emph{The Annals of Applied Probability} {\bfseries 17} (2007) }.

\bibitem{kota2014embedded}
V.K.B.~Kota, \emph{Embedded Random Matrix Ensembles in Quantum Physics}, vol.~884 of \emph{Lecture Notes in Physics}, Springer Cham (2014), \href{https://doi.org/10.1007/978-3-319-04567-2}{10.1007/978-3-319-04567-2}.

\bibitem{Benet:2002br}
L.~Benet and H.A.~Weidenmuller, \emph{{Review of the k body embedded ensembles of Gaussian random matrices}}, \href{https://doi.org/10.1088/0305-4470/36/12/340}{\emph{J. Phys. A} {\bfseries 36} (2003) 3569} [\href{https://arxiv.org/abs/cond-mat/0207656}{{\ttfamily cond-mat/0207656}}].

\bibitem{erdHos2014phase}
L.~Erd\H{o}s and D.~Schr\"oder, \emph{{Phase Transition in the Density of States of Quantum Spin Glasses}}, \href{https://doi.org/10.1007/s11040-014-9164-3}{\emph{Math. Phys. Anal. Geom.} {\bfseries 17} (2014) 441} [\href{https://arxiv.org/abs/1407.1552}{{\ttfamily 1407.1552}}].

\bibitem{Berkooz:2018qkz}
M.~Berkooz, P.~Narayan and J.~Simon, \emph{{Chord diagrams, exact correlators in spin glasses and black hole bulk reconstruction}}, \href{https://doi.org/10.1007/JHEP08(2018)192}{\emph{JHEP} {\bfseries 08} (2018) 192} [\href{https://arxiv.org/abs/1806.04380}{{\ttfamily 1806.04380}}].

\bibitem{Berkooz:2018jqr}
M.~Berkooz, M.~Isachenkov, V.~Narovlansky and G.~Torrents, \emph{{Towards a full solution of the large N double-scaled SYK model}}, \href{https://doi.org/10.1007/JHEP03(2019)079}{\emph{JHEP} {\bfseries 03} (2019) 079} [\href{https://arxiv.org/abs/1811.02584}{{\ttfamily 1811.02584}}].

\bibitem{Berkooz:2024lgq}
M.~Berkooz and O.~Mamroud, \emph{{A cordial introduction to double scaled SYK}}, \href{https://doi.org/10.1088/1361-6633/ada889}{\emph{Rept. Prog. Phys.} {\bfseries 88} (2025) 036001} [\href{https://arxiv.org/abs/2407.09396}{{\ttfamily 2407.09396}}].

\bibitem{Belin:2026ssc}
A.~Belin, Y.~Fu and F.~La~Rocca, \emph{{Microcanonical Operator Spectrum, BPS Chaos and Free Probability Theory}},  \href{https://arxiv.org/abs/2609.22422}{{\ttfamily 2609.22422}}.

\bibitem{Gharibyan:2018jrp}
H.~Gharibyan, M.~Hanada, S.H.~Shenker and M.~Tezuka, \emph{{Onset of Random Matrix Behavior in Scrambling Systems}}, \href{https://doi.org/10.1007/JHEP07(2018)124}{\emph{JHEP} {\bfseries 07} (2018) 124} [\href{https://arxiv.org/abs/1803.08050}{{\ttfamily 1803.08050}}].

\bibitem{Berkooz:2020fvm}
M.~Berkooz, N.~Brukner, V.~Narovlansky and A.~Raz, \emph{{Multi-trace correlators in the SYK model and non-geometric wormholes}}, \href{https://doi.org/10.1007/JHEP09(2021)196}{\emph{JHEP} {\bfseries 09} (2021) 196} [\href{https://arxiv.org/abs/2104.03336}{{\ttfamily 2104.03336}}].

\bibitem{Jia:2019orl}
Y.~Jia and J.J.M.~Verbaarschot, \emph{{Spectral Fluctuations in the Sachdev-Ye-Kitaev Model}}, \href{https://doi.org/10.1007/JHEP07(2020)193}{\emph{JHEP} {\bfseries 07} (2020) 193} [\href{https://arxiv.org/abs/1912.11923}{{\ttfamily 1912.11923}}].

\bibitem{Caceres:2022kyr}
E.~C{\'a}ceres, A.~Misobuchi and A.~Raz, \emph{{Spectral form factor in sparse SYK models}}, \href{https://doi.org/10.1007/JHEP08(2022)236}{\emph{JHEP} {\bfseries 08} (2022) 236} [\href{https://arxiv.org/abs/2204.07194}{{\ttfamily 2204.07194}}].

\bibitem{Altland:2017eao}
A.~Altland and D.~Bagrets, \emph{{Quantum ergodicity in the SYK model}}, \href{https://doi.org/10.1016/j.nuclphysb.2018.02.015}{\emph{Nucl. Phys. B} {\bfseries 930} (2018) 45} [\href{https://arxiv.org/abs/1712.05073}{{\ttfamily 1712.05073}}].

\bibitem{Altland:2024ubs}
A.~Altland, K.W.~Kim, T.~Micklitz, M.~Rezaei, J.~Sonner and J.J.M.~Verbaarschot, \emph{{Quantum chaos on edge}}, \href{https://doi.org/10.1103/PhysRevResearch.6.033286}{\emph{Phys. Rev. Res.} {\bfseries 6} (2024) 033286} [\href{https://arxiv.org/abs/2403.13516}{{\ttfamily 2403.13516}}].

\bibitem{Becker:1995kb}
K.~Becker, M.~Becker and A.~Strominger, \emph{{Five-branes, membranes and nonperturbative string theory}}, \href{https://doi.org/10.1016/0550-3213(95)00487-1}{\emph{Nucl. Phys. B} {\bfseries 456} (1995) 130} [\href{https://arxiv.org/abs/hep-th/9507158}{{\ttfamily hep-th/9507158}}].

\bibitem{Denef:2000nb}
F.~Denef, \emph{{Supergravity flows and D-brane stability}}, \href{https://doi.org/10.1088/1126-6708/2000/08/050}{\emph{JHEP} {\bfseries 08} (2000) 050} [\href{https://arxiv.org/abs/hep-th/0005049}{{\ttfamily hep-th/0005049}}].

\bibitem{Bena:2007qc}
I.~Bena, C.-W.~Wang and N.P.~Warner, \emph{{Plumbing the Abyss: Black ring microstates}}, \href{https://doi.org/10.1088/1126-6708/2008/07/019}{\emph{JHEP} {\bfseries 07} (2008) 019} [\href{https://arxiv.org/abs/0706.3786}{{\ttfamily 0706.3786}}].

\bibitem{Douglas:2002fr}
M.R.~Douglas, S.~Govindarajan, T.~Jayaraman and A.~Tomasiello, \emph{{D branes on Calabi-Yau manifolds and superpotentials}}, \href{https://doi.org/10.1007/s00220-004-1091-x}{\emph{Commun. Math. Phys.} {\bfseries 248} (2004) 85} [\href{https://arxiv.org/abs/hep-th/0203173}{{\ttfamily hep-th/0203173}}].

\bibitem{Douglas:1996yp}
M.R.~Douglas, D.N.~Kabat, P.~Pouliot and S.H.~Shenker, \emph{{D-branes and short distances in string theory}}, \href{https://doi.org/10.1016/S0550-3213(96)00619-0}{\emph{Nucl. Phys. B} {\bfseries 485} (1997) 85} [\href{https://arxiv.org/abs/hep-th/9608024}{{\ttfamily hep-th/9608024}}].

\bibitem{Diaconescu:1997ut}
D.-E.~Diaconescu and R.~Entin, \emph{{A Nonrenormalization theorem for the d = 1, N=8 vector multiplet}}, \href{https://doi.org/10.1103/PhysRevD.56.8045}{\emph{Phys. Rev. D} {\bfseries 56} (1997) 8045} [\href{https://arxiv.org/abs/hep-th/9706059}{{\ttfamily hep-th/9706059}}].

\bibitem{Chang:2024lxt}
C.-M.~Chang, Y.~Chen, B.S.~Sia and Z.~Yang, \emph{{Fortuity in SYK models}}, \href{https://doi.org/10.1007/JHEP08(2025)003}{\emph{JHEP} {\bfseries 08} (2025) 003} [\href{https://arxiv.org/abs/2412.06902}{{\ttfamily 2412.06902}}].

\bibitem{Stanford:2017thb}
D.~Stanford and E.~Witten, \emph{{Fermionic Localization of the Schwarzian Theory}}, \href{https://doi.org/10.1007/JHEP10(2017)008}{\emph{JHEP} {\bfseries 10} (2017) 008} [\href{https://arxiv.org/abs/1703.04612}{{\ttfamily 1703.04612}}].

\bibitem{Heydeman:2020hhw}
M.~Heydeman, L.V.~Iliesiu, G.J.~Turiaci and W.~Zhao, \emph{{The statistical mechanics of near-BPS black holes}}, \href{https://doi.org/10.1088/1751-8121/ac3be9}{\emph{J. Phys. A} {\bfseries 55} (2022) 014004} [\href{https://arxiv.org/abs/2011.01953}{{\ttfamily 2011.01953}}].

\bibitem{Ferrara:1995ih}
S.~Ferrara, R.~Kallosh and A.~Strominger, \emph{{N=2 extremal black holes}}, \href{https://doi.org/10.1103/PhysRevD.52.R5412}{\emph{Phys. Rev. D} {\bfseries 52} (1995) R5412} [\href{https://arxiv.org/abs/hep-th/9508072}{{\ttfamily hep-th/9508072}}].

\bibitem{Strominger:1996kf}
A.~Strominger, \emph{{Macroscopic entropy of N=2 extremal black holes}}, \href{https://doi.org/10.1016/0370-2693(96)00711-3}{\emph{Phys. Lett. B} {\bfseries 383} (1996) 39} [\href{https://arxiv.org/abs/hep-th/9602111}{{\ttfamily hep-th/9602111}}].

\bibitem{Ferrara:1996dd}
S.~Ferrara and R.~Kallosh, \emph{{Supersymmetry and attractors}}, \href{https://doi.org/10.1103/PhysRevD.54.1514}{\emph{Phys. Rev. D} {\bfseries 54} (1996) 1514} [\href{https://arxiv.org/abs/hep-th/9602136}{{\ttfamily hep-th/9602136}}].

\bibitem{Witten:1993yc}
E.~Witten, \emph{{Phases of N=2 theories in two-dimensions}}, \href{https://doi.org/10.1016/0550-3213(93)90033-L}{\emph{Nucl. Phys. B} {\bfseries 403} (1993) 159} [\href{https://arxiv.org/abs/hep-th/9301042}{{\ttfamily hep-th/9301042}}].

\bibitem{Gerchkovitz:2014gta}
E.~Gerchkovitz, J.~Gomis and Z.~Komargodski, \emph{{Sphere Partition Functions and the Zamolodchikov Metric}}, \href{https://doi.org/10.1007/JHEP11(2014)001}{\emph{JHEP} {\bfseries 11} (2014) 001} [\href{https://arxiv.org/abs/1405.7271}{{\ttfamily 1405.7271}}].

\bibitem{Ishtiaque:2017trm}
N.~Ishtiaque, \emph{{2D BPS Rings from Sphere Partition Functions}}, \href{https://doi.org/10.1007/JHEP04(2018)124}{\emph{JHEP} {\bfseries 04} (2018) 124} [\href{https://arxiv.org/abs/1712.02551}{{\ttfamily 1712.02551}}].

\bibitem{Chen:2017fvl}
J.~Chen, \emph{{On exact correlation functions of chiral ring operators in $2d$ $\mathcal{N}=(2, 2)$ SCFTs via localization}}, \href{https://doi.org/10.1007/JHEP03(2018)065}{\emph{JHEP} {\bfseries 03} (2018) 065} [\href{https://arxiv.org/abs/1712.01164}{{\ttfamily 1712.01164}}].

\bibitem{hertling2002ttgeometryfrobeniusmanifolds}
C.~Hertling, \emph{{tt* geometry, Frobenius manifolds, their connections, and the construction for singularities}},  2002.

\bibitem{hertling2008nilpotentorbitsgeneralizationhodge}
C.~Hertling and C.~Sevenheck, \emph{{Nilpotent orbits of a generalization of Hodge structures}},  2008.

\bibitem{Tang_2025}
X.~Tang and J.~Yan, \emph{{CY/LG correspondence for Weil-Petersson metrics and tt⁎ structures}}, \href{https://doi.org/10.1016/j.geomphys.2025.105545}{\emph{Journal of Geometry and Physics} {\bfseries 216} (2025) 105545}.

\bibitem{Green:1987mn}
M.B.~Green, J.H.~Schwarz and E.~Witten, \emph{{SUPERSTRING THEORY. VOL. 2: LOOP AMPLITUDES, ANOMALIES AND PHENOMENOLOGY}} (7, 1988).

\bibitem{Witten:1982im}
E.~Witten, \emph{{Supersymmetry and Morse theory}}, {\emph{J. Diff. Geom.} {\bfseries 17} (1982) 661}.

\bibitem{Hori:2003ic}
K.~Hori, S.~Katz, A.~Klemm, R.~Pandharipande, R.~Thomas, C.~Vafa et~al., \emph{{Mirror symmetry}}, vol.~1 of \emph{Clay mathematics monographs}, AMS, Providence, USA (2003).

\bibitem{Hirzebruch1966}
F.~Hirzebruch, \emph{Topological Methods in Algebraic Geometry}, vol.~131 of \emph{Grundlehren der mathematischen Wissenschaften}, Springer-Verlag, Berlin, Heidelberg, 3rd~ed. (1966).

\bibitem{Lee:2012naa}
S.-J.~Lee, Z.-L.~Wang and P.~Yi, \emph{{BPS States, Refined Indices, and Quiver Invariants}}, \href{https://doi.org/10.1007/JHEP10(2012)094}{\emph{JHEP} {\bfseries 10} (2012) 094} [\href{https://arxiv.org/abs/1207.0821}{{\ttfamily 1207.0821}}].

\bibitem{Green:1987rw}
P.~Green and T.~Hubsch, \emph{{Polynomial Deformations and Cohomology of Calabi-yau Manifolds}}, \href{https://doi.org/10.1007/BF01221257}{\emph{Commun. Math. Phys.} {\bfseries 113} (1987) 505}.

\bibitem{Anderson:2026gvz}
L.B.~Anderson, J.~Gray, E.~Sharpe and A.H.~Zhang, \emph{{GLSMs for non-polynomial deformations and generalized complete intersection Calabi-Yau manifolds}},  \href{https://arxiv.org/abs/2610.12554}{{\ttfamily 2610.12554}}.

\bibitem{Chang:2023gow}
C.-M.~Chang and X.~Shen, \emph{{Disordered $\mathcal{N} = (2, 2)$ supersymmetric field theories}}, \href{https://doi.org/10.21468/SciPostPhys.16.5.140}{\emph{SciPost Phys.} {\bfseries 16} (2024) 140} [\href{https://arxiv.org/abs/2307.08742}{{\ttfamily 2307.08742}}].

\bibitem{Lerche:1989uy}
W.~Lerche, C.~Vafa and N.P.~Warner, \emph{{Chiral Rings in N=2 Superconformal Theories}}, \href{https://doi.org/10.1016/0550-3213(89)90474-4}{\emph{Nucl. Phys. B} {\bfseries 324} (1989) 427}.

\bibitem{Witten:1991zz}
E.~Witten, \emph{{Mirror manifolds and topological field theory}}, {\emph{AMS/IP Stud. Adv. Math.} {\bfseries 9} (1998) 121} [\href{https://arxiv.org/abs/hep-th/9112056}{{\ttfamily hep-th/9112056}}].

\bibitem{ffab916b-4326-37f5-aaf1-66df46a74bac}
P.A.~Griffiths, \emph{On the periods of certain rational integrals: I}, {\emph{Annals of Mathematics} {\bfseries 90} (1969) 460}.

\bibitem{cb71e49b-28cb-3d4d-a302-c5ff3b25456b}
P.A.~Griffiths, \emph{On the periods of certain rational integrals: Ii}, {\emph{Annals of Mathematics} {\bfseries 90} (1969) 496}.

\bibitem{Carlson:1980}
J.A.~Carlson and P.A.~Griffiths, \emph{Infinitesimal variations of {H}odge structure and the global {T}orelli problem},  in \emph{Journ\'ees de G\'eom\'etrie Alg\'ebrique d'Angers, Juillet 1979 / Algebraic Geometry, Angers, 1979}, A.~Beauville, ed., (Alphen aan den Rijn), pp.~51--76, Sijthoff \& Noordhoff (1980).

\bibitem{griffiths1978principles}
P.~Griffiths and J.~Harris, \emph{Principles of algebraic geometry}, Pure and Applied Mathematics. A Wiley-Interscience Publication., John Wiley \& Sons, New York (1978).

\bibitem{Cecotti:1990kz}
S.~Cecotti, \emph{{N=2 Landau-Ginzburg versus Calabi-Yau sigma models: Nonperturbative aspects}}, \href{https://doi.org/10.1142/S0217751X91000939}{\emph{Int. J. Mod. Phys. A} {\bfseries 6} (1991) 1749}.

\bibitem{Vafa:1990mu}
C.~Vafa, \emph{{Topological Landau-Ginzburg models}}, \href{https://doi.org/10.1142/S0217732391000324}{\emph{Mod. Phys. Lett. A} {\bfseries 6} (1991) 337}.

\bibitem{Intriligator:1990ua}
K.A.~Intriligator and C.~Vafa, \emph{{LANDAU-GINZBURG ORBIFOLDS}}, \href{https://doi.org/10.1016/0550-3213(90)90535-L}{\emph{Nucl. Phys. B} {\bfseries 339} (1990) 95}.

\bibitem{Herbst:2004ax}
M.~Herbst and C.-I.~Lazaroiu, \emph{{Localization and traces in open-closed topological Landau-Ginzburg models}}, \href{https://doi.org/10.1088/1126-6708/2005/05/044}{\emph{JHEP} {\bfseries 05} (2005) 044} [\href{https://arxiv.org/abs/hep-th/0404184}{{\ttfamily hep-th/0404184}}].

\bibitem{Gaiotto:2013rk}
D.~Gaiotto and J.~Lamy-Poirier, \emph{{Irregular Singularities in the $H_3^+$ WZW Model}},  \href{https://arxiv.org/abs/1301.5342}{{\ttfamily 1301.5342}}.

\bibitem{Gerchkovitz:2016gxx}
E.~Gerchkovitz, J.~Gomis, N.~Ishtiaque, A.~Karasik, Z.~Komargodski and S.S.~Pufu, \emph{{Correlation Functions of Coulomb Branch Operators}}, \href{https://doi.org/10.1007/JHEP01(2017)103}{\emph{JHEP} {\bfseries 01} (2017) 103} [\href{https://arxiv.org/abs/1602.05971}{{\ttfamily 1602.05971}}].

\bibitem{Gepner:1987qi}
D.~Gepner, \emph{{Space-Time Supersymmetry in Compactified String Theory and Superconformal Models}}, \href{https://doi.org/10.1016/0550-3213(88)90397-5}{\emph{Nucl. Phys. B} {\bfseries 296} (1988) 757}.

\bibitem{Gepner:1987vz}
D.~Gepner, \emph{{Exactly Solvable String Compactifications on Manifolds of SU(N) Holonomy}}, \href{https://doi.org/10.1016/0370-2693(87)90938-5}{\emph{Phys. Lett. B} {\bfseries 199} (1987) 380}.

\bibitem{Berkooz:2021ehv}
M.~Berkooz, A.~Sharon, N.~Silberstein and E.Y.~Urbach, \emph{{Onset of quantum chaos in disordered CFTs}}, \href{https://doi.org/10.1103/PhysRevD.106.045007}{\emph{Phys. Rev. D} {\bfseries 106} (2022) 045007} [\href{https://arxiv.org/abs/2111.06108}{{\ttfamily 2111.06108}}].

\bibitem{Berkooz:2022dfr}
M.~Berkooz, A.~Sharon, N.~Silberstein and E.Y.~Urbach, \emph{{Onset of Quantum Chaos in Random Field Theories}}, \href{https://doi.org/10.1103/PhysRevLett.129.071601}{\emph{Phys. Rev. Lett.} {\bfseries 129} (2022) 071601} [\href{https://arxiv.org/abs/2207.11980}{{\ttfamily 2207.11980}}].

\bibitem{Oganesyan_2007}
V.~Oganesyan and D.A.~Huse, \emph{Localization of interacting fermions at high temperature}, \href{https://doi.org/10.1103/physrevb.75.155111}{\emph{Physical Review B} {\bfseries 75} (2007) }.

\bibitem{Atas_2013}
Y.Y.~Atas, E.~Bogomolny, O.~Giraud and G.~Roux, \emph{Distribution of the ratio of consecutive level spacings in random matrix ensembles}, \href{https://doi.org/10.1103/physrevlett.110.084101}{\emph{Physical Review Letters} {\bfseries 110} (2013) }.

\bibitem{Meh2004}
M.L.~Mehta, \emph{Random Matrices}, 3rd~ed. (2004).

\bibitem{Raz:2025wjw}
A.~Raz and M.~Youssef, \emph{{The late time ramp from chord diagrams in the double-scaled SYK model}}, \href{https://doi.org/10.1007/JHEP06(2026)215}{\emph{JHEP} {\bfseries 06} (2026) 215} [\href{https://arxiv.org/abs/2507.12543}{{\ttfamily 2507.12543}}].

\bibitem{Pastur:1967zca}
L.A.~Pastur and V.A.~Mar{\v{c}}enko, \emph{{Distribution of Eigenvalues for Some Sets of Random Matrices}}, \href{https://doi.org/10.1070/SM1967v001n04ABEH001994}{\emph{Math. USSR Sb.} {\bfseries 1} (1967) 457}.

\bibitem{1991NuPhB.355..455C}
P.~{Candelas} and X.C.~{de la Ossa}, \emph{{Moduli space of Calabi-Yau manifolds}}, \href{https://doi.org/10.1016/0550-3213(91)90122-E}{\emph{Nuclear Physics B} {\bfseries 355} (1991) 455}.

\bibitem{Voisin2002}
C.~Voisin, \emph{Hodge Theory and Complex Algebraic Geometry I}, vol.~76 of \emph{Cambridge Studies in Advanced Mathematics}, Cambridge University Press, Cambridge (2002).

\bibitem{Voisin2003}
C.~Voisin, \emph{Hodge Theory and Complex Algebraic Geometry II}, vol.~77 of \emph{Cambridge Studies in Advanced Mathematics}, Cambridge University Press, Cambridge (2003).

\bibitem{MorrowKodaira1971}
J.~Morrow and K.~Kodaira, \emph{Complex Manifolds}, Holt, Rinehart and Winston, New York (1971).

\bibitem{Kodaira1986}
K.~Kodaira, \emph{Complex Manifolds and Deformation of Complex Structures}, vol.~283 of \emph{Grundlehren der mathematischen Wissenschaften}, Springer-Verlag, New York (1986).

\bibitem{Greene:1996cy}
B.R.~Greene, \emph{{String theory on Calabi-Yau manifolds}},  in \emph{{Theoretical Advanced Study Institute in Elementary Particle Physics (TASI 96): Fields, Strings, and Duality}}, pp.~543--726, 6, 1996 [\href{https://arxiv.org/abs/hep-th/9702155}{{\ttfamily hep-th/9702155}}].

\bibitem{Hubsch:1992nu}
T.~Hubsch, \emph{{Calabi-Yau manifolds: A Bestiary for physicists}}, World Scientific, Singapore (1994).

\bibitem{Bershadsky:1993cx}
M.~Bershadsky, S.~Cecotti, H.~Ooguri and C.~Vafa, \emph{{Kodaira-Spencer theory of gravity and exact results for quantum string amplitudes}}, \href{https://doi.org/10.1007/BF02099774}{\emph{Commun. Math. Phys.} {\bfseries 165} (1994) 311} [\href{https://arxiv.org/abs/hep-th/9309140}{{\ttfamily hep-th/9309140}}].

\bibitem{Wishart1928THEGP}
J.~Wishart, \emph{The generalised product moment distribution in samples from a normal multivariate population}, {\emph{Biometrika} (1928) 32}.

\bibitem{8a873437-53d6-348d-9b7a-19c7327502a1}
P.~Forrester, \emph{Log-Gases and Random Matrices (LMS-34)}, Princeton University Press (2010).

\end{thebibliography}\endgroup

\end{document}